\documentclass[paper, 12pt, letterpaper]{JHEP3}
\usepackage{amsmath}
\usepackage{amssymb}
\usepackage{amsfonts}
\usepackage{mathrsfs}
\usepackage{bbm}
\usepackage{bm}

\usepackage{epsfig,rotating}

\usepackage{longtable,booktabs}

\def\Bbb{\mathbb} \def\C{{\Bbb C}} \def\R{{\Bbb R}} \def\Z{{\Bbb Z}}
  \def\L{{\Bbb L}} \def\P{{\Bbb P}}

\def\bbeta{{\boldsymbol{\beta}}}

\def\p{{\bf p}}

\def\vint{{\int \hspace{-0.47cm} \int}}
\def\Rep{{\rm Rep}}

\def\N{\Bbb{N}}

\def\Hom{\operatorname{Hom}} 
 \def\Spec{\operatorname{Spec}}
  \def\vol {{\rm
    vol}}

\def\Pic{\operatorname{Pic}}

\def\tr{{\rm tr}}
\def\id{{\rm id}}
\def\ker{{\rm ker}}  
\def\im{{\rm im\,}}  
  \def\dim{{\rm dim}}
  
   \def\deg{{\rm deg}}  
  
\def\End{{\rm End}} \def\Aut{{\rm Aut}} 
\def\Spec{{\rm Spec}}

\def\2Cat{{\rm 2Cat}}

\newcommand{\beq}{\begin{equation}} \newcommand{\eeq}{\end{equation}}
\newcommand{\bea}{\begin{eqnarray}} \newcommand{\eea}{\end{eqnarray}}
\newcommand{\beann}{\begin{eqnarray*}}
  \newcommand{\eeann}{\end{eqnarray*}}
\newcommand{\bfig}{\begin{figure}} \newcommand{\efig}{\end{figure}}
\newcommand{\nn}{\nonumber}
\newcommand{\ba}{\begin{array}}\newcommand{\ea}{\end{array}}
\newcommand{\CC}{\mathcal{C}}
\newcommand{\CL}{\mathcal{L}}
\newcommand{\CI}{\mathcal{I}}

\newcommand{\dpar}{\partial}                           
\newcommand{\der}[1]{\frac{\dpar}{\dpar #1}}
\newcommand{\bz}{{\bar{z}}}
\newcommand{\embd}{{\hookrightarrow}}
\newcommand{\bj}{{\bar{\jmath}}}

\newcommand{\derr}[2]{\frac{\dpar #1}{\dpar #2}}        

\title{Generalized Berezin quantization, Bergman metrics and fuzzy Laplacians}

\author{~~~Calin Iuliu Lazaroiu, Daniel McNamee and Christian S{\"a}mann
\\Trinity College Dublin\\
Dublin 2, Ireland\\
calin,~danmc,~saemann@maths.tcd.ie}

\abstract{We study extended Berezin and Berezin-Toeplitz quantization for
  compact K{\"a}hler manifolds, two related quantization procedures which provide
  a general framework for approaching the construction of fuzzy compact K{\"a}hler
  geometries. Using this framework, we show that a particular version of generalized
  Berezin quantization, which we baptize ``Berezin-Bergman quantization'',
  reproduces recent proposals for the construction of fuzzy K{\"a}hler spaces. We
  also discuss how fuzzy Laplacians can be defined in our general framework and
  study a few explicit examples. Finally, we use this approach to propose a general
  explicit definition of fuzzy scalar field theory on compact K{\"a}hler manifolds.

}
\keywords{Non-Commutative Geometry, Differential and Algebraic Geometry}
\preprint{TCDMATH 08-04}

\begin{document}

\section{Introduction}

The quantization of K{\"a}hler manifolds seems to play an increasingly
important role in certain areas of field and string theory. In particular,
recent work on fuzzy geometry, which is partly inspired by string theory,
suggests that certain versions of ``geometric'' quantization provide a
framework for a better understanding of fuzzy spaces. In particular, it was
proposed in \cite{Saemann:2006gf} that a specific quantization procedure leads
to a general definition of fuzzy compact K{\"a}hler manifolds. While
originally formulated in terms of an explicit embedding in projective space,
this procedure has an intrinsic geometric meaning, which we explore and
clarify in the present paper.

It is perhaps not surprising that fuzzy K{\"a}hler geometry turns out to be
intimately related with Berezin quantization \cite{berezin} of compact
K{\"a}hler manifolds, which was studied in \cite{Rawnsley0, Rawnsley1,
Rawnsley2} and more recently in \cite{Bordemann:1993zv}-\cite{RT} in its
Berezin-Toeplitz variant. As we will show, however, the connection involves a
few interesting twists. For example, the proposal of \cite{Saemann:2006gf} is
not ordinary (or classical) Berezin or Berezin-Toeplitz quantization in the
sense of loc.\ cit., but rather a modified ``Berezin-Bergman'' version, which
can itself be viewed as a particular realization of a more general
Berezin-like procedure. To fully clarify the situation, we introduce {\em
generalized} Berezin and Toeplitz quantizations of compact K{\"a}hler
manifolds and show how the proposal of \cite{Saemann:2006gf} fits into this
larger framework.

In classical Berezin quantization \cite{Rawnsley0, Rawnsley1, Rawnsley2}, one
starts with a compact Hodge manifold $(X,\omega)$ (where $\omega$ is the
symplectic form) endowed with a Hermitian holomorphic line bundle $L$ whose
Chern connection has curvature equal to $-2\pi i \omega$. Using the Hermitian
metric $h$ of $L$ and the volume form of $\omega$, one constructs
$L^2$-scalar products $\langle~,~\rangle_k$ on the spaces of holomorphic
sections $E_k:=H^0(L^{\otimes k})$ of the positive tensor powers of $L$. One
then performs Berezin quantization at each sufficiently large level $k$ using the coherent states
of the finite-dimensional Hilbert spaces $(E_k,\langle~,~\rangle_k)$.  The
coherent states define Berezin quantization maps $Q_k:\Sigma_k\rightarrow
\End(E_k)$, where $\Sigma_k$ are finite-dimensional subspaces of ${\cal
C}^\infty(X)$.  A closely related quantization procedure known as Toeplitz
quantization was studied in \cite{Bordemann:1993zv, Schlichenmaier:1996yj,schlichenmaier-1996,
schlichenmaier-1999, schlichenmaierb-1999, karabegov-2000,
Schlichenmaier-2000a}. This prescription has better asymptotic properties and
is related to Berezin quantization via a geometric version of the Berezin
transform. Both quantization prescriptions depend only on the data $(X,L,h)$
-- which determines\footnote{Notice that $(X,\omega,L)$ determines $h$ up to a constant
  factor.}  $\omega$; however, we will often use the redundant
parameterization $(X,\omega,L,h)$ for reasons of notational clarity. 

The extension discussed in the present paper starts with the observation that
the Berezin quantization maps $Q_k$ at each fixed level $k$ depend only on the
holomorphic bundle $L^{\otimes k}$ and on the Hermitian scalar product on its
space of holomorphic sections. Hence the entire procedure can be generalized
by replacing the $L^2$-products $\langle~,~\rangle_k$ with an arbitrary
sequence of Hermitian scalar products $(~,~)_k$ on the spaces $E_k$. This
results in what we call {\em generalized Berezin quantization} of the Hodge
manifold $(X,\omega)$. While classical Berezin quantization depends on the
data $(X,L,h)$, its generalized version depends on $(X,\omega,L)$ 
and on the sequence of scalar products $(~,~)_k$ on the spaces $H^0(L^{\otimes k})$, where
$L$ is a holomorphic line bundle such that $c_1(L)=[\omega]$. This gives a
large class of apparently novel quantizations of $(X,\omega)$. A similar extension exists for
Toeplitz quantization and depends on the same data plus the choice of a Radon
measure on $X$. It is related to the corresponding generalized Berezin
quantization via an extension of the geometric Berezin transform.

Using this framework, we will show that the procedure proposed in
\cite{Saemann:2006gf} amounts to performing generalized Berezin quantization
with respect to a certain series of scalar products on $E_k$ which are induced
in an intrinsic manner from a given scalar product on $E_1=H^0(L)$. This quantization
prescription, which we shall call {\em Berezin-Bergman quantization}, depends
only on the data $(X,\omega,L,(~,~))$ and generally differs from the classical
Berezin quantization based on $(X,\omega, L,h)$. It is intimately related with a certain
sequence of Bergman metrics \cite{Tian} on
$X$ and might be of interest in studies of K{\"a}hler metrics of constant
scalar curvature. Berezin-Bergman quantization has a series of simplifying
features which make it eminently computable. In particular, it is rather
straightforward to determine the associated quantum objects in this
prescription. As an application, we consider a sequence of truncated Laplace operators
inspired by this quantization scheme, which can be used to approximate the
spectrum of the full Laplacian. These truncated Laplacians correspond to the
standard fuzzy Laplacian in the case of complex projective spaces. Explicit
numerical computations are presented for the quadratic and cubic Fermat curves
in $\P^2$ (the complex planar conic and an elliptic curve, respectively).

We will show that classical Berezin and Berezin-Bergman quantization
agree for the case of complex projective spaces, in which they both recover
the usual fuzzy geometry construction. The reason for this agreement is due to
the fact that $\P^n=U(n+1)/\left(U(n)\times U(1)\right)$ is a K{\"a}hler homogeneous space, and that both quantization
prescriptions are compatible with the transitive $U(n+1)$ action. Since they
agree for the only well-studied examples of fuzzy compact K{\"a}hler spaces, 
it follows that both quantization schemes provide potential definitions 
of general `fuzzy compact K{\"a}hler manifolds'. Which of these one chooses to use depends
on the desired asymptotic properties in the classical (i.e.\ large $k$) limit,
which are currently well understood only for classical Berezin
quantization. Thus one could as well choose the latter as a general definition of fuzzy
K{\"a}hler geometry. 

Generalized Berezin-Toeplitz quantization provides a precise framework for ``lifting''
operators from the space of functions to the quantum Hilbert space, an
operation which we call {\em Berezin-Toeplitz lift}. We propose to define the
``fuzzy'' Laplacian of a compact Hodge manifold as the Berezin-Toeplitz lift
of the Laplace operator through the Berezin-Toeplitz quantization of that
manifold. Together with the integral representation of the trace also derived
in the present paper, the Berezin-Toeplitz lift of the Laplacian enables us to
give an explicit definition of fuzzy scalar field theory on arbitrary compact
Hodge manifolds. This construction might be of interest for the fuzzy field
theory community.

The paper is organized as follows. In Section 2, we recall some facts about
polarizations and quantum line bundles, mostly in order to fix our notations
and terminology. We also discuss Bergman metrics and metrized Kodaira
embeddings and introduce a ``relative''
version of Rawnsley's epsilon function \cite{Rawnsley0}, which will prove
useful for our purpose. In Section 3, we introduce the generalized Berezin and
Berezin-Toeplitz quantization defined by a sequence of scalar products 
and explore their basic properties. In particular, we discuss the relation
between the two extended quantization procedures and address the
effect of changing the scalar products. We also discuss the notion of
relatively balanced Bergman metrics, which enters naturally in our set-up, as
well as the lift of linear operators from the space of functions to the
quantization space. Finally, we construct the generalized Berezin
(coherent state) product and give the description of the quantization in
the language of star algebras. In Section 4, we recall the basic properties
of classical Berezin and Berezin-Toeplitz quantization and in particular their
asymptotic behavior, which allows for the construction of the associated
formal deformation quantizations \cite{Schlichenmaier:1996yj,schlichenmaier-1996,schlichenmaierb-1999,karabegov-2000}. We also briefly discuss the
classical quantization of affine and projective spaces, which will be used later. For 
projective spaces, we follow an approach which recovers the formalism used in
\cite{Saemann:2006gf}, showing that the notion of fuzzy
projective spaces \cite{Balachandran} coincides with the classical Berezin
quantization of those spaces. Section 5 gives the general description of
Berezin-Bergman quantization. After presenting the intrinsic formulation, we
show how one can recover the description through embeddings in $\P^n$ and
clarify some of its basic properties. We also show that the Berezin-Bergman
quantization of $\P^n$ coincides with the classical Berezin quantization of the latter,
an accident\footnote{Such a simple relation between Berezin-Bergman and
Berezin quantization should not be expected for general compact Hodge
manifolds.} which is due to the fact that complex projective spaces are
homogeneous K{\"a}hler manifolds. Section 6 takes up the issue of ``quantized harmonic analysis'' in the
framework of classical Berezin quantization. We discuss two ways of
constructing a fuzzy Laplace operator. One is used to compute the approximate
spectrum of harmonic functions on Fermat curves, while the other one appears
in the construction of fuzzy scalar field theory on arbitrary compact Hodge
manifolds.

\section{Polarizations, quantum line bundles and Bergman metrics}

\subsection{Polarizations and quantum line bundles}

Consider a connected compact complex manifold $X$ of complex dimension
$n$. Recall that a {\em polarization} of $X$ is a positive holomorphic line
bundle $L$ over $X$. Given a polarized complex manifold $(X,L)$, there exists a positive
integer $k_0$ such that the tensor powers $L^k:=L^{\otimes k}$ are very ample for all $k\geq
k_0$; in particular, $X$ can be presented as a projective algebraic variety by
the Kodaira embedding determined by $L^k$ for any $k\geq k_0$. 

A K{\"a}hler form $\omega$ on $X$ is called {\em integral} if its cohomology
class $ [\omega]$ belongs to $H^2(X,\Z)$; in this case, $(X,\omega)$ is called
a {\em Hodge manifold}.  Given a polarization $L$ of $X$, the K{\"a}hler form
is called {\em $L$-polarized} if $[\omega]$ equals $c_1(L)$. In this case, $L$
is called a {\em K{\"a}hler polarization} of $(X,\omega)$ and the triple
$(X,L,\omega)$ is called a {\em polarized Hodge manifold}.  It is well-known
that any Hodge manifold $(X,\omega)$ admits K{\"a}hler polarizations;
moreover, the isomorphism classes of K{\"a}hler polarizations for $(X,\omega)$
form a torsor under the Abelian group $\Pic^0(X)$ of degree zero holomorphic
line bundles. In particular, K{\"a}hler polarizations of $(X,\omega)$ are
unique up to isomorphism when $X$ is simply connected. Conversely, a polarized
complex manifold $(X,L)$ admits K{\"a}hler metrics whose K{\"a}hler class
equals $c_1(L)$.

Given a polarized manifold $(X,L)$, there exists a well-known
bijection between $L$-polarized K{\"a}hler metrics on $X$ and homothety
(positive constant prefactor rescaling) classes of Hermitian bundle
metrics on $L$. Recall that this bijection is determined as follows: 

(a) Given a Hermitian metric $h$ on $L$, there exists a unique K{\"a}hler metric on
$X$ whose K{\"a}hler form satisfies $\omega=\frac{i}{2\pi}F$, where $F$ is the curvature of the Chern
connection\footnote{The Chern connection is the unique connection on $(L,h)$
  which is both Hermitian and compatible with the holomorphic structure.} $\nabla$ of $(L,h)$. This K{\"a}hler
metric is automatically $L$-polarized since $c_1(L)=\frac{i}{2\pi}
[F]$. Multiplying $h$ by a positive constant does not change the associated
K{\"a}hler metric. 

(b) Given an $L$-polarized K{\"a}hler form $\omega$, there exists
a Hermitian metric $h$ on $L$ such that $\omega=\frac{i}{2\pi}F$, where $F$ is the curvature of the Chern
connection of $(L,h)$. The metric $h$ is determined by $\omega$ up to multiplication by a positive constant. 
Such a Hermitian line bundle $(L,h)$ is sometimes called a {\em quantum line bundle} for $(X,\omega)$, and
one says that $(X,\omega,L,h)$ is a {\em prequantized Hodge manifold}.  Two
quantum line bundles $(L,h)$ and $(L',h')$ for $(X,\omega)$ are called {\em
equivalent} if there exists an isomorphism $\psi:L\rightarrow L'$ of
holomorphic line bundles such that $\psi^*(h')=h$. Equivalence classes of
quantum line bundles for $(X,\omega)$ form a $\Hom(\pi_1(X),S^1)$-torsor.

Given a quantum line bundle $(L,h)$,
one can endow $L^k$ with the induced metric $h_k=h^{\otimes k}$ and with the
corresponding Chern connection $\nabla_k=\nabla^{\otimes k}$. Then
$\omega=\frac{i}{2\pi k}F_k$, where $F_k=kF$ is the
curvature of $\nabla_k$. A fixed positive measure $\mu$ on $X$ induces a Hermitian scalar product on the space of smooth sections $\Gamma(L^k)$: 
\beq
\label{L2mu}
\langle s_1,s_2\rangle_k^{\mu,h}:=\int_{X}{d\mu ~h_k(s_1,s_2)} ~~.
\eeq
We let $L^2_k(L, h,\mu)$ be the $L^2$-completion of $\Gamma(L^k)$ with respect to
this scalar product. The finite-dimensional subspace $H^0(L^k)\subset
\Gamma(L^k)$ of holomorphic sections inherits a scalar product, which we
denote by the same symbol. The standard choice for $\mu$ is the Liouville
measure determined by the canonical volume form $\frac{\omega^n}{n!}$ of $(X,\omega)$:
\beq
\label{L2}
\langle s_1,s_2\rangle_k^{h}:=\int_{X}{\frac{\omega^n}{n!} ~h_k(s_1,s_2)} ~~,
\eeq
however, it is often useful to work more generally. For example, one has
another natural measure -- namely that defined by the volume form $\Omega\wedge
{\bar \Omega}$ -- when $(X,\omega)$ is algebraically a Calabi-Yau manifold,
(i.e.\ when $c_1(TX)=0$) with holomorphic top form $\Omega$. In these cases,
one is often interested in K{\"a}hler forms $\omega$ in a given integral cohomology
class, which however differ from the K{\"a}hler form
$\omega_{CY}$ of the Calabi-Yau metric in that class. (Thus one has
$[\omega]=[\omega_{CY}]=c$ for some positive class $c\in H^2(X,\Z)$ but
$\omega\neq \omega_{CY}$.) Recall that $\omega_{CY}$ is not explicitly known in practice. 
In such a situation, one has $\frac{\omega^n}{n!}\neq \Omega\wedge \bar{\Omega}=\frac{\omega_{CY}^n}{n!}$.

An {\em automorphism} of a prequantized Hodge manifold $(X,\omega,L,h)$ is
a pair $\gamma:=(\gamma_0,\gamma_1)$ such that $\gamma_0$ is a holomorphic isometry of
$(X,\omega)$ and $\gamma$ is a holomorphic bundle isometry of $(L,h)$ above
$\gamma_0$. In particular, $\gamma_1(x)$ is an isometry from $(L_x,h(x))$ to
$(L_{\gamma_0(x)},h(\gamma_0(x)))$ for all $x\in X$. The automorphisms of  $(X,\omega,L,h)$ form a group which we
denote by $\Aut(X,\omega,L,h)$. This group acts linearly on the space of sections
$H^0(L^k)$ via: 
\beq
\label{rho}
\rho_k(\gamma)(s)=\gamma_1^{\otimes k}\circ s\circ \gamma_0^{-1}~~~~(s\in H^0(L^k))~~.
\eeq
The actions $\rho_k:\Aut(X,\omega,L, h)\rightarrow \End(H^0(L^k))$ are unitary with respect to the $L^2$-scalar product
(\ref{L2mu}) provided that the measure $\mu$ is invariant under the group
$\Aut(X,\omega)$ of holomorphic isometries of $\omega$. This is the case, for
example, when $\mu$ is the Liouville measure defined by $\omega$.  

An automorphism $\gamma$ is called {\em trivial} if $\gamma_0=\id_X$ and
$\gamma_1$ is given by $\gamma_1(x)=(e^{i\alpha})\cdot $ for all $x$, where
$\alpha$ is a real constant. Thus $\Aut(X,\omega,L,h)$ always contains a
$U(1)$ subgroup. The quotient $\Aut(X,\omega,L,h)/U(1)$ is the subgroup
$\Aut_{L,h}(X,\omega)\subset \Aut(X,\omega)$ of those holomorphic isometries $\gamma_0$ of
$(X,\omega)$ which admit a lift $\gamma_1:L\rightarrow L$ such that
$(\gamma_0,\gamma_1)$ is an automorphism of $(X,\omega,L,h)$. Thus we have an
exact sequence of groups \cite{Rawnsley0}: 
\beq
\label{sequence}
1\rightarrow U(1)\rightarrow \Aut(X,\omega,L,h)\rightarrow
\Aut_{L,h}(X,\omega)\rightarrow 1~~.
\eeq
In general, the inclusion $\Aut_{L,h}(X,\omega)\subset \Aut(X,\omega)$ is
strict, i.e.\ not every holomorphic isometry admits a lift. The obstruction to
the existence of such a lift lives in the group $\Hom(\pi_1(X),S^1)$, so in
particular $\Aut_{L,h}(X,\omega)$ equals $\Aut(X,\omega)$ when $X$ is simply
connected. Notice that $\Aut(X,\omega)$ is usually discrete
since a generic Hodge manifold has no continuous holomorphic isometries. The
case usually studied in the fuzzy literature (namely that of rather special 
homogeneous spaces) is highly non-generic in this regard.

\paragraph{Remark.} A holomorphic section $\sigma$ of $L$ which is not identically zero yields a local
frame above the open set $U_{\sigma}:=\{x\in X|\sigma(x)\neq 0\}$. 
With respect to this frame, the Chern connection $\nabla$ 
of $(L,h)$ is given by $\nabla \equiv d+\partial \log
h(\sigma,\sigma)$, where $d=\partial +{\bar \partial}$ and $ \partial$ are the de Rham and Dolbeault operators. 
Its curvature is $F=-\partial {\bar \partial}\log
h(\sigma,\sigma)=-2\pi i\omega$. Hence the function $K_{\sigma}:=-\log h(\sigma,\sigma)$ defines a local K{\"a}hler potential on $U_{\sigma}$:
\beq
\nn
\omega=\frac{i}{2\pi}\partial {\bar \partial} K_{\sigma}~~.
\eeq
Every section $s\in \Gamma(L^k)$ can be written above $U_{\sigma}$ in the form
$s=f \sigma^{\otimes k}$, where $f$ is a smooth complex-valued function on $U_{\sigma}$,
which is holomorphic iff $s$ is holomorphic. When the measure $\mu$ satisfies
$\mu(X\setminus U_\sigma)=0$, this gives
isometries of $\Gamma(L^k)$ and $H^0(L^k)$ with the spaces of smooth, respectively holomorphic functions on $U_{\sigma}$ endowed with
the scalar product: 
\beq
\label{eqn:Usigma_product}
\langle f,g \rangle_{k,\sigma}=\int_{U_{\sigma}}d\mu ~e^{-kK_\sigma} {\bar f}g~~.
\eeq
It follows that $L^2_k(L,h,\mu)$ can be identified with the space
$L^2(U_\sigma, e^{-kK_\sigma}\mu)$.

\subsection{Parameterizing Hermitian bundle metrics and polarized K{\"a}hler
  forms}

Fixing a polarized complex manifold $(X,L)$, let $\L$ be the total
space of $L$ and $\L_0$ be the total space with the graph $o$ of the zero section
removed. Hermitian metrics $h$ on $L$ are uniquely determined by their square norm
functions ${\hat  h}\in {\cal C}^\infty(\L_0, \R_+)$: 
\beq
\nn
{\hat h}(q):=h(q,q)~~,~~q\in \L~~. 
\eeq
These are smooth non-negative functions on $\L$, strictly positive on $\L_0$ and having the property ${\hat h}(cq)=|c|^2{\hat
  h}(q)$ for all $q\in \L$ and all $c\in \C$ (this property
implies ${\hat h}|_o=0$). The set  ${\rm Met}(L)$ of
Hermitian metrics on $L$ can be identified with the set of all such functions
on $\L$ and thus forms an infinite-dimensional convex cone in ${\cal C}^\infty(\L,\R)$.
As a consequence, $L$-polarized K{\"a}hler metrics are parameterized by rays in
this real cone. If we fix a reference metric $h_0$ on $L$, then any other
metric $h$ is described by the smooth positive function $\phi=\frac{{\hat
    h}}{{\hat h}_0}$ on $X$, and we find that ${\rm
  Met}(L)$ can also be identified with  ${\cal C}^\infty(X,\R_+^*)$. Taking
the logarithm $\psi=\log \phi$, this gives bijections between ${\rm Met} (L)$ and
${\cal C}^\infty(X,\R)$, as well as between the set of $L$-polarized K{\"a}hler metrics
and the space $\{\psi\in {\cal C}^\infty(X,\R)| \psi(x)=0\}$, where $x$ is any fixed
point of $X$.

In this paper, we will use a slightly different parameterization 
in the case when $L$ is very ample. For any $q\in \L_0$, we let ${\hat q}:H^0(L)\rightarrow \C$ be
the linear functional (called {\em evaluation functional}) defined through:
\beq
\label{hatq}
s(\pi(q))={\hat q}(s)q~,~~~s\in H^0(L)~,
\eeq
where $\pi:\L\rightarrow X$ is the bundle projection.  We have the obvious
property $\widehat{cq}=\frac{1}{c}{\hat q}$ for all non-vanishing complex numbers
$c$. The very ampleness of $L$ implies ${\hat q}\neq 0$ for all $q\in \L_0$.

A Hermitian scalar product $(~,~)$ on the finite-dimensional space $H^0(L)$ induces a scalar product
on the dual space $H^0(L)^*=\Hom_\C(H^0(L),\C)$, which allows us to consider the
Hermitian metric $h_B$ on $L$ whose square norm function is given
by:
\beq
\label{hatbergman}
{\hat h}_B(q)=\frac{1}{||{\hat q}||^2}~~(q\in \L_0)~~
\eeq
(and ${\hat h}_B|_o=0$). This is called the {\em Bergman metric}\footnote{The
  name of these metrics honors the work of the mathematician Stefan Bergman.} \cite{Tian} on $L$ defined by the scalar product
$(~,~)$. Since we now have a reference Hermitian metric on $L$, we can
describe any other metric $h$ via the positive function:
\beq
\label{eqn:relepsilon}
\epsilon:=\frac{{\hat h}}{{\hat h}_B}\in {\cal C}^\infty(X,\R_+^*)~~,
\eeq
which we call the {\em epsilon function of $h$ relative to $(~,~)$}:
\beq
\label{he}
h(q,q)=\epsilon(\pi(q))h_B(q,q)~~.
\eeq 
Thus Hermitian metrics on $L$ are parameterized by their relative epsilon functions, once
we fixed a scalar product on $H^0(L)$.

The relative epsilon function defined above depends on $h$ and on the scalar
product chosen on $H^0(L)$, and is a generalization of the more familiar
object considered in \cite{Rawnsley0,Rawnsley1,Rawnsley2}. To make contact
with the latter, notice that fixing $h$ gives a distinguished choice of a
scalar product on $H^0(L)$, namely the $L^2$-product $\langle~,~\rangle$
defined by $h$ and by the Liouville measure of the associated K{\"a}hler form
$\omega$. The epsilon function of $h$ with respect to this $L^2$-scalar
product depends on $h$ only (remember that $\omega$ is determined by $h$), and
will be called the {\em absolute epsilon function} of $h$. The latter is the
epsilon function considered in \cite{Rawnsley0, Rawnsley1, Rawnsley2}.

The $L$-polarized K{\"a}hler metric on $X$ determined by $h_B$ is
called the {\em Bergman metric on $X$ induced by $(~,~)$}. Its K{\"a}hler form is
denoted by $\omega_B$. The K{\"a}hler form $\omega$ determined by the
Hermitian bundle metric (\ref{he}) takes the form: 
\beq
\nn
\omega=\omega_B-\frac{i}{2\pi}\partial{\bar \partial}\log \epsilon~~,
\eeq
so as expected we have $\omega=\omega_B$ iff the relative epsilon function of $h$ is
constant. Since $\omega$ determines $h$ up to multiplication by a constant, it
also determines the relative epsilon function of the latter up to the same ambiguity.
We will see below that the $L$-polarized Bergman metrics are those metrics
induced on $X$ by pulling-back Fubini-Study metrics through the Kodaira
embedding $i:X\hookrightarrow \P [H^0(L)^*]$ determined by the very ample line
bundle $L$, where the Fubini-Study metric being pulled-back is determined by the
scalar product on $H^0(L)^*$.

\paragraph{Remarks.} 1. Let $n+1:=\dim_\C H^0(L)$ and pick an arbitrary basis
$s_0\ldots s_n$ of $H^0(L)$. Setting $G_{ij}:=(s_i,s_j)$, we have:
\beq
\nn
||{\hat q}||^2=\sum_{i,j=0}^{n}{G^{ij}\overline{{\hat q}(s_i)}{\hat
    q}(s_j)}~~(q\in \L_0)~~,
\eeq
where $G^{ij}$ are the entries of the inverse matrix to $(G_{ij})$:
\beq
\nn
\sum_{j=0}^n G^{ij}G_{jk}=\delta_{ik}~~.
\eeq
The norm square with respect to the bundle Bergman metric determined by $(~,~)$ takes the form: 
\beq
\nn
h_B(q,q)=\frac{1}{\sum_{i,j=0}^n G^{ij}\overline{{\hat q}(s_i)}{\hat
    q}(s_j)} ~~(q\in \L_0)~~,
\eeq
while the epsilon function relative to $(~,~)$ of an arbitrary Hermitian
metric $h$ on $L$ is given by:
\beq
\nn
\epsilon(x)=\sum_{i,j=0}^n {G^{ij}h(x)(s_i(x),s_j(x))}~~.
\eeq
The Hermitian metric $h$ is given as follows in terms of its relative epsilon function: 
\beq
\nn
h(q,q)=\epsilon(x)h_B(q,q)=\frac{\epsilon(x)}{\sum_{i,j=0}^n {G^{ij}\overline{{\hat q}(s_i)}{\hat
    q}(s_j)}}~~.
\eeq

2. Bergman bundle metrics on $L$ are in bijection with Hermitian
products on $H^0(L)$, which form the non-compact homogeneous space
$U(n+1,\C)\setminus GL(n+1,\C)$ under the action of
$GL(n+1,\C)\simeq GL(H^0(L))$. They are extremely
special in the set of all Hermitian bundle metrics on $L$. Correspondingly,
$L$-polarized Bergman metrics on $X$ are extremely special among $L$-polarized K{\"a}hler
metrics.

3. The $L^2$-scalar product on $H^0(L)$ defined by $h_B$ and by
the volume form of $\omega_B$:
\beq
\nn
\langle s,t\rangle=\int_{X} \frac{\omega_B^n}{n!}h_B(s,t)~~(s,t\in H^0(L))
\eeq 
{\em need not} coincide with the scalar product $(~,~)$ which parameterizes
$h_B$. If they do, one says that the scalar product $(~,~)$ and
associated Bergman bundle and manifold metrics $h_B$, $\omega_B$ are {\em balanced}
\cite{Donaldson:2001aa}. It is clear that $\omega_B$ is balanced
iff its {\em absolute} epsilon function is constant;
Hermitian line bundles $(L,h_B)$ endowed with balanced bundle metrics were
called {\em regular} in \cite{Rawnsley1, Rawnsley2}. It was shown in
\cite{Donaldson:2001aa} that a balanced scalar
product on $H^0(L)$ is unique up to a constant scale factor if it exists, so $L$-polarized balanced metrics
on $X$ are at most unique.  A polarized complex manifold
$(X,L)$ is called balanced if $H^0(L)$ admits a balanced scalar product.  
When $L$ is very ample, it is known (see e.g.\ \cite{PS, Wang-2004}) that $(X,L)$ is balanced iff 
its Kodaira embedding $i(X)$ is Chow-Mumford stable in the projective space $\P[H^0(L)^*]$.

\subsection{Bergman metrics from metrized Kodaira embeddings}

Let $X$ be a compact complex manifold. By the Kodaira embedding theorem, a very ample line bundle $L$ gives  a
holomorphic embedding $i:X\hookrightarrow \P V$, where $V=E^*$ and $E:=H^0(L)$ is the
space of holomorphic sections of $L$, whose complex dimension we denote by
$n+1$.  The embedding allows us to view
$X$ as a regular projective variety in $\P V$, whose homogeneous coordinate
ring $R(X,L)=\oplus_{k\geq 0}{H^0(L^k)}$ is generated in degree one. In particular, $L$ and the pull-back
$i^*(H)$ of the hyperplane bundle $H:={\cal O}_{\P V}(1)$
are isomorphic as holomorphic line bundles.  

Conversely, if we are given any smooth projective variety $X$ in a projective
space $\P V $ whose vanishing ideal $I(X)$ is generated in degrees greater
than one, then the restriction ${\cal O}_X(1)={\cal
  O}_{\P V}(1)|_X$ is very ample and the embedding $X\embd \P V$ can be viewed as the Kodaira embedding determined by
this restriction. The space of holomorphic sections of ${\cal O}_X(1)$
identifies with the vector space $E=V^*$.

A {\em metrized Kodaira embedding} is a Kodaira embedding determined by a very
ample line bundle $L$ on $X$ together a fixed choice of a Hermitian scalar
product $(~,~)$ on its space of holomorphic sections $E:=H^0(L)$. For such
embeddings, the scalar product on $E$ induces a scalar product on $V=E^*$,
which makes $\P V $ into a (finite-dimensional) projective Hilbert space. The
latter carries the Fubini-Study metric\footnote{Recall that homogeneous
K{\"a}hler metrics on $\P V $ are in bijection with Hermitian scalar products
on $E$ taken up to constant rescaling, and these are the Fubini-Study
metrics. They are all related by $PGL(E)$-transformations.} determined by the
scalar product.  Its K{\"a}hler form is given by:
\beq
\nn
\pi^*(\omega_{FS})(v)=\frac{i}{2\pi}\partial {\bar \partial}\log(||v||^2)~~,
\eeq
where $\pi:V \rightarrow \P V$ is the canonical projection while
$||~||$ is the norm induced on $V=E^*$.  There exists a one to one
correspondence between metrized Kodaira embeddings of $X$ and holomorphic embeddings in
  finite-dimensional projective Hilbert spaces such that the vanishing ideal
  of the embedding is generated in degrees greater than one. 

The Fubini-Study metric admits the hyperplane bundle $H$ as a quantum  line bundle, when the latter is endowed with
the Hermitian bundle metric $h_{FS}$ induced from $E$. Since $L\simeq
i^*(H)$ as holomorphic line bundles, the pull-back
$i^*(h_{FS})$ defines a Hermitian metric $h_B$ on $L$. The latter coincides with the Bergman bundle metric determined by
$(~,~)$. The pulled-back K{\"a}hler form $\omega_B=i^*(\omega_{FS})$ admits $(L,h_B)$
as a quantum line bundle, and coincides with the Bergman K{\"a}hler form
determined by $(~,~)$. It follows that Bergman metrics on $X$ coincide with
pull-backs of Fubini-Study metrics via metrized Kodaira embeddings.

\paragraph{Remark.} A choice of basis $z_0\ldots z_n$ for $E=V^*$ allows us to
express $v\in V$ as: $v=\sum_{i=0}^n{v_ie_i}$, where $(e_i)$ is the basis of $V$
dual to $(z_i)$ and $v_i=z_i(v)$. This gives an identification of $V$ with the space
$\C^{n+1}$ endowed with the scalar product given by $\langle
u,v\rangle=\sum_{i,j=0}^n G^{ij}{\bar u}_i v_j$, where the $G^{ij}$ are given as above. 
Then $\P V$ identifies with $\P^n$ endowed with the Fubini-Study metric
defined by this scalar product. It is customary to choose an orthonormal basis,
in which case the Fubini-Study metric takes the familiar form in homogeneous
coordinates. In this case, the freedom of choosing the scalar product $(~,~)$
is replaced by the freedom of acting with $PGL(n+1,\C)$ transformations on the
homogeneous coordinates of $\P^n$. 

\section{Generalized Berezin and Toeplitz quantization}

Two related general methods for quantizing compact Hodge manifolds are
provided by the so-called Berezin and Berezin-Toeplitz quantization, which
were studied in \cite{Rawnsley0,Rawnsley1, Rawnsley2} and
\cite{Bordemann:1993zv,schlichenmaier-1996,
schlichenmaier-1999,schlichenmaierb-1999,karabegov-2000, Schlichenmaier-2000a,
RT}. This quantization scheme realizes ideas going back to \cite{berezin} in a
modified and extended form. In this approach, one starts with a prequantized
Hodge manifold $(X,\omega,L,h)$ and considers the sequence of Hermitian vector
spaces $(E_k:=H^0(L^k),\langle~,~\rangle_k)$ for $k\geq k_0$, where $k_0$ is
a positive integer $k$ such that $L^k$ is very ample for all $k\geq k_0$. The Hermitian
scalar products $\langle~,~\rangle_k$ are taken to be the $L^2$-products
(\ref{L2}) induced by $h$ and by the Liouville measure of $\omega$. At every
level $k$, the Hermitian structure makes $H^0(L^k)$ into a reproducing kernel
Hilbert space, and in particular allows one to introduce coherent vectors,
which are special holomorphic sections of $L^k$ parameterized by the points of
$X$. Using these vectors, one defines Berezin symbol maps
$\sigma_k:\End(E_k)\rightarrow {\cal C}^\infty(X)$, which turn out to be
injective due to compactness of $X$. The inverses on the images $\Sigma_k$ of
these maps provide bijections $Q_k:\Sigma_k\rightarrow
\End(E_k)$ which are known as {\em Berezin quantization maps}. The collection
$(Q_k)_{k\geq k_0}$ of such maps constitutes the {\em classical Berezin
quantization} of $(X,\omega)$ induced by the quantum line bundle $(L,h)$.

A fundamental problem in this approach is to describe the asymptotic behavior
of $Q_k$ for large $k$. One relevant problem is whether the sequence $Q_k$
defines in some manner a formal deformation quantization of $X$, and to
identify the corresponding formal star product. It turns out that these
questions can be answered quite elegantly by considering a variation of
Berezin's approach, which is known as classical Berezin-Toeplitz or simply
{\em classical Toeplitz quantization}. This modified quantization prescription consists of replacing
$Q_k$ by the so-called Toeplitz quantization maps $T_k:{\cal
C}^\infty(X)\rightarrow \End(E_k)$, which are constructed as integral
operators with the help of the coherent state projector. The asymptotic
behavior of $T_k(f)$ can be controlled using results of de Monvel, Guillemin
and Sj{\"o}strand \cite{MG, MS}, allowing one to prove \cite{Schlichenmaier:1996yj,schlichenmaier-1996,schlichenmaierb-1999,karabegov-2000}
that Toeplitz quantization gives rise to a formal star product and thus to a
formal deformation quantization of $(X,\omega)$. Since Berezin and Toeplitz
quantization turn out to be related via a general version of the Berezin
transform (which corresponds to a ``change of operator ordering''), this also
allows one to construct a formal star product corresponding to Berezin
quantization \cite{karabegov-2000} (see \cite{RT} for a different approach).

The construction of classical Berezin and Toeplitz quantizations can be
generalized by considering an arbitrary sequence of scalar products $(~,~)_k$
on the spaces $H^0(L^k)$ instead of the $L^2$-products (\ref{L2}).  This leads
to what we call {\em generalized Berezin and Toeplitz quantizations}. In this
section, we discuss the basic properties of the resulting quantization
schemes. As in the classical case, an important question -- which we do not
attempt to settle here -- concerns the asymptotic behavior of these
generalized quantizations for large $k$, which will of course depend markedly
on the choice of scalar products.

When studying the situation at each fixed level $k\geq k_0$, the replacement
$L\rightarrow L^{k}$ allows us to work with a very ample line bundle $L$ while
dropping the index $k$ from the notation. Let us therefore fix a compact
complex manifold $X$, a very ample line bundle $L$ on $X$ and a 
Hermitian scalar product $(~,~)$ on the vector space $E:=H^0(L)$, whose
dimension we denote by $N+1$. In the following we will sometimes
consider an arbitrary basis $s_0\ldots s_{N}$ of $H^0(L)$. In this
case, we let $G$ be the Hermitian positive matrix with entries $G_{ij}:=(
s_i, s_j)$ and $G^{ij}$ be the entries of the inverse matrix
$G^{-1}$. Any section $s\in E$ can be expanded as:
\beq
\nn
s=\sum_{i,j=0}^{N}{G^{ij}( s_j,s) s_i}~~.
\eeq

\subsection{Coherent states}

Given $q\in \L_0$, consider the evaluation functional ${\hat q}$ on
$E$ defined in (\ref{hatq}). By Riesz's theorem, there exists a unique holomorphic section
$e_q\in E$ such that $( e_q, s) ={\hat q}(s)$ for all
$s\in  E$. Direct computation gives the explicit expression:
\beq
\nn
e_q=\sum_{i,j=0}^{N}{G^{ji}\overline{{\hat q}(s_i)} s_j}~~,
\eeq
which implies: 
\beq
\nn
||e_q||^2=\sum_{i,j=0}^{N}{G^{ij}\overline{{\hat q}(s_i)} {\hat q}(s_j)}~~.
\eeq
Notice that $e_q$ cannot be the zero section, since that would imply that all sections of
$L$ vanish at $x=\pi(q)$, which is impossible since $L$ is very ample. The
element $e_q$ of $E$ is called the {\em Rawnsley coherent vector} \cite{Rawnsley0} defined by
$q$. Also notice that $e_q$ depends only on the scalar product chosen on $E$. 

If $q'$ is another non-vanishing element of the fiber $L_x$, then 
$q'=cq$ for some non-vanishing complex number $c$ and we have
$e_{q'}=\frac{1}{\bar c}e_q$. It follows that the complex line $l_x:=\langle
e_q\rangle=\C e_q\subset E$ depends only on the point $x\in X$.  This can be interpreted as follows. Let
$\bar{L}$ be the line bundle obtained by reversing the complex structure of all
fibers\footnote{ Thus the fiber $\bar{L}_x$ coincides with $L_x$ as an additive
group, but is endowed with the external multiplication with scalars given by $\alpha
*u=\bar{\alpha} u$ for all $\alpha\in \C$ and all $u\in L_x$. The
identity map becomes an antilinear involution when viewed as a map from $L_x$
to ${\bar L}_x$. This gives an involution 
between $L$ and $\bar{L}$, which we denote by an overline.}; this 
is a holomorphic line bundle over the complex manifold
${\bar X}$ obtained by reversing the complex structure of $X$. The scaling
property of coherent vectors implies that the element $e_x:={\bar q}\otimes
e_q\in \bar{L}_x\otimes H^0(L)$ depends only on the point $x\in X$. 
The scalar product on $H^0(L)$ extends to a sesquilinear map 
taking $[\bar{L}_x\otimes E]\times [\bar{L}_y\otimes E]$ into ${\bar
  L}_x\otimes {\bar L}_y$. In particular, the combination $K(x,y)=(
e_x,e_y)$ defines a holomorphic section $K$ of the external tensor
product $\bar{L}\boxtimes \bar{L}$ (which is a
holomorphic line bundle over $X\times {\bar X}$). This is the {\em reproducing
  kernel} of the finite-dimensional Hilbert space $(H^0(L),
(~,~))$. Also notice that the vector $e_q$ gives a well-defined element
$[e_x]$ of the projective space $\P E$,  which depends antiholomorphically on $x\in X$. This is called the
{\em Rawnsley  coherent state} at $x$.  Thus we have an antiholomorphic embedding 
$j: X\rightarrow \P E $, called the {\em coherent state embedding} (cf. \cite{berceanu}); it can be
viewed as dual to the metrized Kodaira embedding. 

Rawnsley's {\em coherent projectors} are the orthoprojectors on the lines $l_x\subset E$:
\beq
\label{eqn:Rawnsley_proj}
P_x:=\frac{|e_q) ( e_q|}{( e_q|e_q)}~~~(q\in L_x\setminus \{0\})~~.
\eeq
They depend only on $L$, on the point $x\in X$ and on the scalar product chosen on
$E$. Given a linear operator $C\in \End(E)$, its {\em lower Berezin symbol} is the
function $\sigma(C):X\rightarrow \C$ given by: 
\beq
\label{symbol}
\sigma(C)(x):=\tr(CP_x)=\frac{( e_q|C|e_q)}{( e_q|e_q)}~~(q\in L_x\setminus \{0\})~~.
\eeq
This gives a linear map $\sigma:\End(E)\rightarrow {\cal
  C}^\infty(X)$, whose image we denote by $\Sigma$. Notice that $\sigma$ and
$\Sigma$ depend only on $L$ and on the scalar product $(~,~)$ chosen on
$E$. The obvious relation:
\beq
\nn
\sigma(C^\dagger)=\overline{\sigma(C)}~~
\eeq
implies that $\Sigma$ is closed under complex conjugation, i.e.\ $\bar{\Sigma}=\Sigma$.
Also notice that $\Sigma$ contains the constant unit function $1_X=\sigma(\id_E)$.

\subsection{Generalized Berezin quantization}

It was shown in \cite{Rawnsley2} that the Berezin symbol map
$\sigma:\End(E)\rightarrow {\cal C}^\infty(X)$ is injective when
$(~,~)$ is the $L^2$-scalar product defined by a Hermitian metric
on $L$ and by the Liouville measure of the associated
K{\"a}hler form. We show below that the Berezin symbol changes as in
(\ref{sigma_change}) when changing the scalar product. This implies that
$\sigma$ is in fact injective for an arbitrary scalar product on $E$. Hence the
corestriction $\sigma|^\Sigma:\End(E)\rightarrow \Sigma$ is a linear isomorphism and
we can associate an operator on $E$ to every function $f\in \Sigma$ via the
{\em generalized Berezin quantization map} $Q=(\sigma|^\Sigma)^{-1}:\Sigma\rightarrow \End(E)$:
\beq
\label{eqn:gen_berezin_q}
Q(f):=\sigma^{-1}(f)~~~~\forall f\in \Sigma~~. 
\eeq
The extension from the case of \cite{Rawnsley0, Rawnsley1, Rawnsley2} is simply that we allow for an arbitrary scalar
product on $E$. The Berezin quantization map depends only on $L$ and on the choice of this scalar product. 
It satisfies the relations:
\beq
\nn
Q(\bar{f})=Q(f)^\dagger~~,~~Q(1_X)=\id_E~~.
\eeq

\paragraph{The Berezin star algebra.} The {\em Berezin  product} $\diamond:\Sigma\times \Sigma\rightarrow \Sigma$
is defined via the formula:
\beq
\label{Ber_product}
f\diamond g:=\sigma(Q(f)Q(g))\Leftrightarrow Q(f\diamond g)=Q(f)Q(g)~~.
\eeq
Together with the usual complex conjugation of functions $f\rightarrow
\bar{f}$, it makes $\Sigma$ into a unital finite-dimensional associative
$*$-algebra. The Berezin quantization map gives an isomorphism of
$*$-algebras:
\beq
\nn
Q:(\Sigma,\diamond,\bar{~})\rightarrow (\End(E),\circ,\dagger)~~.
\eeq
Recall that  $(\End(E),\circ,\dagger, ||~||_{HS})$ is a $B^*$-algebra\footnote{A $B^*$
  algebra is a Banach  ($||xy||\leq ||x||\,||y||$) $*$-algebra in which the identity $||x^*||=||x||$ is satisfied.}
with non-degenerate trace given by the usual trace of operators. It follows
that the induced linear map (called the {\em Berezin trace}): 
\beq
\label{strace}
\vint{f}:=\tr~Q(f)~~(f\in \Sigma)
\eeq
is a nondegenerate trace on the Berezin star algebra $(\Sigma,\diamond,\bar{~})$:
\begin{eqnarray}
\vint{\bar f}&=&\overline{\vint f}\nn\\
\vint f\diamond g &=&\vint g\diamond f~~\nn\\
\vint f\diamond g =0~~,\ &\forall& g\in \Sigma \Rightarrow f=0~~.\nn
\end{eqnarray}
Moreover, the scalar product on $\Sigma$ (called the {\em Berezin scalar product}) obtained by transporting the
Hilbert-Schmidt product: 
\beq
\label{diamond_product}
\prec f, g\succ_B:=\langle Q(f), Q(g)\rangle_{HS}=\tr\left(Q(f)^\dagger Q(g)\right)
\eeq
coincides with the scalar product induced by the Berezin trace:
\beq
\nn
\prec f, g\succ_B=\vint {\bar f}\diamond g~~.
\eeq
Since $(\End(E),\circ,\dagger, ||~||_{HS})$ is a $B^*$-algebra, the norm
defined by the Berezin product satisfies:
\begin{eqnarray}
||f\diamond g||_B &\leq& ||f||_B\,||g||_B\nn\\
||{\bar f}||_B&=&||f||_B~~.\nn
\end{eqnarray}
Thus $(\Sigma,\diamond,\bar{~},||~||_B)$ is a $B^*$-algebra with non-degenerate
trace, and $(Q,\sigma)$ are mutually inverse isomorphisms of $B^*$-algebras
with trace. Notice that $||1_X||_B=||\id_E||_{HS}=N+1$.

\paragraph{The push and pull of linear operators.} The isomorphism $Q$ allows us to transport $\C$-linear operators between
$\Sigma$ and $\End(E)$. Given a linear operator ${\cal O}:\Sigma\rightarrow \Sigma$, define its {\em Berezin push} 
${\cal O}^B: \End(E)\rightarrow \End(E)$ via: 
\beq
\label{eqn:Bpush}
{\cal O}^B:=Q\circ {\cal O}\circ \sigma~~\Leftrightarrow Q\circ {\cal O}={\cal
 O}^B\circ Q~~.
\eeq
Given a linear operator  ${\cal V}:\End(E)\rightarrow \End(E)$, define its
{\em Berezin pull} though: 
\beq
\label{eqn:Bpull}
{\cal V}_B:=\sigma\circ {\cal V}\circ Q~~\Leftrightarrow \sigma \circ {\cal
  V}={\cal V}_B \circ \sigma~~.
\eeq
The operations of Berezin push and pull are mutually inverse
linear isomorphisms between $\End_\C(\Sigma)$ and $\End_\C(\End(E))$. They are
well-behaved with respect to the Berezin scalar product on $\Sigma$ in the
sense that the following identities hold:
\begin{eqnarray}
\label{push_pull_ids}
\prec f, {\cal O}(g)\succ_B~~~~~&=&\langle Q(f), {\cal O}^B (Q(g))\rangle_{HS}~~\nn\\
\langle C_1, {\cal V}(C_2)\rangle_{HS}&=&\prec \sigma(f), {\cal V}_B(\sigma(g))\succ_B~~.
\end{eqnarray}
In particular, the Berezin push of a $\prec ~,~\succ_B$-Hermitian operator is
$\langle~,~\rangle_{HS}$-Hermitian and the Berezin pull of a
$\langle~,~\rangle_{HS}$-Hermitian operator is $\prec ~,~\succ_B$-Hermitian.

\paragraph{The squared two point function.} For later reference, 
define the squared two-point function $\Psi\in {\cal
  C}^\infty(X\times X,\R_+)$ of coherent states: 
\beq
\label{two_point}
\Psi(x,y):=\tr(P_xP_y)=\sigma(P_y)(x)=\sigma(P_x)(y)=\frac{|(
  e_x|e_y)|^2}{||e_x||^2||e_y||^2}\geq 0~~.
\eeq
This function is symmetric and non-negative on $X\times X$:
\beq
\nn
\Psi(x,y)=\Psi(y,x)~~\forall x,y\in X~~
\eeq
and vanishes at points $(x,y)$ such that $e_x$ is orthogonal to $e_y$. The
vanishing divisor of $\Psi$ is known as the \emph{polar divisor} \cite{berceanu}. 

\paragraph{Behavior under automorphisms.} Recall that the group $\Aut(X,\omega,L,h)$ acts linearly on $E$ (see
eq. (\ref{rho})). It is easy to check the relation:
\beq
\nn
\rho(\gamma^{-1})^\dagger (e_q)=e_{\gamma(q)}~~.
\eeq
Let us assume that the action $\rho$ is $(~,~)$-unitary:
\beq \nn
\rho(\gamma)^\dagger=\rho(\gamma)^{-1}~~.
\eeq
Then the relation above becomes:
\beq
\nn
\rho(\gamma)(e_q)=e_{\gamma(q)}
\eeq
and the Rawnsley projectors satisfy:
\beq
\label{rho_proj}
P_{\gamma_0(x)}=\rho(\gamma)P_x\rho(\gamma)^{-1}~~.
\eeq
In particular, the square two-point function (\ref{two_point}) is invariant:
\beq
\nn
\Psi(\gamma_0(x),\gamma_0(y))=\Psi(x,y)~~
\eeq
and the Berezin symbol map is $\Aut(X,\omega,L,h)$-equivariant:
\beq  \nn
\sigma(\rho(\gamma)C\rho(\gamma)^{-1})(x)=\sigma(C)(\gamma_0^{-1}(x))~~~~~(C\in \End(E))~~.
\eeq
If we let ${\hat \rho}=\rho^*\otimes_\C \rho$ be the representation induced by $\rho$ on $\End(E)$:
\beq  \nn
{\hat \rho}(\gamma)(C)=\rho(\gamma)C\rho(\gamma)^{-1}~~~~,
\eeq
then we can write the equivariance property above as follows: 
\beq
\label{sigma_equiv}
\sigma \circ {\hat \rho}(\gamma)=\tau(\gamma_0)\circ \sigma~~~~~(\gamma\in \Aut(X,\omega,L,h))~~.
\eeq
Here $\tau$  is the natural action of
$\Aut_{L,h}(X,\omega)$ on ${\cal C}^\infty(X)$:
\beq
\label{eqn:tau}
\tau(\gamma_0)(f)=f\circ \gamma_0^{-1}~~,
\eeq
which preserves the symbol space $\Sigma=\im \sigma$ as a consequence of
(\ref{sigma_equiv}):
\beq \nn
\tau(\gamma_0)(\Sigma)=\Sigma~~.
\eeq
We will sometimes view $\tau$ as a representation of $\Aut(X,\omega,L,h)$ via the
morphism $\Aut(X,\omega,L,h)\rightarrow \Aut_{L,h}(X,\omega)$ (see
(\ref{sequence})), without indicating this explicitly. 

\noindent The properties above imply that the Berezin quantization map is equivariant as well:
\beq
\label{Q_equiv}
{\hat \rho}(\gamma)\circ Q=Q\circ \tau(\gamma_0)~~~~~(\gamma\in \Aut(X,\omega,L,h))~~.
\eeq
Finally, the Berezin scalar product satisfies: 
\beq
\nn
\prec \tau(\gamma_0)(f),\tau(\gamma_0)(g)\succ_B=\prec f,g\succ_B~~~~(f,g\in
\Sigma,~~\gamma_0\in \Aut_{L,h}(X,\omega))~~,
\eeq
which shows that the representation of $\Aut_{L,h}(X,\omega)$ induced by $\tau$ on the invariant
subspace $\Sigma\subset {\cal C}^\infty(X)$ is unitary with respect to
$\prec ~,~\succ_B$. The Berezin trace (\ref{strace}) and the Berezin product are
also $\Aut_{L,h}(X,\omega)$-invariant: 
\beq
\nn
\vint\circ \tau(\gamma_0)=\vint
\eeq
and: 
\beq
\nn
\tau(\gamma_0)(f)\diamond\tau(\gamma_0)(g)=\tau(\gamma_0)(f\diamond g)~~.
\eeq

\subsection{Changing the scalar product in generalized Berezin quantization}

Let us consider what happens when we change the scalar product. 
An arbitrary Hermitian scalar product $(~,~)'$ on $E$ has the form: 
\beq
\label{productprime}
( s,t)'=( As,t )=( s, At )
\eeq
with $A$ a $(~,~)$-Hermitian positive-definite matrix\footnote{Of
  course, $A^{-1}$ and thus $A$ are also Hermitian and strictly positive with
  respect to $(~,~)'$.}. The coherent states with respect to the new product $(~,~)'$ are given by:
\beq
\label{eprime}
e_q'=A^{-1}e_q~~(q\in L_x\setminus \{0\})~~,
\eeq
while the new Rawnsley projectors take the form: 
\beq
\label{Pprime}
P'_x=\frac{1}{\sigma(A^{-1})(x)}A^{-1}P_x~~(x\in X)~~.
\eeq
The symbol $\sigma(A^{-1})(x)=\frac{( e_q|A^{-1}|e_q)}{(
  e_q|e_q)}$ of $A^{-1}$ computed with respect to
$(~,~)$ and the symbol $\sigma'(A)(x)=\frac{(
  e'_q|A|e'_q)'}{( e'_q|e'_q)'}$ of $A$ computed with
respect to $( ~,~)'$ are related by:
\beq
\label{rel1}
\sigma(A^{-1})(x)=\frac{1}{\sigma'(A)(x)}~~.
\eeq
Notice that $\sigma(A)$ and $\sigma'(A)$ are strictly positive smooth
functions on $X$. Given an operator $C$, we have more generally:
\beq
\label{sigma_change}
\sigma'(C)=\frac{\sigma(CA^{-1})}{\sigma(A^{-1})}
\eeq
and:
\beq
\label{sigma_change2}
\sigma(C)=\frac{\sigma'(CA)}{\sigma'(A)}~~.
\eeq
Let $Q'$  be the Berezin quantization map defined by $(~,~)'$ and
$\Sigma'\subset {\cal  C}^\infty(X)$ be the image of $\sigma'$. Equation (\ref{sigma_change}) shows that
\beq
\nn
\Sigma'=\frac{1}{\sigma(A^{-1})}\cdot \Sigma=\left\{\frac{1}{\sigma(A^{-1})}\cdot f~|~f\in \Sigma \right\}
\eeq
and that:
\beq
\nn
Q'(f)=Q(\sigma(A^{-1})f) A~~~\forall f\in \Sigma'~~.
\eeq

\paragraph{Proposition.} The Berezin quantizations defined by two different
scalar products on $E$ agree iff the operator $A$ is proportional to the identity,
i.e.\ iff the two scalar products are related by a constant scale factor. In
this case, the coherent states differ by a constant homothety and the coherent
projectors are equal.

\paragraph{Proof.}

The quantizations will agree iff $\sigma'(C)=\sigma(C)$ for all $C\in
  \End(E)$. Using relation (\ref{sigma_change2}), this implies
  $\sigma(CA)=\sigma(C)\sigma(A)$ for all
  $C\in \End(E)$. Taking the complex conjugate and replacing $C$ by
  $B^\dagger$, we also find $\sigma(AB)=\sigma(A)\sigma(B)$ for all $B$.  Thus
  $\sigma(AB)=\sigma(BA)$ for all $B$, which implies that $A$ commutes with
  all operators on $E$ since $\sigma$ is injective. Thus $A=\lambda \id_E$
  (with $\lambda>0$) by  Schur's Lemma, i.e.\ the two scalar products differ by
  a positive constant rescaling. Conversely, it is clear that such a rescaling does not affect
  the Berezin symbol map. The last statement follows from relations (\ref{eprime})
  and (\ref{Pprime}).

\

\paragraph{Remarks.} 1. The Hermitian conjugate $C^\oplus$ of a linear operator $C\in
\End(E)$ with respect to $(~,~)'$ takes the form: 
\beq
\nn
C^\oplus=A^{-1}C^\dagger A~~.
\eeq
This allows one to check the identity $(P'_x)^\oplus=P'_x$ by direct computation. The
property $(P'_x)^2=P'_x$ also follows directly from the definition of the symbol
$\sigma(A^{-1})$.

2. The operator $\CI:=A^{1/2}$ is an isometry from the Hilbert
space $(E,(~,~))$ to the Hilbert space $(E,(~,~)')$:
\beq
\label{tf}
( \CI u,\CI v)=( u,v)'~~.
\eeq
In general, this operator is non-local in $x\in X$. The operators ${\tilde
  P}_x:=\CI P_x\CI^{-1}=A^{1/2}P_xA^{-1/2}$ are orthoprojectors in
$(E,(~,~)')$, but they are {\em not} the Rawnsley projectors of
$(~,~)'$. The reason for this is that the coherent states with respect to the
latter scalar product have changed, and thus we have to
$(~,~)'$-orthoproject onto the {\em new} coherent states. More generally, any
invertible operator $\CI\in GL(E)$ can be used to define a new
scalar product on $E$ via (\ref{tf}). The decomposition $C=UA$ with
$A=(\CI^\dagger \CI)^{1/2}$ and $U=\CI A^{-1}$ shows that $( u,v)'$ has
the form (\ref{productprime}) and that it depends only on the positive
operator $A$. Of course, the space of scalar products on $E$ can be identified
with the homogeneous space $U(N+1,\C)\setminus GL(N+1,\C)$. Taking into
account the previous proposition,  we find that the different Berezin quantizations associated with the line
bundle $L$ are parameterized by the points of the homogeneous space $(SU(N+1,\C)\times
\C^*) \setminus GL(N+1,\C)$. 

3. We can also parameterize the new scalar product $(~,~)'$ by the symbol
$a=\sigma(A^{-1})$, i.e.\ by strictly positive smooth functions $a\in
\Sigma$. Then different Berezin quantizations based on $L$ are parameterized
by equivalence classes of such functions under rescaling by a positive
constant.  We have $\Sigma'=\frac{1}{a}\Sigma$ and $Q'(f)=Q(af)Q(a)$ as well
as $Q'\circ \Phi=Q$, where the map $\Phi:\Sigma\rightarrow \Sigma'$ is given
by $\Phi(f)=\frac{\beta(af)}{a}$. Here, $\beta$ is the Berezin transform of
the quantization with respect to the original scalar product $(~,~)$ as
defined in Section 3.7.

\subsection{Integral representations of the scalar product}

Let $L$, $E=H^0(L)$ and $(~,~)$ be as above.
Fixing a positive Radon measure $\mu$ on $X$, we consider the problem of representing the scalar product
$(~,~)$ on $E$ as the $L^2$-product induced by $\mu$ and
by a Hermitian metric $h$ on the line bundle $L$. Such representations will be
used below when discussing generalized Toeplitz quantization. 

Recall that any Hermitian bundle metric $h$ on $L$ can be parameterized by its
epsilon function relative to $(~,~)$: 
\beq
\label{epsilon}
\epsilon(x):=h(x)(q,q)||e_q||^2=\sum_{i,j=0}^{N}{G^{ij}h(x)(s_i(x),s_j(x))}~~.
\eeq
Here, $q$ is any non-zero vector in the fiber $L_x$, and the
right hand side is independent of the choice of $q$. We use the fact that the
value $h(x)$ of the metric on the fiber $L_x$ is uniquely determined by
$h(q,q)$. Conversely, any positive smooth function $\epsilon:X\rightarrow (0,+\infty)$
determines a Hermitian metric on $L$ via this formula, namely\footnote{The corresponding metric is well
defined since $e_{cq}=\bar{c}^{-1}e_q$ implies $h(x)(cq,cq)=|c|^2 h(x)(q,q)$
for all non-vanishing complex numbers $c$.} $h(x)(q,q)=\frac{\epsilon(x)}{||e_q||^2}$.  
This parameterization allows us to describe Hermitian metrics on $L$ through
positive smooth functions on $X$, provided that we have fixed a scalar product
$(~,~)$ on $E$. 

The scalar product $(~,~)$ on $E$ coincides with the
$L^2$-product induced by $\mu$ and $h$ if and only if the following identity
holds for any two holomorphic sections $s,t$ of $L$:
\beq
\nn
( s, t)=\int_Xd\mu(x)~h(x)(s(x),t(x))~~.
\eeq
It is easy to see that  the right hand side equals $\int_X{d\mu(x)  \epsilon(x)( s|P_x|t)}$. 
This implies the following:

\paragraph{Proposition.} The scalar product $(~,~)$ on $E$ coincides
with the $L^2$-scalar product induced by $(\mu,h)$ iff the relative epsilon function of
the pair $(h,(~,~))$ satisfies the identity
\beq
\label{completeness}
\int_{X}{d\mu(x) {\epsilon}(x) P_x}=\id_E~~,
\eeq
i.e.\ iff the coherent states defined by $(~,~)$ form an overcomplete set with respect to the
measure $\mu_{\epsilon}=\epsilon \mu$. 

\

\noindent Since the Berezin symbol map is injective, equation (\ref{completeness}) is
equivalent with the following Fredholm equation of the first kind:
\beq
\nn
\int_{X} d\mu(y)\Psi(x,y)\epsilon(y)=1~~~~(x\in X)~~.
\eeq

\noindent Combining everything, we obtain:

\paragraph{Proposition.} There exists a bijection between:

(a) pairs $(\mu,h)$ such that $\mu$ is a positive Radon measure on $X$ and $h$ is a Hermitian metric on $L$

(b) triples $(\mu,(~,~), \epsilon)$ such that $\mu$ is a positive Radon measure
on $X$, $(~,~)$ is a Hermitian scalar product on $H^0(L)$
and $\epsilon$ is a non-negative solution of the integral operator equation
(\ref{completeness}), where $P_x$ are the Rawnsley projectors determined by the
coherent states defined by $(~,~)$.

\

When the scalar product on $E$ is fixed, equation (\ref{completeness}) can be
viewed as a constraint on the pairs $(\mu,h)$ which allow for an integral
representation of the scalar product. Taking the trace shows that an epsilon 
function satisfying (\ref{completeness}) is normalized to total mass $N+1$
with respect to the measure $\mu$:
\beq
\nn
\int_{X} d\mu ~ \epsilon=N+1~~.
\eeq
In particular, when $\epsilon$ is a constant function, then its value must be given by
$\epsilon=\frac{N+1}{\mu(X)}$. Also notice that equations (\ref{symbol}) and
(\ref{completeness}) imply the following integral representation for the trace
on $\End(E)$:
\beq
\label{trace_identity}
\tr(C)=\int_{X}d\mu(x)~\epsilon(x)\sigma(C)(x)~~.
\eeq
Here $\sigma$ is the Berezin symbol map defined by the scalar product $(~,~)$.

If $\mu$ and the scalar product on $E$ are fixed, condition
(\ref{completeness}) can be written in a basis of $E$ as a system of 
inhomogeneous linear integral equations for $\epsilon$ (which in turn
determines $h$):
\beq
\label{sys}
\int_Xd\mu~\epsilon(x) \frac{\overline{{\hat q}(s_i)}
  \hat{q}(s_j)}{\sum_{i,j=0}^{N}{G^{ij}
\overline{{\hat q}(s_i)} {\hat q}(s_j)}} = G_{ij}~~.
\eeq
Taking the complex conjugate we see that only $\frac{1}{2}(N+1)(N+2)$ of these
equations are independent. It is clear that (\ref{sys}) admits an infinity of
  solutions $\epsilon$, so there is an infinity of Hermitian metrics 
$h$ on $L$ which allow us to represent the scalar product $(~,~)$ as
an $L^2$-product with respect to $\mu$. Each such metric $h$ also defines an $L^2$-scalar product
on the space $\Gamma(L)$ of {\em smooth} sections by formula (\ref{L2mu}),
which extends the given scalar product on $E$. If we let $L^2(\mu,h)$
be the Hilbert space obtained by completing $\Gamma(L)$ with respect to the
associated norm, we find an isometric embedding of $E$ into $L^2(\mu,h)$.
Hence any solution of (\ref{sys}) provides a realization of $H^0(L)$
as a finite-dimensional subspace of an infinite-dimensional Hilbert space.

\paragraph{Remark. } When considering quantization with an integral
representation of the scalar product on the space of holomorphic sections, one
has to deal with three different scalar products on the space of
functions. Indeed, the measure $\mu$ on $X$ defines a scalar product
$\prec~,\succ$ on ${\cal C}^\infty(X)$:
\beq
\label{mu_product}
\prec f,g \succ:=\int_{X}{d\mu {\bar f}g}~~,
\eeq
which extends to the natural scalar product on the space $L^2(X,\mu)$. 

On the other hand, the measure $\mu_\epsilon=\mu \epsilon$ appearing in the
overcompleteness relation (\ref{completeness}) defines its own $L^2$-scalar product
$\prec~,~\succ_{\epsilon}$ on ${\cal C}^\infty(X)$:
\beq
\label{epsilon_product}
\prec f,g\succ_\epsilon=\int{d\mu~ \epsilon {\bar f} g}~~.
\eeq
This extends to the natural scalar product of the space
$L^2(X,\mu_\epsilon)$. We have:
\beq
\prec f,g\succ_{\epsilon}:=\prec f,M_{\epsilon} g\succ~~,\nn
\eeq
where $M_\psi:\CC^\infty(X)\rightarrow \CC^\infty(X)$, $M_\psi(f)=\psi f$ denotes the operator of
multiplication with a smooth function $\psi$. Notice that $M_\psi$ is
$\prec~,~\succ$-Hermitian when $\psi$ is real valued, and $\prec~,~\succ$-positive when
$\psi$ is everywhere positive. 

Finally, the Berezin symbol space $\Sigma$ carries the Berezin scalar product
$\prec~,~\succ_B$, which is induced from the Hilbert-Schmidt scalar product of $\End(E)$ via the Berezin
quantization map (see eq. (\ref{diamond_product})). Hence $\Sigma$ is endowed with
three different scalar products, namely the Berezin product and the
restrictions of the products (\ref{mu_product}) and
(\ref{epsilon_product}). It is often the case that various linear operators on
$\Sigma$ are self-adjoint with respect to one of these products but not with
respect to the others. 

Equation (\ref{trace_identity}) provides an integral representation of the
Berezin trace:
\beq
\label{vint}
\vint{f}=\tr~Q(f)=\int_X d\mu(x) \epsilon(x) f(x)~~(f\in \Sigma)~~,
\eeq
which in turns gives the following representation of the Berezin
scalar product:
\beq
\label{induced_sp}
\prec f, g\succ_B =\vint {\bar f}\diamond g=\langle Q(f),Q(g)\rangle_{HS}=\int_X d\mu(x)
\epsilon(x) ({\bar f}\diamond g)(x)~~.
\eeq

\subsection{The relative balance condition}

\paragraph{Definition.} We say that a
  scalar product on $E$ is $\mu$-{\em balanced} if equation (\ref{completeness})
  admits the constant solution $\epsilon=\frac{N+1}{\mu(X)}$, i.e.\ if the
  following condition is satisfied:
\beq
\nn
\int_{X}d\mu(x)P_x=\frac{\mu(X)}{N+1}\id_E~~.
\eeq
\

\noindent Hence the scalar product is $\mu$-balanced iff the matrix
$G$ satisfies the system of equations: 
\beq
\nn
\frac{N+1}{\mu(X)}\int_Xd\mu~ \frac{\overline{{\hat q}(s_i)}
  \hat{q}(s_j)}{\sum_{i,j=0}^{N}{G^{ij}
\overline{{\hat q}(s_i)} {\hat q}(s_j)}} = G_{ij}~~.
\eeq

For a $\mu$-balanced scalar product, the overcompleteness property  of coherent
states takes the form $\frac{N+1}{\mu(X)}\int_{X}{d\mu P_x}=\id_E$. The Hermitian
metric $h$ on $L$ has epsilon function $\epsilon=\frac{N+1}{\mu(X)}$
and therefore is given by: 
\beq
\nn
h(x)(q,q)= \frac{N+1}{\mu(X)}\frac{1}{||e_q||^2}=\frac{N+1}{\mu(X)}\frac{1}{\sum_{i,j=0}^{N}{G^{ij}
\overline{{\hat q}(s_i)} {\hat q}(s_j)}}~~.
\eeq

Let $\omega_h$ be the $L$-polarized K{\"a}hler form on $X$ determined by a
Hermitian scalar product $h$ on $L$, and let $\mu_h:=\mu_{\omega_h}$ be
the Liouville measure on $X$ defined by $\omega_h$. We say that
$(~,~)$ is {\em balanced} if it is $\mu_h$-balanced. This boils
down to the system of equations: 
\beq
\nn
\frac{N+1}{\mu_h(X)}\int_Xd\mu_{h}~ \frac{\overline{{\hat q}(s_i)}
  \hat{q}(s_j)}{\sum_{i,j=0}^{N}{G^{ij}
\overline{{\hat q}(s_i)} {\hat q}(s_j)}} = G_{ij}~~,
\eeq
where $h$ is determined by:
\beq
\nn
h(x)(q,q)=\frac{N+1}{\mu_h(X)}\frac{1}{\sum_{i,j=0}^{N}{G^{ij}
\overline{{\hat q}(s_i)} {\hat q}(s_j)}}~~.
\eeq
This is the case considered in \cite{Tian,Donaldson:2001aa},
which was already mentioned in the previous section. 

\subsection{Generalized Toeplitz quantization}

Let us consider the case when $(~,~)$ is the $L^2$-scalar product on $E$
determined by a fixed measure $\mu$ on $X$ and a Hermitian scalar product $h$ on
$L$. As we saw in the previous section, this can always be achieved by
some pair $(\mu,h)$. Consider the embedding $E\subset L^2(\mu,h)$. Since the scalar product on
$L^2(\mu,h)$ restricts to the original product on $E$, we will denote it
by the same symbol $(~,~)$. Interpreting the coherent vectors $e_q$ as
elements\footnote{This means that we view the holomorphic sections
   of $L$ as smooth sections of $L$, which in particular are elements
  of $L^2(\mu,h)$.} of $E$, the orthogonal projector $P_x$ of $E$ becomes the orthogonal
projector $\Pi_x$ of $L^2(\mu,h)$ onto the one-dimensional subspace of
$L^2(\mu,h)$ defined by $e_x$. Hence equation (\ref{completeness})
becomes: 
\beq
\label{comp}
\int_X d\mu(x) {\epsilon}(x)\Pi_x=\Pi~~,
\eeq
where $\Pi$ is the orthoprojector of $L^2(\mu,h)$ onto $E$. We can now
define the Toeplitz operator $T(f)\in \End(E)$ associated to a smooth complex function $f\in
{\cal C}^\infty(X)$:  
\beq
\label{eqn:gen_toep}
T(f)(s)=\Pi(fs)~~\forall s\in E~~.
\eeq
Using equation (\ref{comp}), we find the integral expression: 
\beq
\nn
T(f)=\int_Xd\mu(x) {\epsilon} (x) f(x) P_x~~.
\eeq

The map $T:{\cal C}^\infty(X)\rightarrow \End(E)$ will be called the
{\em generalized Toeplitz quantization} defined by ($L$, $\mu$, $h$). 
It satisfies $T(\bar{f})=T(f)^\dagger$ and $T(1_X)=\id_E$. Notice that $T(f)$
depends in an essential manner on the measure
$\epsilon \mu $, which is constrained only by condition
(\ref{completeness}). One should contrast this with the generalized Berezin
quantization, which is uniquely determined by the scalar product $(~,~)$. 
Since $h$ characterizes $L$-polarized K{\"a}hler forms, this quantization
is useful for an amply polarized Hodge manifold $(X,\omega, L)$ endowed with a
natural measure $\mu$. Two basic examples are when $\mu$ is the K{\"a}hler
volume form or when $X$ is algebraically Calabi-Yau and $\mu$ is the volume
form defined by the holomorphic top form.

\paragraph{Behavior under automorphisms.} Let us assume that the measure $\mu$
and the scalar product $(~,~)$ on $E$ are invariant under the automorphism group of
$(X,\omega,L,h)$ . Then it is easy to see that the relative epsilon function
(\ref{epsilon}) is $\Aut_{L,h}(X,\omega)$-invariant:
\beq
\nn
\epsilon(\gamma_0(x))=\epsilon(x)~~~~~(\gamma_0\in \Aut_{L,h}(X,\omega)).
\eeq
Together with relation (\ref{rho_proj}), this shows that generalized Toeplitz
quantization is $\Aut(X,$ $\omega,L,h)$-equivariant: 
\beq
\nn
T\circ \tau={\hat \rho}\circ T~~,
\eeq
i.e.:
\beq
\rho(\gamma)T(f)\rho(\gamma)^{-1}=T(f\circ \gamma_0^{-1})~~.
\eeq

\subsection{Relation between generalized Berezin and Toeplitz quantization}

Fixing the scalar product $(~,~)$, we can consider the associated
Berezin quantization as well as any of the Toeplitz quantizations based on an
integral representation of this scalar product defined by a compatible pair
$(\mu,h)$. The relation is given by the {\em generalized Berezin transform},  
the linear map $\beta:{\cal C}^\infty(X)\rightarrow \Sigma$ defined through:
\beq
\nn
\beta:=\sigma\circ T~~,
\eeq
i.e.:
\beq
\label{Berezin_tf}
\beta(f)(x)=\int_{X}{d\mu(y)~\epsilon(y)\Psi(x,y)f(y)}~~,
\eeq
where $\Psi$ is the squared two-point function (\ref{two_point}). 

\

\noindent Adapting an argument of \cite{schlichenmaier-1999}, we obtain the following:

\paragraph{Proposition.}The linear maps $T:{\cal C}^\infty(X)\rightarrow \End(E)$ and
$\sigma:\End(E)\rightarrow {\cal C}^\infty(X)$ are adjoint to each other with
respect to the scalar product $\prec~,~\succ_{\epsilon}$ on
$L^2(X,\mu_\epsilon)$ and the Hilbert-Schmidt scalar product on $\End(E)$. In particular:

(1) $T$ is surjective since $\sigma$ is injective.

(2) The Berezin transform $\beta=\sigma \circ T:{\cal C}^\infty(X)\rightarrow {\cal
  C}^\infty(X)$ is a $\prec~,~\succ_\epsilon$-non-negative Hermitian operator whose image coincides
with $\Sigma$.

(3) We have $\ker T=\ker \beta=\Sigma^{\perp_\epsilon}=\{f\in {\cal C}^\infty(X)|\prec
f,g\succ_\epsilon=0~~\forall g\in \Sigma\}$ and thus $\beta(\Sigma)\subset \Sigma$
and $T(\Sigma)=\End(E)$.

\

\paragraph{Proof.} We have:
\begin{equation}\nn
\langle T(f),C\rangle_{HS}=\tr(T(f)^\dagger
C)=\tr(T({\bar f})C)=\int_{X}d\mu(x)\epsilon(x){\bar f}(x)\sigma(C)=\prec
f,\sigma(C)\succ_{\epsilon}~~.
\end{equation}
If $C$ is $\langle~,~\rangle_{HS}$-orthogonal to $\im T$, it
follows that $\prec f,\sigma(C)\succ_{\epsilon}=0$ for all $f\in {\cal
  C}^\infty(X)$, which implies $\sigma(C)=0$. Therefore $C=0$ by injectivity of $\sigma$. The rest is
obvious. 

\

\noindent We also have: 
\paragraph{Proposition.}(cf.\ \cite{schlichenmaier-1999}) The Berezin operator $\beta:{\cal
  C}^\infty(X)\rightarrow {\cal C}^\infty(X)$ is a contraction with respect to
the sup norm $||f||_\infty=\sup_{x\in X}|f(x)|$, i.e.
\beq
\nn
||\beta(f)||_\infty\leq ||f||_\infty~~\forall f\in {\cal C}^\infty(X)~~.
\eeq

\

The proof is elementary and virtually identical with that in
\cite{schlichenmaier-1999}. It follows that all eigenvalues of $\beta$ are contained in the
interval $[0,1]$. Notice that $\beta$ has the form $\beta=\beta|_\Sigma \pi$
where $\beta|_\Sigma$ is a positive contraction in the finite-dimensional subspace
$\Sigma$ and $\pi$ is the orthoprojector on $\Sigma$. Of course, $\beta$ is
also a contraction with respect to the $L^2$-norm $||~||_\epsilon$ defined by
the measure $\mu_\epsilon=\epsilon \mu$. 

\

Since $T=Q\circ \beta$, the Toeplitz quantization of $f$
is related to the Berezin quantization of $\beta(f)$: 
\beq
\nn
T(f)=Q(\beta(f))~~.
\eeq
After restriction to $\Sigma$, we have a commutative diagram of bijections: 
\beq
\nn
\begin{array}{ccc}
\Sigma & \stackrel{T|_\Sigma}{\longrightarrow} & \End(E) \\
\beta|_\Sigma\downarrow ~~~~~& ~~ &\parallel \\
\Sigma & \stackrel{Q}{\longrightarrow} &\End(E) 
\end{array}
\eeq
where $\beta$ and $T$ depend on the measure $\mu_\epsilon$ but $Q$ and
$\Sigma$ depend only on the scalar product $(~,~)$.  Thus Toeplitz
quantizations associated with different $L^2$-representations of the scalar
product $(~,~)$ on $E$ give different integral descriptions of the Berezin
quantization $Q$ defined by this product. Each Toeplitz quantization is
equivalent with $Q$ via the corresponding Berezin transform.

\paragraph{Remark.} When  $\mu$ and $(~,~)$ are $\Aut_{L,h}(X,\omega)$ and
$\Aut(X,\omega,L,h)$-invariant, respectively, the equivariance properties of $\sigma$ and
$T$ (see eqs. (\ref{sigma_equiv}) and (\ref{Q_equiv})) imply that
the Berezin map $\beta=\sigma\circ T$ commutes with the natural action of $\Aut_{L,h}(X,\omega)$ on ${\cal
  C}^\infty(X)$:
\beq \nn
\beta\circ \tau(\gamma_0)=\tau(\gamma_0)\circ \beta~~~~~(\gamma_0\in \Aut_{L,h}(X,\omega)).
\eeq
In this case, the epsilon function is constant and the action $\tau$ is
unitary on ${\cal C}^\infty(X)$ with respect to each of the $L^2$-scalar products (\ref{mu_product}) and
(\ref{epsilon_product}), while its restriction to $\Sigma$ is also unitary with
respect to the Berezin scalar product.  

\subsection{The Berezin-Toeplitz lift of linear operators} 

Recall that the Toeplitz quantization map $T$ is the Hermitian conjugate $\sigma^\dagger$ of the Berezin symbol map
with respect to the scalar products $\langle~,~\rangle_{HS}$ and
$\prec~,~\succ_{\epsilon}$. This implies that
the Hermitian conjugate $\sigma^\oplus$ of the symbol map with respect to the scalar products
$\langle~,~\rangle_{HS}$ and $\prec~,~\succ$ is given by: 
\beq
\nn
\sigma^\oplus=T \circ M_{\frac{1}{\epsilon}}~~.
\eeq
In particular, we find:
\beq
\nn
\prec \sigma(C_1),\sigma(C_2)\succ=\langle C_1, (\sigma^\oplus\circ
\sigma)(C_2)\rangle_{HS}~~~~(C_1,C_1\in \End(E))
\eeq
i.e.:
\beq
\nn
\prec f,g\succ=\langle Q(f), \sigma^\oplus(g)\rangle_{HS}~~~~(f,g\in \Sigma)~~.
\eeq

\noindent Let us fix a linear operator ${\cal D}:\CC^\infty(X)\rightarrow\CC^\infty(X)$.

\paragraph{Definition.} The {\em Berezin-Toeplitz lift} ${\hat {\cal D}}$ of
${\cal D}$ is the following linear operator on $\End(E)$: 
\beq
\label{lift}
{\hat {\cal D}}:=\sigma^\oplus \circ {\cal D}\circ \sigma=T\circ
M_{\frac{1}{\epsilon}}\circ {\cal D}\circ \sigma:\End(E)\rightarrow \End(E)~~.\nn
\eeq
Explicitly, we have:
\beq
\label{lift_int}
{\hat {\cal D}}(C)=\int_{X} d\mu(x) ({\cal D}\sigma(C))(x)P_x~~~~~(C\in \End(E))~~.
\eeq

The operation of taking the Berezin-Toeplitz lift gives a linear surjection from $\End_\C({\cal C}^\infty(X))$
to $\End_\C(\End(E))$. The identities:
\begin{eqnarray}
\label{ident}
\prec \sigma(C_1),{\cal D} \sigma(C_2)\succ&=&\langle C_1,{\hat {\cal
    D}}(C_2)\rangle_{HS}~~~~~~(C_1,C_2\in \End(E))~~\nn\\
\prec f,{\cal D}(g)\succ~~~~~&=&\langle Q(f),{\hat {\cal
    D}}(Q(g))\rangle_{HS}~~~~~~~(f,g\in \Sigma)~~
\end{eqnarray}
show that the Berezin-Toeplitz lift is well-behaved with respect to the
$L^2$-product $\prec~,~\succ$ on functions defined by $\mu$. This
should be compared with eqs. (\ref{push_pull_ids}) for the
Berezin push and pull. In particular, the Berezin-Toeplitz lift ${\hat {\cal
D}}$ is $\langle~,~\rangle_{HS}$-Hermitian when ${\cal D}$ is
$\prec~,~\succ$-Hermitian and $\langle~,~\rangle_{HS}$-positive when
${\cal D}$ is $\prec~,~\succ$-positive.

The Berezin -Toeplitz lift of the identity operator $\id_{{\cal
    C}^\infty(X)}$ is given by:
\beq
\nn
\nu:=\widehat{
  \id_{{\cal C}^\infty(X)}}=\sigma^{\oplus}\circ \sigma=T\circ M_{\frac{1}{\epsilon}}\circ \sigma~~
\eeq 
and generally does not coincide with the identity operator on $\End(E)$. The integral expression (\ref{lift_int}) gives:
\beq
\nn
\nu(C)=\int_{X} d\mu(x)\sigma(C)(x)P_x=\int_{X} d\mu(x)P_x\tr(P_xC)~~.
\eeq

\noindent Define the {\em modified Berezin  transform} $\beta_{mod}=\beta\circ
M_{\frac{1}{\epsilon}}:{\cal C}^\infty(X)\rightarrow \Sigma$ as follows:
\beq
\label{mod_Berezin_tf}
\beta_{mod}(f)(x)=\int_{X}{d\mu(y)\Psi(x,y) f(y)}~~,
\eeq
where $\Psi$ is the squared two-point function (\ref{two_point}). 
This is obtained by formally replacing $\epsilon$ with $1$ in
(\ref{Berezin_tf}).  Notice that $\im\beta_{mod}=\Sigma$.

\paragraph{Definition.} The {\em Berezin-Toeplitz transform} ${\cal
  D}_\diamond:\Sigma\rightarrow \Sigma$ of ${\cal D}$ is the Berezin pull (\ref{eqn:Bpull})
of ${\hat {\cal D}}$: 
\beq
\label{diamond}
{\cal D}_\diamond:={\hat {\cal D}}_B=\sigma\circ {\hat {\cal D}}\circ
Q=\beta\circ M_{\frac{1}{\epsilon}}\circ {\cal D}|_\Sigma=\beta_{mod}\circ {\hat {\cal D}}|_\Sigma~~.
\eeq
Explicitly, we have:
\beq
\nn
({\cal D}_\diamond f)(x)=\int_{X}d\mu(y)\Psi(x,y)({\cal D}f)(y)~~~~~(f\in \Sigma)~~.
\eeq

The last equation in (\ref{ident}) shows that the bilinear form of ${\cal D}_\diamond$ with respect to the Berezin
scalar product equals the bilinear form of ${\cal D}$ with respect to the
$L^2$-scalar product induced by $\mu$:
\beq
\nn
\prec f,{\cal D}(g)\succ=\prec f, {\cal D}_\diamond (g)\succ_B
    ~~(f,g\in \Sigma)~~.
\eeq
In particular, ${\cal D}_\diamond$ is $\prec~,~\succ_B$-Hermitian iff
${\cal D}$ is $\prec~,~\succ$-Hermitian and
$\prec~,~\succ_B$-positive iff ${\cal D}$ is $\prec~,~\succ$-positive. The
Berezin-Toeplitz transform $\id_\diamond=\nu_B$ of the identity operator $\id_{{\cal
    C}^\infty(X)}$ coincides with the modified Berezin transform: 
\beq
\nn
\id_\diamond=\beta_{mod}~~.
\eeq

\subsection{Changing the scalar product in generalized Toeplitz quantization}

Let us consider what happens when we change the scalar product on $E$ while
keeping $h$ and $\mu$ fixed. Equations (\ref{Pprime}) and (\ref{completeness})
give:
\beq
\label{primecomp}
\int_{X} d\mu (x) \epsilon(x) \sigma(A^{-1})(x) P'_x=A^{-1}~~,
\eeq
i.e.:
\beq
\nn
( s,t)=\int_{X} d\mu (x) \epsilon(x) \sigma(A^{-1})(x) \frac{(
  s|e'_q) '( e'_q|t)'}{( e'_q|e'_q)'}~~,
\eeq
where the scalar product $( s,t)$ on the left hand side is 
the original (unprimed) product. Relations (\ref{productprime}),
(\ref{eprime}) and (\ref{epsilon}) show that the epsilon function of the pair $(h,(~,~)')$ is given by:
\beq
\label{rel2}
\epsilon'(x)=\epsilon(x)\sigma(A^{-1})~~,
\eeq
so (\ref{primecomp}) takes the form:
\beq
\nn
\int_{X} d\mu (x) \epsilon'(x) P'_x=A^{-1}~~,
\eeq
which allows us to express the original scalar product as:
\beq
\nn
( s,t)=( s,A^{-1}t)'=\int_{X} d\mu (x) \epsilon'(x)( s|P'_x|t)'=
\int_{X} d\mu (x) \epsilon'(x)\frac{\overline{{\hat q}(s)}\hat{q}(t)}{( e'_q|e'_q)'}~~.
\eeq
The original epsilon function can be recovered from equations (\ref{rel2}) and
(\ref{rel1}):
\beq
\nn
\epsilon(x)=\sigma'(A)\epsilon'(x)~~,
\eeq
while the original Rawnsley coherent states can be recovered as $e_q=Ae'_q$. In
principle, this allows us to recover the original Toeplitz quantization from
knowledge of $(~,~)'$. We can define an operator: 
\beq
\nn
T'(f):=\int_{X} d\mu (x) \epsilon'(x) f(x)P'_x~~,
\eeq
which satisfies $T'(f)^\oplus=T({{\bar f}})$ as well as:
\beq
\nn
\tr(AT'(f))=\int_{X} d\mu (x) \epsilon(x)f(x)=\tr(T(f))
\eeq
and: 
\beq
\nn
T'(1_X)=A^{-1}~~.
\eeq

In practice, one is often interested in the case when $(~,~)$ is the $L^2$-scalar product $\langle~,~\rangle^{\mu,h}_1$ defined by $\mu$ and $h$:
\beq
\nn
(s,t):=\langle s,t\rangle^{\mu,h}_1=\int_Xd\mu(x) h(x)(s(x),t(x))~~.
\eeq
In such a situation one might be able to compute the coherent states and
epsilon function with respect to another scalar product $(~,~)'$
on $E$. Then the expressions above allow one to recover the Toeplitz
quantization with respect to the $L^2$-scalar product of $(\mu,h)$. 

We can also ask about relating the Berezin quantization $Q'$ defined by $(~,~)'$
to the Toeplitz quantization defined by the $L^2$-scalar product $\langle~,~\rangle^{\mu,h}_1$. The two quantizations are related by the map
$\beta:=\sigma'\circ T:{\cal  C}^\infty(X)\rightarrow \Sigma'$:
\beq
\nn
T(f):=Q'(\beta(f))~~.
\eeq
We have the integral expression
\beq
\nn
\beta(f)(x)=\int_{X}d\mu(y)\epsilon(y) f(y) \lambda(x,y)~~,
\eeq
where
$\lambda(x,y)=\sigma'(P_y)(x)=\frac{\sigma(A^{-1}P_y)(x)}{\sigma(A^{-1})(x)}=\frac{\tr(A^{-1}P_yP_x)}{\tr(A^{-1}P_x)}=
\frac{\sigma(A^{-1}(y))}{\sigma(A^{-1}(x))}\sigma(P'_y(x))$. Notice that
$\lambda(x,y)$ need not equal $\lambda(y,x)$.

\subsection{Extension to powers of $L$}

We can easily extend everything by replacing the very ample line bundle $L$ with any of its
positive powers $L^k:=L^{\otimes k}$ ($k\geq 1$). In this case,
generalized Berezin quantization requires a choice of Hermitian scalar
products $(~,~)_k$ on each of the finite-dimensional vector 
spaces $E_k:=H^0(L^k)$. Accordingly, we have coherent states
$e_x^{(k)}\in E_k$ and Rawnsley projectors $P_x^{(k)}$, as well as surjective Berezin
symbol maps $\sigma_k:\End(E_k)\rightarrow {\cal C}^\infty(X)$ whose images we
denote by $\Sigma_k$. The inverse of the corestrictions $\sigma_k|^{\Sigma_k}$
define a sequence of Berezin quantization maps $Q_k:\Sigma_k\rightarrow
\End(E_k)$. The entire construction depends crucially on the precise sequence
of Hermitian scalar products $(~,~)_k$ chosen for the spaces
$E_k$.

\section{Classical Berezin and Toeplitz quantization of compact\newline Hodge manifolds}

Classical Berezin and Toeplitz quantization of prequantized Hodge
manifolds $(X,\omega,$ $L,h)$ arises as a particular case of the generalized
constructions discussed above. In the classical set-up, one fixes an integer $k_0$
such that $L^{k_0}$ is very ample. For each integer $k\geq k_0$, we apply the general
construction for the ample line bundle $L^k$ endowed
with the Hermitian scalar product $h_k:=h^{\otimes k}$ and with the
$L^2$-scalar product $\langle~,~\rangle_k$ on $E_k:=H^0(L^k)$ induced by $h_k$
and by the Liouville measure $\mu_\omega$ defined by $\omega$, see eq. (\ref{L2}).

\subsection{Classical Berezin and Toeplitz quantization}

Applying Berezin quantization with the choices above leads to
bijective symbol maps $\sigma_k:E_k\rightarrow \Sigma_k\subset {\cal
  C}^\infty(X)$ and quantization maps $Q_k:\Sigma_k\rightarrow
\End(E_k)$. We also have surjective Toeplitz quantization maps $T_k:{\cal  C}^\infty(X)\rightarrow \End(E_k)$
given by $T_k(f):=\Pi_k (f\cdot)$, where $\Pi_k:L^2_k(L,
h,\mu_\omega)\rightarrow E_k$ is the orthoprojector on the subspace of
holomorphic sections. The surjective Berezin transforms $\beta_k=\sigma_k\circ T_k:{\cal
  C}^\infty(X)\rightarrow \Sigma_k$ relate the two
quantizations via $T_k=Q_k\circ \beta_k$. The maps $\beta_k$ and $T_k$ have equal 
kernel $\Sigma_k^{\perp_{\epsilon_k}}\subset {\cal C}^\infty(X)$ and the restrictions of $\beta_k$ and $T_k$
give linear isomorphisms. For each $k\geq k_0$, we have a commutative diagram of
bijections: 
\beq
\nn
\begin{array}{ccc}
\Sigma_k & \stackrel{T_k|_\Sigma}{\longrightarrow} & \End(E_k) \\
\beta_k|_\Sigma\downarrow ~~~~~& ~~ &\parallel \\
\Sigma_k & \stackrel{Q_k}{\longrightarrow} &\End(E_k) 
\end{array}
\eeq
where $\beta_k|_{\Sigma_k}$ is a strictly positive self-adjoint contraction. 

It is convenient to consider the Hilbert space: 
\beq
\label{eqn:hilbert_direct_sum}
{\cal E}_X:=\overline{\oplus}_{k=0}^\infty (E_k,\langle~,~\rangle_k)~~.
\eeq
Recall that the completed direct sum ${\cal E}_X$ consists of all infinite
sequences $s=\sum_{k=0}^\infty{s_k}$ with $s_k\in E_k$ which satisfy
the condition: 
\beq
\nn
\sum_{k=0}^\infty{\langle s_k, s_k\rangle_k}<\infty~~.
\eeq
It is endowed with the scalar product:
\beq
\label{eqn:direct_sum_sp}
\langle s,t\rangle_X=\sum_{k=0}^\infty{ \langle s_k, t_k\rangle_k}~~.
\eeq
Then the $E_k$ become closed subspaces of ${\cal E}_X$ and the scalar products
$\langle s_k, t_k\rangle_k$ coincide with the restriction of
$\langle~,~\rangle_X$. 

Similarly, we let $H_k:=L^2_k(L^k,\mu,h)$ be the $L^2$-completion of $\Gamma(L^k)$
and ${\cal H}_X:=\overline{\oplus}_{k=0}^\infty H_k$ be the Hilbert direct
sum of $H_k$. Then ${\cal E}_X$ is
a closed subspace in ${\cal H}_X$ and the orthoprojector $\Pi$ on the former
decomposes as:
\beq
\label{Szego}
\Pi=\sum_{k=0}^\infty{\Pi_k}~~,
\eeq
where $\Pi_k$ is the orthoprojector on $E_k$ inside $H_k$. The
projector (\ref{Szego}) is sometimes called the {\em Szeg{\"o} projector}. 

\paragraph{Remark.} Consider the total space ${\Bbb S}$ of the unit circle bundle of $L^*$. This is a
Cauchy-Riemann (CR) manifold of CR-codimension one, whose CR-structure is induced by its
obvious embedding as a real hypersurface in the total space $\L$ of $L^*$. Moreover, it is the boundary of
the total space ${\Bbb D}$ of the unit disk bundle of $L^*$, which is known to be a
strictly pseudoconvex domain in $\L$. The K{\"a}hler form $\omega$ of $X$ induces a contact
one-form $\alpha$ on ${\Bbb S}$ such that the pull-back of $\omega$
through the projection of the circle bundle equals $d\alpha$. The {\em Hardy space} is the Hilbert space of all CR-holomorphic functions on
${\Bbb S}$, endowed with the $L^2$-scalar product induced by the volume form
$\alpha\wedge (d\alpha)^{\dim X}$ of the contact form $\alpha$. This is the
usual Hardy space of boundary values of holomorphic functions on the domain ${\Bbb D}$.
It is well-known that the Hardy space is isometric to the Hilbert space
${\cal E}_X$. Because of this, we will identify the two and sometimes
refer to the latter as the Hardy space. Similarly, the space of
$L^2$-functions on ${\Bbb S}$ identifies with ${\cal H}_X$. 

\

\subsection{The formal star products and associated deformation quantizations}

Let $\{~,~\}$ be the Poisson bracket defined by $\omega$, $h$ be a formal variable and consider the $\C[[h]]$-module ${\cal
  C}^\infty(X)[[h]]$ of formal power series with smooth function
coefficients. Recall that a normalized formal star product on $X$ is a $\C[[h]]$-bilinear map $\star:{\cal C}^\infty(X)[[h]]\times {\cal
  C}^\infty(X)[[h]]\rightarrow {\cal C}^\infty(X)[[h]]$ on this module
such that:

(a) $\star$ is associative

(b) The coefficients of the formal expansion\footnote{Notice that these completely
  determine the star product.}: 
\beq
\nn
f\star g=\sum_{n=0}^\infty{h^j C_j(f,g)}~~~~(f,g\in {\cal C}^\infty(X))
\eeq
are bi-differential operators satisfying the identities: 

1. $C_0(f,g)=fg$

2. $C_1(f,1)=C_1(1,f)=0$

3. $C_1(f,g)-C_1(g,f)=\frac{i}{2\pi}\{f,g\}$

\noindent Notice that we use an expansion in $h$ rather than $\hbar=\frac{h}{2\pi}$ due to our
convention for integral symplectic forms ($[\omega]\in H^2(X,\Z)$), as
required by agreement with the Bohr-Sommerfeld condition.

\paragraph{The Toeplitz deformation quantization.} 

It was shown in \cite{schlichenmaierb-1999} that there exists a unique normalized formal star product $\star_T$
on $X$ (known as the Toeplitz star product) whose coefficients $C_j$ have the property: 
\beq
\nn
||T_k(f)T_k(g)-\sum_{j=0}^{N}\frac{1}{k^j}T_k(C_j(f,g))||_k=K_p(f,g)\frac{1}{k^N}
\eeq
for all $m$ and all sufficiently large $k$. Here, $||~||_k$ is the operator norm on
$\End(E_k)$ and the $K_p(f,g)$ are constants which depend only on $p$ and
$f,g$. This can be interpreted as an asymptotic expansion:
\beq
\label{as}
T_k(f)T_k(g)\sim \sum_{j=0}^{\infty}\frac{1}{k^j}T_k(C_j(f,g))~~{\rm
  for}~~k\rightarrow \infty~~,
\eeq
where the right hand side formally corresponds to $T_k(f\star_T g)$ at
$h=\frac{1}{k}$ (here $h=2\pi\hbar$). It should be stressed that the formal star product $\star_T$ captures the entire
asymptotic expansion (\ref{as}), which includes information from {\em all}
values of $k\geq k_0$. The Toeplitz
star product has `anti-separation of variables' in the sense that $a\star_T g=f\star_T b=0$
whenever $a$ is antiholomorphic and $b$ is holomorphic. 

\paragraph{Remark.} One has \cite{Bordemann:1993zv}:
\beq
\nn
||T_k(f)T_k(g)-T_k(fg)||_k =O\left(\frac{1}{k}\right)~~{\rm  for}~~k\rightarrow \infty~~
\eeq
as well as:
\beq
\nn
||ik[T_k(f),T_k(g)]-T_k(\{f,g\})||_k=O\left(\frac{1}{k}\right)~~{\rm  for}~~k\rightarrow \infty~~
\eeq
and:
\beq
\nn
||f||_\infty-\frac{C_f}{k}\leq ||T_k(f)||_k \leq ||f||_\infty 
\eeq
for all $f\in {\cal C}^\infty$ and some constant $C_f$ depending only on
$f$. In particular, one finds $\lim_{k\rightarrow \infty}
||T_k(f)||=||f||_\infty$. These properties imply that
the continuous field of $C^*$-algebras on the set
$I=\{\frac{1}{k}|k\in \N^*\}\cup \{0\}$ given by $A_\frac{1}{k}:=(\Sigma_k,
||~||_k)$, with $A_0:=({\cal C}^\infty(X), ||~||_\infty)$ and
section $\frac{1}{k}\rightarrow T_k$, $T_0:=\id_{{\cal C}^\infty(X)}$
forms a `strict quantization' in the sense of Rieffel \cite{Rieffel} though not
a `strict deformation quantization'.

\paragraph{The Berezin deformation quantization.}

It was further shown in \cite{karabegov-2000} that the Berezin transform $\beta_k$ has an
asymptotic expansion $\beta_k\sim\sum_{r=0}^\infty{\frac{1}{k^r}\beta_{r}}$ with $\beta_0=1$. This 
allows one to define an automorphism of the $\C[[h]]$-module ${\cal
  C}^\infty(X)[[h]]$, known as the {\em formal Berezin transform}, via:
\beq
\label{eqn:formalberezin}
\bbeta=\sum_{r=0}^\infty{\beta_{r}h^r}~~.
\eeq
The Berezin star product $\star_B$ is the formal normalized star product obtained
from $\star_T$ via the formal Berezin transform, $f\star_Bg=\bbeta(\bbeta^{-1}(f)\star_T
\bbeta^{-1}(g))$. Again, it should be stressed that $\star_B$ contains information
from all powers $L^k$, $k\geq k_0$. We have the relation: 
\beq
\nn
\bbeta(f\star_T g)=\bbeta(f)\star_B \bbeta(g)~~.
\eeq
The Berezin star product has `separation of variables' in the sense of \cite{Karabegov-1999}, i.e.\ one has $a\star_Bg=f\star_B b=0$ whenever $a$ and $b$ are
holomorphic and antiholomorphic functions, respectively \cite{schlichenmaier-1996}.

\subsection{Relation with geometric quantization} 

In Konstant-Souriau geometric quantization of K{\"a}hler manifolds \cite{geomq},
one considers the linear maps $\Theta_k:{\cal
  C}^\infty(X)\rightarrow \End(E_k)$ given by: 
\beq
\label{eqn:ksgq}
\Theta_k(f):=\Pi_k \theta_k(f) ~~,
\eeq
where 
\beq 
\theta_k(f):=-i\nabla^{(k)}_{X^k_f}-f\cdot ~~. \nn
\eeq
$X_f$ is the Hamiltonian vector field
defined by the smooth function $f$ with respect to the symplectic form
$k\omega$ and $\nabla^{(k)}$ is the Chern connection on $L^k$. This procedure corresponds to using
the so-called {\em complex polarization}. One has
the following relation \cite{Tuynman:1987jc}:
\beq
\nn
\Theta_k(f)=T_k\left(f-\frac{1}{2k}\Delta f\right)~~\forall f\in {\cal
  C}^\infty(X)~~,
\eeq
where $\Delta$ is the Laplace operator of $(X,\omega)$ -- at least on compact symmetric spaces. Notice that we use
conventions in which $\Theta_k(f)^\dagger=\Theta_k({\bar f})$.

\subsection{The Berezin product or coherent state star product}

In the vast majority of the literature on fuzzy geometry, the ``star product''
used is the Berezin product $\diamond_k:\Sigma_k\times\Sigma_k\rightarrow \Sigma_k$
introduced in Section 3.2:
\begin{equation}
\label{eqn:Bproduct_k} 
f\diamond_k g\ :=\
\sigma_k(Q_k(f)Q_k(g))~,~~~f,g\in\Sigma_k~~.
\end{equation} 
This product is also called the {\em coherent state star product}, as
$\sigma_k(C)=\tr(C P^{(k)}_x)$ is determined by the coherent states. It is associative by definition and the algebra
$(\Sigma_k,\diamond_k,\bar{~})$ is isomorphic as a $*$-algebra to
$(\End(E_k),\circ,\dagger)$ with the Berezin quantization map $Q_k$ providing
the isomorphism.

Note that the Berezin product is {\em not} a formal star product as it is
defined only on $\Sigma_k$, instead of $\CC^\infty(X)[[h]]$. However, it
has been shown \cite{Rawnsley2} that in the case of flag manifolds, there is a
formal differential star product on the set
$\Sigma_\bullet:=\bigcup_{k=0}^\infty\Sigma_k$, which agrees with the asymptotic
expansion of the Berezin products on $\Sigma_k$ for certain
$h=\frac{1}{k}$.

As an example, consider the Berezin quantization of $(\P^n,\omega_{FS})$ with
the prequantum line bundle $H^k$, where $H$ is again the hyperplane line
bundle. If we normalize the homogeneous coordinates on $\P^n$ by demanding
that $|z|=1$, we obtain the particularly simple form \cite{Kurkcuoglu:2006iw}:
\begin{equation} 
\nn
f\diamond_k
g=\sum_{i_1,...,i_k}\left(\frac{1}{k!}\der{z_{i_1}}...\der{z_{i_k}}f\right)\left(\frac{1}{k!}\der{\bz_{i_1}}...\der{\bz_{i_k}}g\right)~.
\end{equation} 
A different form of the Berezin product corresponding to a finite sum resembling
the first terms in an expansion of a formal star product can be written down
in the real setting, using the embedding $\P^n\embd\R^{(n+1)^2-1}$
\cite{Balachandran}.

\subsection{The quantization of affine spaces}

Classical Berezin-Toeplitz quantization can be extended to the non-compact
case\footnote{In fact, historically this was the original class of examples.}
upon replacing the space of holomorphic sections of the quantum line bundle
$L$ with the subspace of those holomorphic sections which are square
integrable with respect to an appropriately weighted version of the Liouville
measure $\frac{\omega^n}{n!}$. In particular, this can be applied to the case
of complex affine spaces, where the weight is provided by the global
K{\"a}hler potential, leading to the well-known construction of the Bargmann
representation of the bosonic Fock space. In this subsection, we recall this
construction in order to fix our notations for later use.

Let us start with a few remarks about the coordinate-free description. If $V$
is an $n+1$-dimensional complex vector space, the algebra $B:=\C[E]$ of polynomial
functions over the dual space $E:=V^*=\Hom_\C(V,\C)$ is the symmetric algebra associated with $V$:
\beq
\label{eqn:symmetric_algebra}
B=\C[E]=\oplus_{k=0}^\infty E^{\odot k}=\oplus_{k=0}^\infty (V^*)^{\odot k}~~.
\eeq
We let $B_k:=E^{\odot k}\subset B$. As an algebraic variety, the affine space over $V$ is the affine spectrum
$A(V)=\Spec B$ of this algebra. A choice of basis $e_0\ldots e_n$ for $V$
allows us to define coordinate functionals $z_j\in E=\Hom_\C(V,\C)$ such
that $z_j(v)=v_j$ for all $v=\sum_{j=0}^n{v_je_j}\in V$. Thus $(z_j)$ is the basis of $E$ dual to the given basis of $V$. Once a basis of
$V$ has been fixed, we can write the elements of $B$ as polynomial
functions over $V$:
\beq
\nn
f=\sum_{|\p|={\rm bounded}}{a_\p\chi_\p}~~,
\eeq
where $\p=(p_0\ldots p_n)\in \N^{n+1}$, $|\p|:=\sum_{j=0}^n{p_j}$ and:
\beq
\label{c}
\chi_\p:=z^\p:=z_0^{p_0}\ldots z_n^{p_n}~~.
\eeq
We denote the symmetrized tensor product $\odot$ by
juxtaposition. As a function on $V$, we have:
\beq
\nn
f(v)=\sum_{|\p|}a_\p v_0^{p_0}\ldots v_n^{p_n}~~.
\eeq
If we use the given basis to identify $V$ with $\C^{n+1}$, then $v$ identifies
with the vector $(v_0\ldots v_n)$ and we obtain the polynomial function
$f(v_0\ldots v_n)=\sum_{|\p|}a_\p v_0^{p_0}\ldots v_n^{p_n}$ on $\C^{n+1}$. In
this case, $B$ identifies with the polynomial algebra $\C[v_0\ldots v_n]$
in $n+1$ variables, which is the coordinate ring of $\C^{n+1}$. 

Given a Hermitian scalar product $(~,~)$ on $E$, we have an induced product on
$V$ (denoted by the same symbol) and can chose the basis $e_0\ldots e_n$ to be orthonormal with respect
to this induced product. In this case, the basis $z_0\ldots z_n$ of $E$ is
also orthonormal and the scalar product of two elements of $V$ can be written as:
\beq
\nn
(u,v)={\bar z}(u)\cdot z(v)=\sum_{j=0}^n{\overline{z_j(u)} z_j(v)}=\sum_{j=0}^n{\overline{u_j}v_j}~~,
\eeq
where $z(v):=(z_0(v)\ldots z_n(v))=(v_0\ldots v_n)$. Notice that
$||v||^2=|z(v)|^2=\sum_{j=0}^n{|z_j(v)|^2}$. The scalar product induces a flat Hermitian metric on $V$ whose K{\"a}hler form:
\beq
\nn
\omega=\frac{i}{2\pi}\sum_{j=0}^{n}{dz_j\wedge d{\bar z}_j}
\eeq
corresponds to the standard symplectic form on the underlying real vector
space $V_\R$ of $V$ if we set
$z_j=\frac{1}{\sqrt{2}}(q_j+ip_j)$, where $q_j,p_j\in \Hom_\R(V,\R)$ are real
coordinates on $V_\R$:
\beq
\nn
\omega=\frac{1}{2\pi}\sum_{j=0}^{n}{dq_j\wedge dp_j}~~.
\eeq
Since $\frac{\omega^{n+1}}{(n+1)!}=\frac{1}{(2\pi)^{n+1}}dq_0\wedge dp_0\wedge \ldots \wedge
dq_n\wedge dp_n$, the associated Liouville measure is the scaled Lebesgue
measure $d\mu=\frac{1}{(2\pi)^{n+1}}d^{n+1}q d^{n+1} p$ on
$V_\R$. The K{\"a}hler form is polarized with respect to the trivial line
bundle  ${\cal O}=V\times \C$, whose unit section we denote by $s_0=1$ (this
is just the unit constant function on $V$). ${\cal O}$ becomes a quantum line
bundle when endowed with the Hermitian metric $h$ given by:
\beq
\nn
{\hat h}(v):=h(v)(s_0(v),s_0(v)):=e^{-|z(v)|^2}=e^{-||v||^2}~~.
\eeq
The unit section gives the global K{\"a}hler potential:
\beq
\nn
K(v)=-\log {\hat h}(v)=|z(v)|^2=||v||^2~~.
\eeq
The holomorphic sections of ${\cal O}$ are simply the entire functions
$f$ on $V$, since every such section can be written as $s_f=fs_0$.
The $L^2$-scalar product (subsequently to be referred to as the \emph{Bargmann product}) is: 
\beq
\label{flat}
\langle f,g\rangle_B:=\langle s_f, s_g\rangle= \int_{V}d\mu(v)
  e^{-|z(v)|^2} {\bar f}(v)g(v)=\int_{V}d\nu(v) {\bar f}(v)g(v)~~,
\eeq
where:
\beq
\nn
d\nu(v)=e^{-|z(v)|^2}d\mu(v)=\frac{1}{(2\pi)^{n+1}}e^{-\frac{1}{2}\sum_{j=0}^n{(q_j^2+p_j^2)}}~~d^{n+1}qd^{n+1}p
\eeq
is the weighted Lebesgue measure, which is normalized to unit total mass: 
\beq
\nn
\int_{V}{d\nu}=1~~.
\eeq
The space of square integrable holomorphic sections of ${\cal O}$ is the well-known weighted
Bargmann space ${\cal B}(V):=L^2_{\rm hol}(V, d\nu)$ of
$\nu$-square integrable entire functions on $V$, which contains the algebra $B=\C[E]$ of polynomial functions as a dense subspace. The Bargmann space
carries the unitary representation of the $n+1$-th Weyl group with creation and annihilation operators given by:
\beq
\nn
({\hat a}_i^\dagger f)(v) :=z_i(v) f(v)~~,~~({\hat a}_i f)(v) =\partial_i f(v)~~,
\eeq
where $\partial_i=\frac{\partial}{\partial e_i}$ is the directional derivative
along $e_i$. We have:  
\beq
\nn
[{\hat a}_i, {\hat a}_j^\dagger]=\delta_{ij}~~. 
\eeq
The normalized vacuum vector is the constant unit function $|0\rangle:=1$. For every tuple $\p=(p_0\ldots
p_n)\in \N^{n+1}$, let $\p!:=p_0!\ldots p_n!$. 
We have: 
\beq
\label{chi}
~~||\chi_\p||_B= \sqrt{\p!}~~,
\eeq
where $\chi_\p$ are the monomials (\ref{c}). The normalized occupation vectors are given by:
\beq
\label{occ}
|\p\rangle=\frac{1}{\sqrt{\p!}}\chi_\p=\frac{({\hat a}^\dagger)^\p}{\sqrt{\p!}}|0\rangle~~.
\eeq
They are the common eigenvectors of the particle number operators ${\hat
  N}_i={\hat a}_i^\dagger {\hat a}_i$:
\beq
\nn
{\hat N}_i|\p\rangle=p_i|\p\rangle~~. 
\eeq
An entire function: 
\beq
\label{Taylor}
f=\sum_{\p\in \N^{n+1}}{c_\p
  \chi_\p}=\sum_{\p\in \N^{n+1}}{c_\p \sqrt{\p!} |\p\rangle} ~~~(c_\p\in \C)~~
\eeq
belongs to ${\cal B}$ iff 
\beq
\nn
\langle f|f\rangle_B=\sum_{\p\in \N^{n+1}}{\p ! |c_\p|^2} <\infty~~.
\eeq

Defining the total particle number operator ${\hat N}=\sum_{i=1}^{n}{{\hat
    N}_i}$, relation (\ref{occ}) shows that its eigenspace of eigenvalue $k$
coincides with $B_k$. We have the orthogonal decomposition ${\cal B}=\overline{\oplus}_{k=0}^\infty
B_k$ (completed direct sum) with $B_k=\ker({\hat N}-k)$. It is easy to see
that ${\cal B}$ is unitarily isomorphic with the bosonic Fock space ${\cal
  F}_s(E)=\overline{\oplus}_{k=0}^\infty E^{\odot k}$ over
the finite-dimensional Hilbert space $(E,(~,~))$. Under this identification,
$|\p\rangle$ becomes the orthonormal basis of the Fock space canonically
associated  with the orthonormal basis  $(z_0\ldots z_n)$ of $E$. 

Since ${\cal O}$ has a global nowhere vanishing section (the unit section $s_0$), we
can consider Rawnsley's coherent vectors with respect to $q=s_0(v)=1\in {\cal O}_v$. These are the usual Glauber vectors: 
\beq
\nn
|v\rangle=e^{\sum_{i=0}^n{\bar
    v}_i{\hat a}_i^\dagger}|0\rangle=\sum_{\p}{\frac{\overline{v}^\p}{\sqrt{\p!}}|\p\rangle}~~,
\eeq
where $\overline{v}^\p=\overline{v_0}^{p_0}\ldots \overline{v_n}^{p_n}$. One has the well-known identity:
\beq
\nn
f(v)=~\langle v|f\rangle_B~~~~(f\in {\cal B})~~.
\eeq
We have:
\beq
\nn
\hat{a}_i|v\rangle={\bar v}_i|v\rangle~~,~~\langle u|v\rangle_B=e^{(v,u)}~~. 
\eeq
In particular $|v\rangle$ has norm $e^{||v||^2}$.  The reproducing
kernel is the well-known Bergman kernel: 
\beq
\nn
K_B(u,v)=\frac{\langle u|v\rangle}{\sqrt{\langle u|u\rangle \langle
    v|v\rangle}}=e^{-\frac{1}{2}(|u|^2+|v|^2)+(v,u)}~~.
\eeq
The Rawnsley projector is\footnote{Recall that $\langle v|_B$ stands for the
  linear functional $\langle v|_B(\psi):=\langle v|\psi\rangle_B$ which is
  Riesz dual to the  vector $|v\rangle$. Since Riesz duality depends on the choice
  of scalar product on ${\cal B}$, we use an underscript $B$ on
  bra vectors.} $P_v=\frac{1}{\langle v| v\rangle_B}|v\rangle \langle
v|_B=e^{-||v||^2} |v\rangle \langle v|_B$. The epsilon function is constant and
equal to one: 
\beq
\nn
\epsilon_{\C^{n+1}}={\hat h}(v)\langle v|v\rangle_B=1~~. 
\eeq
The decomposition of the identity takes the form: 
\beq
\nn
\int_{V} d\mu(v) P_v=1\Leftrightarrow \int_{V} d\mu(v)
e^{-||v||^2}|v\rangle \langle v|_B=1\Leftrightarrow 
\int_{V} d\nu(v) |v\rangle \langle v|_B=1 ~~.
\eeq 

\paragraph{Toeplitz quantization of $A(V)$.}The Toeplitz quantization of $f\in {\cal C}^\infty(V,\C)$ is given by:
\beq
\label{plane_T}
T(f)=\int_{V}d\mu(v) f(v)P_v=\int_{V} d\mu(v) e^{-||v||^2} f(v)|v\rangle \langle v|_B~~.
\eeq
In particular, we have $T(z_i)={\hat  a}_i^\dagger$ and $T({\bar
    z}_i)={\hat a}_i$.  When $f$ is a polynomial in $z$ and $\bar{z}$, (\ref{plane_T})
obviously reduces to the anti-Wick prescription:
\beq
\nn
T(f)=\vdots f({\hat a}^\dagger, {\hat a})\vdots~~,
\eeq
where the triple dots indicate antinormal ordering. In this case, $T$ is not
surjective due to the infinite-dimensionality of the Bargmann space. 

\paragraph{Berezin quantization of $A(V)$.}

The Berezin symbol map $\sigma:{\cal L}({\cal B})\rightarrow {\cal
  C}^\infty(V,\C)$ is defined on the algebra ${\cal L}({\cal B})$ 
of bounded operators in the Bargmann space. The symbol of a bounded operator $C$ takes the form: 
\beq
\nn
\sigma(C)(v)=e^{-||v||^2}\langle v|C|v\rangle_B~~,
\eeq
while the Berezin transform $\beta(f)=\sigma\circ T$ is given by: 
\beq
\nn
\beta(f)(u)=\int_{V} d\mu(v) f(v)e^{-||u-v||^2}~~.
\eeq
Thus $\beta=\frac{1}{2^{n+2}}e^{-4\Delta}$, where $\Delta$ is the Laplacian on
  $V_\R$; this is the heat kernel up to normalizations.  

The symbol map gives rise to the Berezin quantization
$Q:\Sigma\rightarrow {\cal L}({\cal B})$, where $\Sigma\subset {\cal
  C}^\infty(\C^{n+1})$ is the image of $\sigma$. We have $Q(z_i)={\hat a}_i^\dagger$ and
$Q({\bar z}_i)={\hat a}_i$. For a polynomial function
$f(z,{\bar z})$, we find: 
\beq
\nn
Q(f)=:f({\hat a}^\dagger, {\hat a}):~~,
\eeq
where the double dots indicate normal ordering. Hence both quantization
prescriptions send $z_i$ into ${\hat a}_i^\dagger$ and ${\bar z}_i$ into
${\hat a}_i$, but Toeplitz quantization corresponds to anti-Wick ordering,
while Berezin quantization corresponds to Wick ordering. 

\paragraph{Truncated coherent vectors.} For later use, consider the expansion of Glauber's
coherent vectors in components of fixed total particle number:
\beq
\nn
|v\rangle=\sum_{k=0}^\infty |v,k \rangle~~,
\eeq
where the `truncated coherent vectors'
\beq
\nn
|v,k\rangle:=\frac{1}{k!}\left(\sum_{i=0}^n{\bar
  v}_i{\hat a}_i^\dagger\right)^k|0\rangle=\sum_{|\p|=k}{\frac{\bar{v}^\p}{\sqrt{\p!}}|\p\rangle}
\eeq
satisfy 
\begin{eqnarray}
{\hat N}|v,k\rangle&=&k|v,k\rangle\nn\\
\langle u,k|v,l\rangle_B&=&\delta_{k,l}\frac{1}{k!}[(v,u)]^k~~. \nn
\end{eqnarray}
In particular, we have $\langle v,k|v,k\rangle_B=\frac{1}{ k!}||v||^{2k}$. Since
$[\hat{N},{\hat a}_i]=-1$ and $[\hat{N},{\hat a}_i^\dagger]=+1$, we find: 
\beq
\label{truncated}
{\hat a}_i |v,k\rangle=\bar{z_i}|v,k-1\rangle~~.
\eeq
Notice that $|\lambda v,k\rangle={\bar \lambda}^k|v,k\rangle$ for any $\lambda
\in \C$, and therefore the ray $\C^*|\lambda v,k\rangle$ depends only on the
image $[v]$ of $v$ in the projective space $\P V=(V\setminus
\{0\})/\C^*$.

\subsection{The quantization of complex projective spaces}

We next consider the case of complex projective spaces, which has been studied
extensively in the literature on Berezin quantization.  Using the
`truncated coherent vectors' of the previous subsection, we will show that the (yet to be defined)
Berezin-Bergman quantization of $\P^n$ coincides with its Berezin
quantization. In particular, the fuzzy version of $\P^n$ considered in
\cite{Saemann:2006gf} coincides with its well-known Berezin
quantization\footnote{In
particular, the fuzzy two-sphere of \cite{Madore} is simply the
classical Berezin quantization of $\P^1$.}. To make contact with the 
formalism used in \cite{Saemann:2006gf}, we will use the fact that the homogeneous
coordinate ring of $\P^n$ can be identified with the affine coordinate ring
of $\C^{n+1}$, provided that the latter is endowed with the canonical grading $\deg
z_i=1$. It follows that both the Bargmann space of $\C^{n+1}$ and the
Hardy space of $\P^n$ are Hilbert space completions of the ring of polynomials
in $n$ complex variables, albeit with respect to
different scalar products: the former uses the scalar product induced by the
flat metric on $\C^{n+1}$, while the latter uses the scalar product induced by
the Fubini-Study metric on $\P^n$. According to our general discussion, the
relation between the restriction of these scalar products to $B_k$ should be provided by
isomorphisms $A_k$ which relate the Berezin quantizations of $\C^{n+1}$ and $\P^n$. 
In the case at hand, $A_k$ will be proportional
to $1_{B_k}$, with a $k$-dependent proportionality constant. This implies
that the Berezin quantizations of $\P^n$ and $\C^{n+1}$ agree at
any fixed level $k$. Such an extremely simple relation is not to
be expected in general, but is a consequence of the fact that $\P^n$ is a
homogeneous space, which is why any reasonable quantization procedure for this
space leads to the same result.

Consider an $n+1$-dimensional complex vector space $V$ and its dual $E=V^*$ as
in the previous subsection. As an algebraic variety, the projective space
$\P V=(V\setminus \{0\})/\C^*$ over $V$ is given by ${\rm Proj} B$, where $B=\C[E]$ is viewed as a graded
algebra with respect to the obvious grading. For any vector $v\in V$, we let $[v]$ be the corresponding point in
$\P V$. Recall that the tautological bundle $\tau:={\cal  O}_{\P(V)}(-1)$ has a fiber
equal to the line $\tau_v=\C v\subset V$ above the point $[v]\in \P V$. The dual bundle $H:={\cal
  O}_{\P(V)}(1)$ is the hyperplane bundle, which is very ample. Any functional 
$s\in B_k=(V^*)^{\odot k}$ determines and is determined by a holomorphic
section $S\in H^0(H^k)$, namely: 
\beq
\nn
S([v])=s|_{\tau_v^{\otimes k}}\in (\tau_v^*)^{\odot k}~~~~,
\eeq
so $H^0(H^k)$ identifies with $B_k$. Hence the graded algebra $B$ identifies with the
homogeneous coordinate ring $\oplus_{k=0}^\infty H^0(H^k)$ of $\P V$ with
respect to $H$.  In particular, $H^0(H)$ identifies with $B_1=E$. 

As in the previous section, a basis $(e_0\ldots e_n)$ of $V$ determines
coordinate functionals $z_j\in E$. The corresponding holomorphic sections
$Z_j\in H^0(H)$ are the homogeneous coordinates of $\P V$ associated with the
given basis. The homogeneous polynomials (\ref{occ}) for $|\p|=k$ provide a basis for $B_k$. We have: 
\beq
\nn
\dim B_k=N_k+1=\frac{(n+k)!}{n!k!}~~. 
\eeq

Fixing a scalar product  $(~,~)$ on $E$, we have an induced scalar product on
$V$ and take the basis $(e_j)$ to be orthonormal. We endow the hyperplane bundle with the Hermitian metric $h_{FS}$ specified by:
\beq
\label{eqn:hypermetric}
h_{FS}([v])(Z_j([v]),Z_j([v]) )=\frac{|z_j(v)|^2}{|z(v)|^2}=\frac{|v_j|^2}{||v||^2}~~.
\eeq
The associated K{\"a}hler metric on $\P V$ is the Fubini-Study metric
determined by $(~,~)$, whose K{\"a}hler form is given by:
\beq
\nn
\omega_{FS}([v])=\frac{i}{2\pi}\partial {\bar \partial}\log
|z(v)|^2=\frac{i}{2\pi}\partial {\bar \partial}\log ||v||^2~~.
\eeq
We endow $H^k$ with the tensor product metric $h_{FS}^k=h_{FS}^{\otimes k}$,  which satisfies: 
\beq
\nn
h_{FS}^k([v])(S([v]),S([v]))=\frac{|s(v)|^2}{||v||^{2k}}~~\forall s\in B_k~~.
\eeq
As measure on $\P V$, we use the Liouville measure of the volume form $\frac{\omega_{FS}^n}{n!}$. In
particular, we have:
\beq
\nn
\vol_{\omega_{FS}}(\P V)=\frac{1}{n!}~~.
\eeq
The space $H^0(H^k)=B_k$ carries the associated $L^2$-product:
\beq
\label{eqn:hyperproduct}
\langle s_1,s_2\rangle_k:=\langle S_1,S_2\rangle^{h^k_{FS}}_k=\int_{\P V}{\frac{\omega_{FS}^n}{n!} h^k_{FS}(S_1,S_2)}~~.
\eeq 
The monomials $\chi_\p$ (such that $|\p|=k)$ provide an orthogonal but not orthonormal basis of $B_k$ with
respect to this product. Direct computation gives: 
\beq
\nn
||\chi_\p||_k=\sqrt{\frac{\p!}{(n+k)!}}~~~~(|\p|=k)~~.
\eeq
Comparing with (\ref{chi}), we find: 
\beq
\nn
\frac{||\chi_\p||_k}{||\chi_\p||_B}=\frac{1}{\sqrt{(n+k)!}} ~~~~(|\p|=k)~~.
\eeq
where $||\chi_\p||_B$ is computed with respect to the scalar product of the Bargmann space ${\cal B}({\C^{n+1}})$. It
follows that the Hardy product of the projective Hilbert space $(\P V,\omega_{FS})$ is related to the Bargmann product by a constant rescaling: 
\beq
\label{products}
\langle s, t\rangle_k= \frac{1}{(n+k)!}\langle s,t\rangle_B~~~\forall s,t\in B_k~~.
\eeq
Let $({\cal  E}(\P V),\langle~,~\rangle_{\P V})=\overline{\oplus}_{k=0}^\infty{(B_k,\langle~,~\rangle_k)}$ 
denote the Hardy space of $\P V$. It can be
identified with the space of those entire functions (\ref{Taylor}) on
$V$ whose coefficients satisfy:
\beq
\nn
\langle f|f\rangle_{\P V}=\sum_{\p\in \N^{n+1}}{\frac{\p !}{(n+|\p|)!} |a_\p|^2} <\infty~~.
\eeq
Since $(n+\p)! \geq 1$, we can view ${\cal B}(V)$
as a closed subspace of ${\cal E}(\P V)$. (Of course,  the Bargmann scalar product
does not agree with the restriction of the Hardy product.) The  operator
$W:{\cal  E}(\P V)\rightarrow {\cal
  B}(V)\subset {\cal  E}(\P V)$  given  by: 
\beq
\label{eqn:isometry}
(Wf)(z)=\sum_{\p\in \N^{n+1}}{\sqrt{(n+|\p|)!} a_\p  z^\p}~~
\eeq
provides an isometry between $({\cal  E}(\P V),\langle~,~\rangle_{\P V})$ and $({\cal
  B}(V),\langle~,~\rangle_B)$. Defining $A:=W^2$, we have $A_k:=A|_{B_k}=(n+|\p|)! 1_{B_k}$ and:
\beq
\nn
\langle s, t\rangle_{\P V}=\langle As,t\rangle_B~~.
\eeq

Since $\P V$ is a homogeneous space, its coherent states can be extracted by
the well-known method due to Perelomov \cite{Perelomov:1971bd}. The unitary group $U(V)$ of the Hermitian vector space $(V,(~,~))$ acts unitarily on $B_k$ via
$\hat{U}_k(f)(v)=f(U^{-1}v)$, i.e.\ ${\hat U}_k|v, k\rangle =|U v, k\rangle$. The
representation is irreducible and isomorphic with the $k$-fold symmetric representation. 
One can construct Perelomov coherent states \cite{Perelomov:1971bd} of this action
as orbits of a given non-vanishing state in the projective Hilbert space
$\P (B_k)$. Let us start with the state defined by the vector
$|v_0,k\rangle\in B_k$, where $v_0$ is any fixed non-vanishing vector of
$V$. Then the ray $\C^*|v_0,k\rangle\in B_k$ has stabilizer
$U(v_0^\perp)\times U(1)$ in $U(V)$. It follows that the Perelomov states are parameterized by
points of the homogeneous space $U(V)/ (U(v_0^\perp)\times U(1))$, which
coincides with $\P V$.  Since ${\hat U}_k|z_0,k\rangle=|Uz_0,k\rangle$, the Perelomov state at $[z]$
coincides with the ray $\C^*|z,k\rangle$. Hence Perelomov's coherent projectors take the form: 
\beq
\label{eqn:perelomov}
P^{(k)}_{[v]}:=\frac{|v,k\rangle{}\langle v,k|_B}{\langle v,k|v,k\rangle_B}=k!
\frac{|v,k\rangle\langle v,k|_B}{||v||^{2k}}~~.
\eeq
These projectors are unaffected by the constant rescaling of the scalar
product when translating between the Bargmann  and Hardy metrics on $B_k$. 
Since the Fubini-Study metric is invariant on the homogeneous space
$\P V$, the Liouville form determined by its volume form $\frac{\omega_{FS}^n}{n!}$ provides
an invariant measure. Thus Perelomov's theory implies the overcompleteness property:
\beq
\label{Pcompleteness}
\frac{N_k+1}{\vol(\P V)}\int_{\P V}{\frac{\omega_{FS}^n}{n!}P^{(k)}_{[v]}}=P_k~~,
\eeq
where $P_k$ is the orthoprojector on $B_k$ in ${\cal B}(V)$ and the normalization constant in front of the integral has been identified
by taking the trace. Since both Rawnsley's and Perelomov's coherent projectors satisfy (\ref{Pcompleteness}),
they  determine a reproducing kernel for $B_k$, so
they must agree with each other if one uses scalar products on
this space differing only by a constant rescaling. It follows that the $P^{(k)}_{[v]}$ coincide with the Rawnsley projectors of
$(B_k,\langle~,~\rangle_k)$ while the rays $\C^*|v,k\rangle$ are Rawnsley's coherent
states. Rawnsley's coherent vectors take the form: 
\beq
\label{eqn:rawnsleyproj}
e^{(k)}_{v}=(n+k)! |v,k\rangle~~(v\in E\setminus \{0\})~~,
\eeq
where the prefactor is due to relation (\ref{products}) between the Hardy
and Bargmann scalar products. The reproducing kernel for $(B_k, \langle~,~\rangle_k)$ is given by: 
\beq
\nn
K_k(u,v)=\langle e^{(k)}_u|e^{(k)}_v\rangle_{k}=(n+k)!\langle
u,k|v,k\rangle_B=\frac{(n+k)!}{k!}(u\cdot {\bar v})^k~~.
\eeq
The epsilon function is constant:
\beq
\label{proj_epsilon}
\epsilon_k^{\P V}=\frac{N_k+1}{\vol (\P V)}=\frac{(n+k)!}{k!}~~. 
\eeq
In particular, we recover the well-known fact that $k\omega_{FS}$ is balanced
for all $k$. The overcompleteness property (\ref{Pcompleteness}) can also be written as: 
\beq
\nn
(n+k)!\int_X  \frac{\omega_{FS}^n}{n!} \frac{|v,k\rangle\langle v,k|_B}{||v||^{2k}}=P_k~~,
\eeq
where we used $\vol(\P V)=\frac{1}{n!}$ and the identity
$k!(N_k+1)=\frac{(n+k)!}{n!}$.

It will prove convenient to consider the functions: 
\beq
\label{f}
f_{IJ}:=\frac{{\bar z}^I z^J}{|z|^{2m}}=\frac{{\bar z}_0^{i_0}\ldots
  {\bar z}_n^{i_n} z_0^{j_0}\ldots z_n^{j_n}}{|z|^{2m}}\in {\cal
  C}^\infty(\P V, \C)~~,
\eeq
where $ I=(i_0\ldots i_n),J=(j_0\ldots j_n)\in
\N^{n+1}$ and $|I|=|J|=m$ and where we set $f_{IJ}=1$ for $m=0$. The linear
span ${\cal S}(\P V)$ over
$\C$ of these functions is a unital $*$-subalgebra of the $C^*$-algebra $({\cal
  C}^\infty(\P V), ||~||_\infty)$ (recall that $||~||_\infty$ is the sup
norm). Let ${\cal S}_m(\P V)$ be the subspace spanned by those $f_{IJ}$ with
$|I|=|J|=m$ (notice that ${\cal S}_0=\C$). The set of functions (\ref{f})  with a fixed $m$ gives a basis
for ${\cal S}_m(\P V)$ so in particular we have $\dim_\C {\cal S}_m(\P V)=N_m+1$. 
For any $l=0\ldots n$, let $\Delta_l\in \N^{n+1}$ be given by
$\Delta_l(i)=\delta_{il}$. The obvious relation:
\beq
\nn
f_{IJ}=\sum_{l=0}^n f_{I+\Delta_l, J+\Delta_l}
\eeq
shows that ${\cal S}_m(\P V)\subset {\cal S}_{m+1}(\P V)$ for all $m\geq 0$, so that
${\cal S}(\P V)=\cup_{m=0}^\infty{\cal S}_m(\P V)$ is a filtered $*$-algebra. Notice that
${\cal S}(\P V)$ is generated as a $*$-algebra by the elements $f_{ij}=\frac{{\bar  z}_iz_j}{|z|^2}\in {\cal
  S}_1$.

\paragraph{Proposition.} The $*$-subalgebra ${\cal S}(\P V)$ is dense in the
$C^*$-algebra $( {\cal  C}^\infty(\P V), ||~||_\infty)$. 

\paragraph{Proof.} Given a point $[v]\in \P V$, there exists an index $i=0\ldots
n$ such that $z_i(v)\neq 0$. In particular
$f_{ii}([v])=\frac{|z_i(v)|^2}{|z(v)|^2}\neq 0$. It follows that ${\cal S}_1$
separates points. Since ${\cal S}_1$ generates  ${\cal S}$ as a $*$-algebra,
the conclusion follows from the Stone-Weierstra{\ss} theorem.

\paragraph{Toeplitz quantization of $\P V$.} The Toeplitz quantization map
$T_k:{\cal C}^\infty(\P V,\C)\rightarrow
\End(B_k)$ takes the form: 
\beq
\nn
T_k(f)=\frac{N_k+1}{\vol(\P V)}\int_{\P V} \frac{\omega_{FS}^n}{n!}
f(x)P^{(k)}_x=(n+k)!\int_{\P V}  \frac{\omega_{FS}^n}{n!}
f([v]) \frac{|v,k\rangle\langle v,k|_B}{||v||^{2k}}~~.
\eeq
This map is surjective since $\P V$ is compact. Using relations (\ref{truncated}), we find:
\beq
\nn
T_k(f_{IJ})= \frac{(n+k)!}{n!}\int_{\P V} \omega_{FS}^n
  \frac{{\hat a}^I|v,k+d\rangle\langle v,k+d|_B
  ({\hat a}^\dagger)^J}{|z|^{2(k+m)}}=
\frac{(n+k)!}{(n+k+m)!} {\hat a}^I P_{k+m}({\hat a}^\dagger)^J~~.
\eeq
In particular, we have: 
\beq
\nn
T_k(f_{ij})=\frac{1}{n+k+1}{\hat a}_i{\hat a}_j^\dagger~~.
\eeq

\paragraph{Berezin quantization of $\P V$.} The Berezin symbol map $\sigma_k:\End(B_k)\rightarrow {\cal
  C}^\infty(\P V, \C)$ takes the form: 
\beq
\nn
\sigma_k(C)([v])=\frac{\langle v , k|C|v, k\rangle}{ \langle v, k|v,
  k\rangle}~~~~\forall C\in \End(B_k)~~.
\eeq
This map  is injective and its inverse on the image $\Sigma_k(\P V):=\im
\sigma_k$ defines the Berezin quantization $Q_k:\Sigma_k(\P V)\rightarrow
\End(B_k)$, which is a linear isomorphism. For the functions (\ref{f}), we find: 
\beq
\nn
Q_k(f_{IJ})=\frac{(k-m)!}{k!}P_k({\hat a}^\dagger)^I {\hat a}^JP_k
\eeq
and in particular:
\beq
\nn
Q_k(f_{ij})=\frac{1}{k}\hat{a}_j^\dagger \hat{a}_i~~.
\eeq
Notice that $Q_k(f_{IJ})$ vanishes for $m\geq k$. Since the operators
${\hat f}_{IJ}=P_k({\hat a}^\dagger)^I {\hat a}^JP_k$ with $m:=|I|=|J|=k$ provide a basis
for $\End(B_k)$, it follows that the image $\Sigma_k(\P V)$ of the Berezin symbol
map coincides with ${\cal S}_k(\P V)$:
\beq
\nn
\Sigma_k(\P V)={\cal S}_k(\P V)~~\forall k\geq 1~~.
\eeq
It follows that $\Sigma_k(\P V)$ provides a weakly exhaustive filtration of $( {\cal
  C}^\infty(\P V),||~||_\infty)$:
\beq
\nn
\overline{\cup_{k=1}^\infty \Sigma_k(\P V)}={\cal    C}^\infty(\P V)~~.
\eeq
The Berezin transform $\beta_k:{\cal C}^\infty(\P V)\rightarrow \Sigma_k(\P V)$ takes
the form: 
\beq
\nn
\beta_k(f)(v)=\sigma_k(T_k(f))=\frac{(n+k)!}{k!}\int_{\P V}\frac{\omega_{FS}^n}{n!}(u)\left(\frac{|(u, v)|}{||u||||v||}\right)^{2k}~~.
\eeq
Notice that again Berezin and Toeplitz quantizations use Wick and anti-Wick
orderings, respectively. An extension of this quantization providing access to vector bundles over quantized $\P^n$ has been presented in \cite{Dolan:2006tx}.

\section{Berezin-Bergman quantization}

In this section we discuss a generalized Berezin quantization procedure which
clarifies the proposal of \cite{Saemann:2006gf}. This prescription, which we
call Berezin-Bergman quantization, is relevant for compact complex manifolds endowed
with a Bergman metric. 

Let $(X,L)$ be a polarized compact complex manifold and assume that $L$ is
very ample with $\dim_\C H^0(L)=n+1$. We let $E_k:=H^0(L^k)$ and $\dim_\C
E_k=M_k+1$ (thus $M_1=n$). The homogeneous coordinate ring
$R(X,L)=\oplus_{k=0}^\infty{H^0(L^k)}=\oplus_{k=0}^\infty E_k$ of $X$ with respect to $L$ is
generated in degree one and we have an isomorphism of graded algebras: 
\beq
\label{phi}
\phi:R\stackrel{\sim}{\rightarrow}B/I~~,
\eeq
where $B=\oplus_{k=0}^\infty{E_1^{\odot k}}$ is the symmetric algebra over
the vector space $E_1:=H^0(L)$ and $I$ is a graded ideal in $B$ generated in degrees
$\geq 2$. The algebra $B$ can be identified with the algebra of
polynomial functions on $E_1^*$, and thus with the coordinate ring $\C[E_1^*]$ of
the affine space $\Spec B$ over $E_1^*\simeq \C^{n+1}$. As a graded algebra, it is also the
homogeneous coordinate ring of the projective space $\P[E_1^*]$. The Kodaira embedding $i:X\hookrightarrow
\P(E_1^*)$ defined by $L$ presents $X$ as a projective variety in $\P[E_1^*]$,
whose vanishing ideal equals $I$, and whose homogeneous coordinate ring equals
$R$. Writing $B=\oplus_{k=0}^\infty{B_k}$ and $I=\oplus_{k=0}^\infty{I_k}$,
the homogeneous components satisfy $I_k\subset B_k$ as well as: 
\beq
\nn
E_k\simeq B_k/I_k~~.
\eeq

Let us now consider a scalar product $(~,~)_1$ on $E_1$ and the associated
Bergman metric on $X$, whose K{\"a}hler form we denote by $\omega$. We also let
$h$ be the induced Bergman Hermitian scalar product on $L$. For every
$k\geq 1$, we have {\em two} natural ways to induce a scalar product on
$H^0(L^k)$. The first choice is to take the $L^2$-product:
\beq
\nn
\langle s,t\rangle_k=\int_{X}\frac{\omega^n}{n!}h_k(s,t)~~,
\eeq
where $h_k=h^{\otimes k}$. Performing generalized Berezin quantization with respect to
this sequence of products leads to the classical Berezin-Toeplitz theory
discussed in the previous section.

The second choice is as follows. The product $(~,~)_1$ on $E_1$ induces a
scalar product $(~,~)_B$ on the symmetric algebra $B=\oplus_{k=0}^\infty {E_1^{\odot k}}$ via the prescription:
\beq
\label{Bproduct}
( s_1\odot \ldots s_k,t_1\odot \ldots t_l)_B=\frac{1}{k!}\delta_{k,l}\sum_{\sigma\in
  S_k}{(s_1,t_{\sigma(1)})_1\ldots (s_k,t_{\sigma(k)})_1}~~,
\eeq
where $S_k$ is the symmetric group on $k$ letters and $s_i,t_i\in E_1$. Notice
that the completion of 
$B=\oplus_{k=0}^\infty {E_1^{\odot k}}$ with respect to the
product (\ref{Bproduct}) is the bosonic Fock space over the $n+1$-dimensional
Hilbert space $(E_1,(~,~)_1)$. Of course, this can also be viewed as the
Bargmann space over $V=E_1^*$, which appeared in the quantization of the affine
space $A[V]$. Thus we can view $B$ as embedded in the Bargmann space
${\cal B}(V)$, and (\ref{Bproduct}) is the restriction of the Bargmann product
(\ref{flat}) to $B$.

Using the scalar product (\ref{Bproduct}), we can identify $E_k\simeq B_k/I_k$ with the orthogonal
complement $I_k^\perp=\{s\in B_k|(s,t)_B=0~~\forall t\in I_k\}$
of $I_k$ in $B_k$. This identification gives a scalar product $(~,~)_k$ on
$E_k$, which is induced by the restriction of $(~,~)_B$ to
$I_k^\perp$.  To state this precisely, notice that the Kodaira embedding $i:X\hookrightarrow
\P[E_1^*]$ defined by $L$ allows us to identify $B_k$ with the space of holomorphic sections of $H^k$,
where $H$ is the hyperplane bundle $H={\cal O}_{\P[E_1^*]}(1)$:
\beq
\nn
B_k=H^0(H^k)~~.
\eeq
Furthermore, the homogeneous component $I_k$ of the vanishing ideal $I$ can
be identified with the kernel of the pull-back map (restriction) on sections
$i^*_k:H^0(H^k)=B_k\rightarrow H^0(L^k)=E_k$:
\beq
\nn
I_k:=\ker~ i_k^*~~.
\eeq
Since $i_k^*$ is surjective, it induces an isomorphism
$\psi_k:I_k^\perp\rightarrow E_k$, whose inverse
$\phi_k:=\psi_k^{-1}:E_k\rightarrow I_k^\perp \simeq B_k/I_k$ we can take as the homogeneous
$k$-component of (\ref{phi}). We define $(~,~)_k$ as
follows:
\beq
\label{induced_product}
(s,t)_k:=\alpha_k( \phi_k(s),\phi_k(t))_B~~,
\eeq
where the scaling constants: 
\beq
\nn
\alpha_k:=\frac{\vol_{\omega}(X)}{\vol_{\omega_{FS}}(\P V)}\frac{N_k+1}{M_k+1}
\eeq
are chosen for later convenience. Here, $\vol_{\omega_{FS}}(\P V)=\frac{1}{n!}$. 

\paragraph{Definition.} The {\em Berezin-Bergman quantization} of $(X,L)$
determined by the scalar product $(~,~)_1$ on $H^0(L)$ is the generalized
Berezin quantization performed with respect to the sequence of scalar
products $(~,~)_k$ on $H^0(L^k)$ defined in (\ref{induced_product}). 

\

Using the orthogonal decomposition $B_k=I_k\oplus I_k^\perp$, let us
pick a $(~,~)_B$-orthonormal basis $S_0\ldots S_{N_k}$ of $B_k$ such that
$S_0\ldots S_{M_k}$ is a basis of $I_k^\perp$ and such that $S_{M_k+1}\ldots S_{N_k}$ is
a basis of $I_k$. Then $i^*_k(S_j)=0$ for $j>M_k$ and the sections
$s_j:=\frac{1}{\sqrt{\alpha_k}} i^*_k(S_j)$
(with $j=0\ldots M_k$) give an orthonormal basis of the space
$(E_k,(~,~)_k)$. The epsilon function of $\P V$ at level $k$ takes the form:
\beq
\nn
\epsilon_k^{\P V}([v])=\sum_{j=0}^{N_k} h_{FS}^k ([v])(S_j([v]),
S_j([v]))=\frac{N_k+1}{\vol_{\omega_{FS}}(\P V)}~~.
\eeq
Restricting this identity to $X$ shows that the epsilon function of the pair
$(h_k, (~,~)_k)$ is constant on $X$:
\beq
\nn
\epsilon_k(x)=\sum_{j=0}^{M_k} h_k(s_j(x),s_j(x))=\frac{M_k+1}{\vol_{\omega}(X)}~~~~(x\in X)~~.
\eeq
In particular, the induced scalar product $(~,~)_k$ coincides with the $L^2$-scalar
product $\langle~,~\rangle_k$ iff the epsilon function of the latter is constant, i.e.\ iff
$k\omega$ is balanced. 

We now consider the generalized Berezin quantization of $(X,\omega,L,h)$ with respect to the sequence of induced
scalar products $(~,~)_k$ on $E_k=H^0(L^k), k\geq 1$. We
have surjective Berezin symbol maps $\sigma_k:\End(E_k)\rightarrow {\cal
  C}^\infty(X)$ whose images we denote by $\Sigma_k$, and associated
quantization maps $Q_k:=(\sigma_k|_{\Sigma_k})^{-1}:\Sigma_k\rightarrow
\End(E_k)$. 

Let $\Lambda_k$ be the orthogonal projector of $B_k$ onto $I_k^\perp$ with
respect to the product (\ref{Bproduct}). It is easy to see that the Rawnsley coherent states of $E_k$ with respect
to $( ~,~)_k$ are given by:
\beq
\nn
e^{(k)}_v:=\frac{1}{\alpha_k} i_k^*(\Lambda_k |v,k\rangle)=\frac{1}{\alpha_k}i^*(|v,k\rangle) ~~\forall v\in
C(X)\setminus\{0\} \subset V~~,
\eeq
where $C(X)=\Spec(B/I)\subset V$ is the affine cone over $X$. The
second form follows from the fact that the component $(1-\Lambda_k)|z,k\rangle$ along $I_k$ vanishes for $z\in C(X)$, so that: 
\beq
\nn
\Lambda_k |v,k\rangle=|v,k\rangle~~{\rm for}~~v\in C(X)~~.
\eeq
The Rawnsley projectors take the form:
\beq
\label{eqn:rawnsleyproj_bb}
P^{(k)}_{[v]}=i^*_k\circ \frac{ |v,k\rangle \langle v,k|_B}{\langle
  v,k|v,k\rangle_B}\circ \phi_k ~~~~([v]\in X\subset \P V)~~,
\eeq
while the Berezin symbol of an operator $C\in \End(E_k)$ is given by:
\beq
\label{induced_sigma}
\sigma_k(C)([v])=\frac{\langle
  v,k|{\tilde C}|v,k\rangle_B}{\langle
  v,k|v,k\rangle_B}~~,
\eeq
where ${\tilde C}:=\phi_k\circ C\circ i_k^*=\phi_k\circ C\circ
\phi_k^{-1}\circ \Lambda_k\in \End(B_k)$ is the transport of $C$ to an
operator on $B_k$ through the isomorphism $\phi_k:E_k\rightarrow I_k^\perp\subset B_k$. 

Recall that the operators ${\hat f}_{IJ}:=P_k({\hat a}^\dagger)^I{\hat a}^JP_k$ 
(where $|I|=|J|=k$) form a basis of $\End(B_k)$. Thus space
$\End(I_k^\perp)$ is spanned by the operators $\Lambda_k{\hat
  f}_{IJ}\Lambda_k$ and $\End(E_k)$ is spanned by: 
\beq
\nn
{\hat f}'_{IJ}:=i_k^*\circ {\hat
  f}_{IJ}\circ \phi_k~~(|I|=|J|=k) ~~.
\eeq
We have ${\tilde {\hat f}'_{IJ}}:=\Lambda_k {\hat
  f}_{IJ}\Lambda_k$. Applying (\ref{induced_sigma}) to these operators, we find: 
\beq
\nn
\sigma_k({\hat f}'_{IJ})([v])=\frac{\langle
  v,k|{\hat
  f}_{IJ}|v,k\rangle_B}{\langle
  v,k|v,k\rangle_B}=f_{IJ}([v])~~ ~~~~([v]\in X)~~.
\eeq
It follows that:
\beq
\nn
Q_k(f_{IJ}|_X)={\hat f}'_{IJ}~~,
\eeq
where $f_{IJ}|_X:=f_{IJ}\circ i\in {\cal C}^\infty(X)$. We conclude that
$\Sigma_k(X)$ is spanned by the restrictions $f_{IJ}|_X$ with
$|I|=|J|=k$. Since ${\cal S}_k(\P V)\subset {\cal S}_{k+1}(\P V)$, we have
$\Sigma_k(X)\subset \Sigma_{k_+1}(X)$. The union ${\cal S}(X):=\cup_{k\geq 0}\Sigma_k(X)$
(where $\Sigma_0:=\C$ consists of the constant functions on $X$) is a
filtered unital $*$-subalgebra of the $C^*$-algebra $({\cal C}^\infty(X),||~||_\infty)$. Since
the $*$-algebra ${\cal S}(\P V)$ is generated by ${\cal S}_1(\P V)$, it follows
that ${\cal S}(X)$ is generated by $\Sigma_1(X)$ as a $*$-algebra. Moreover,
$\Sigma_1(X)$ separates the points of $X$ since $X$ can be viewed as a subset of
$\P V$ and since ${\cal S}_1(X)$ separates the points of the latter. It follows
that ${\cal S}(X)$ is dense in  $({\cal C}^\infty(X),||~||_\infty)$.

\paragraph{Remark.} Let $I$ be generated by $p$ homogeneous polynomials
$F_1\ldots F_p$ of degrees at least two. Since ${\hat a}_i$ act on the Bargmann space
as multiplication by $z_i$, we have the linear decomposition $I=\sum_{l=1}^p{\im F_l({\hat a})}$, where
$\im$ denotes the image of a linear operator and all operators are taken to act
in the space $B$. It follows that:
\beq
\nn
I^\perp=\cap_{l=1}^p{\ker {\bar F}_l({\hat a}^\dagger)}~~,
\eeq
where ${\bar f}$ is the polynomial in $z_0\ldots z_n$ obtained by conjugating
all {\em coefficients} of $f$. Since ${\hat a}^\dagger_i$ act as
$\partial_i$, the operators in the right hand
side are holomorphic differential operators: 
\beq
\nn
 {\bar F}_l({\hat a}^\dagger)={\bar F}_l(\partial_0\ldots \partial_n)
\eeq
and we find that $I^\perp$ is the graded vector space of those solutions $s$ of the
system of the following linear holomorphic partial differential equations with constant coefficients: 
\beq
\label{sys1}
{\bar F}_l(\partial_0\ldots \partial_n)s(v_0\ldots v_n)=0~~,
\eeq
which are homogeneous polynomials in $v_0\ldots v_n$. The graded components
$I_k^\perp\subset B_k$ can be obtained by restricting to homogeneous
polynomials of degree $k$, which amounts to imposing the condition: 
\beq
\label{sys2}
Gs=ks~~,
\eeq 
where $G=\sum_{i=0}^n{v^i\partial_i}$ is the Euler operator. These equations
determine the unique extension of a section $s\in H^0(X,L^k)$ to a section of
$H^0(\P V,H^k)$ lying in $I_k^\perp$. The point is that the system
(\ref{sys1}), (\ref{sys2}) has a unique polynomial solution which has prescribed 
behavior on the affine cone $C(X)\subset V$.

\paragraph{Relation with fuzzy geometry.}

A procedure for defining fuzzy versions of compact Hodge manifolds $X$ was
proposed in \cite{Saemann:2006gf}. It is clear from the above that:

\

{\em The proposal of \cite{Saemann:2006gf} amounts to defining the fuzzy version of
  $X$ as the generalized Berezin quantization of $X$ with respect to the sequence
  of induced scalar products $(~,~)_k$ on the spaces
  $E_k:=H^0(L^k)$. This quantization prescription agrees with classical
  Berezin quantization at a fixed level $k$ iff $k\omega$ is balanced.}

\

\noindent It is also easy to prove the following:

\paragraph{Proposition.} The Berezin-Bergman quantization of $\P V$ coincides
with its classical Berezin quantization.

\paragraph{Proof.} The vanishing ideal $I$ is zero in this case, so the Berezin-Bergman scalar product $(~,~)_k$ of section 5 coincides with the restriction of the
Bargmann product $\langle~,~\rangle_B$ (\ref{flat}) to $E_k$ (recall (\ref{Bproduct}) the intermediate scalar product $(~,~)_B$ induced by $(~,~)_1$). Since $\langle~,~\rangle_B$ coincide with the $L^2$-product $\langle~,~\rangle_k$ up to a constant scale factor, the results of Section 3 show that the generalized Berezin quantization based on $(~,~)_k$ is equivalent to that based on $\langle~,~\rangle_k$. The first quantization is the Berezin-Bergman quantization, while the second is the classical Berezin quantization of $\P V$.

\section{Harmonic analysis on quantized Hodge manifolds}

In this section, we discuss the construction of a ``fuzzy'' Laplace operator on
Berezin quantized compact Hodge manifolds. There are two obvious choices: the
Berezin push and the Berezin-Toeplitz lift of the ordinary Laplace operator. We
use the former to calculate the approximate spectrum of the Laplace operator
for two examples and employ the latter to define fuzzy scalar field theories on
arbitrary compact Hodge manifolds. Let us fix a compact prequantized Hodge
manifold $(X,\omega,L,h)$.

\subsection{`Quantizing' the classical Laplacian}

Let us take $\mu$ to be the Liouville measure defined by the  K{\"a}hler form
$\omega$ and fix some generalized Berezin quantization of $(X,\omega)$. 
Recall (\ref{mu_product}) that the space $\CC^\infty(X)$ carries the $L^2$-scalar product induced by the K{\"a}hler metric: 
\beq
\label{omega_product}
\prec f,g\succ=\int_X\frac{\omega^n}{n!}{\bar f}g~~
\eeq
as well as the scalar products (\ref{epsilon_product}):
\beq
\nn
\prec f,g\succ_{\epsilon_k}=\int_X\frac{\omega^n}{n!}\epsilon_k {\bar f}g~~,
\eeq
On the other hand, the Berezin algebras $(\Sigma_k,\diamond_k)$ 
(where $\Sigma_k=\sigma(\End(E_k))$ with $E_k=H^0(L^k)$) carry the Berezin scalar products
(\ref{induced_sp}) induced by the trace (\ref{vint}):
\beq
\label{d_product}
\prec f,g\succ_B=\vint_k {\bar f}\diamond_k g=\int_{X}{\frac{\omega^n}{n!} \epsilon_k
  f\diamond_k g}=\langle Q_k(f),Q_k(g)\rangle_{HS}~~.
\eeq
The Laplace operator $\Delta$ of $(X,\omega)$ is Hermitian and positive with
respect to the ordinary $L^2$-scalar product (\ref{omega_product}) on functions, which is the realization
of (\ref{mu_product}) in the case at hand. 

\paragraph{The Berezin push.}

Considering the symbol spaces $\Sigma_k$, we define truncated Laplacians $\Delta_k:\Sigma_k\rightarrow \Sigma_k$ via: 
\beq
\nn
\Delta_k=\pi_k\circ \Delta|_{\Sigma_k}~~,\nn
\eeq
where $\pi_k$ is the $\prec~,~\succ$-orthoprojector on $\Sigma_k$ inside
${\cal C}^\infty(X)$. Explicitly, let the K{\"a}hler form be given by $\omega=\omega_{i\bj} d z^i\wedge d\bz^\bj$ in some
local coordinates $z^i$ defined on a Zariski open set and let $\Sigma_k$ be spanned by
the functions $e_i\in {\cal C}^\infty(X)$ with $i=0,...,{N_k} $. The Laplace operator of $(X,\omega)$ takes the form:
\begin{equation} \nn
 \Delta f\ =\ \omega^{i\bj}\dpar_i\dpar_\bj f~~,\nn
\end{equation}
and the orthoprojector $\pi_k:\CC^\infty(X)\rightarrow \Sigma_k$ is given by the map:
\begin{equation}
\label{L2-projection}
 \pi_k(f)=\sum_{i=0}^{N_k} e_i \int_X \frac{\omega^n}{n!} \bar{e}_i f~~(f\in
 {\cal C}^\infty(X))~~.
\end{equation}

Notice that $\Delta_k$ is $\prec~,~\succ$-self-adjoint and positive on $\Sigma_k$. 
The truncated Laplacian need not be Hermitian with respect to the Berezin scalar
product, so the Berezin push $\Delta_k^B=Q_k\circ \Delta_k\circ \sigma_k$ may
generally fail to be Hermitian with respect to the Hilbert-Schmidt scalar
product on $\End(E_k)$. It follows that the Berezin push $\Delta_k^B$ does
{\em not} provide a good general notion of ``fuzzy Laplacian''. In fuzzy field theory,
the fuzzy Laplacian is used when building the kinetic term for scalar fields
in the ``fuzzified'' field action, which is defined on $\End(E_k)$. Since the
natural scalar product on that space is the Hilbert-Schmidt product, the
kinetic term should be specified by an operator which is
$\langle~,~\rangle_{HS}$ -Hermitian and positive. 

\paragraph{The Berezin-Toeplitz lift.}

The discussion of Section 3.8 shows that the Berezin-Toeplitz lift (\ref{lift}) of the Laplacian:
\beq
\nn
{\hat {\Delta}}_k:=T_k\circ M_{\frac{1}{\epsilon_k}}\circ \Delta \circ \sigma_k:\End(E_k)\rightarrow \End(E_k)
\eeq
is a positive Hermitian operator on $(\End(E_k),\langle~,~\rangle_{HS})$. Moreover, the Berezin-Toeplitz transform
(\ref{diamond}):
\beq
\nn
\Delta_{\diamond_k}=\beta_{mod,k}\circ \Delta|_{\Sigma_k}=\beta_k\circ
M_{\frac{1}{\epsilon_k}}\circ\Delta|_{\Sigma_k}:\Sigma_k\rightarrow \Sigma_k
\eeq 
is Hermitian and positive-definite with respect to the Berezin product
(\ref{d_product}). We will view ${\hat {\Delta}}_k$ (equivalently, $\Delta_{\diamond_k}$)
as the definition of the ``fuzzy Laplacian'' at level $k$. Explicitly, we
have: 
\beq
\nn
{\hat \Delta}_k(C)=\int_X\frac{\omega^n}{n!}(x) P^{(k)}_x(\Delta
\sigma_k(C))(x)~~~~(C\in \End(E_k))~~
\eeq
and:
\beq
\nn
\Delta_{\diamond_k}(f)(x)=\int_{X}\frac{\omega^n}{n!}(y)\Psi_k(x,y)(\Delta f)
(y)~~~~(f\in \Sigma_k)~~,
\eeq 
where we used the integral expression (\ref{mod_Berezin_tf}) for the modified Berezin transform:
\beq
\nn
\beta_{mod,k}(f)(x):=\int_{X}\frac{\omega^n}{n!}(y)\Psi_k(x,y)f(y)~~.
\eeq
We are using $\Psi_k$ which is the squared two-point function (\ref{two_point}) at level $k$:
\beq
\nn
\Psi_k(x,y):=\sigma_k(P^{(k)}_x)(y)=\sigma_k(P^{(k)}_y)(x)=\tr\left(P^{(k)}_xP^{(k)}_y\right)~~.
\eeq

\subsection{The case of K{\"a}hler homogeneous spaces}

Recall that a K{\"a}hler manifold $(X,\omega)$ is called a K{\"a}hler
homogeneous space when its group of holomorphic isometries $\Aut(X,\omega)$
acts transitively on $X$. In this case, the equivariance properties of Berezin
and Berezin-Toeplitz quantization allow one to use representation-theoretic
arguments in order to extract further information. Let us consider the case of 
simply connected coadjoint orbits $X=G/H$, where we can take $G$ to be a compact simple Lie group. This includes
the case of projective spaces $\P^n$, which  corresponds to the choice
$G=SU(n+1)$. In this situation $G=\Aut_{L,h}(X,\omega)=\Aut(X,\omega)$. This class of spaces has been studied
in great detail, so we will only make a few basic remarks for use
in the next subsection. 

Consider the classical Berezin and Berezin-Toeplitz quantizations of such a
space. The general discussion of Section 3 shows that all objects associated
with these quantization schemes are $G$-equivariant.  In particular,
$G$-invariance of the (absolute) epsilon function implies that it is constant
and given by: \beq \nn \epsilon_k=\frac{N_k+1}{\vol_\omega(X)}~~, \eeq where
$N_k+1=\dim E_k$. The unitary representation $\rho_k$ of $G$ in $E_k$ gives
the decomposition: 
\beq \nn E_k=\oplus_{\theta\in \Rep(G)}R_\theta\otimes_\C
E_k(\theta)~~, \eeq
 where $\Rep(G)$ is the discrete set of unitary
finite-dimensional irreps of the compact Lie group $G$, $R_\theta$ are the
corresponding $G$-modules and $E_k(\theta)$ are (possibly zero) Hermitian
vector spaces.  Their dimensions $m_k(\theta):=\dim E_k(\theta)\geq 0$ are the
corresponding multiplicities.  Of course, only a finite number of irreps
appear with non-zero multiplicity in the sum above, i.e.\ we have
$E_k(\theta)=0$ except for a finite number of values of $\theta$. The
Hermitian vector space $(\End(E_k),\langle~,~\rangle_{HS})=E_k^*\otimes E_k$
carries the unitary representation ${\hat \rho}_k=\rho_k^*\otimes \rho_k$,
whose orthogonal decomposition into irreducibles we write as: \beq
\label{End_decomp}
\End(E_k)=\oplus_{\theta\in \Rep(G)}R_\theta\otimes_\C W_k(\theta)~~.  \eeq
The equivariance property (\ref{Q_equiv}) of the bijection
$Q_k:\Sigma_k\rightarrow \End(E_k)$ shows that the Berezin quantization map is
an isomorphism between the unitary $G$-modules $(\Sigma_k,$ $ \prec~,~\succ_B,\tau|_{\Sigma_k})$ and $(\End(E_k),\langle~,~\rangle_{HS},{\hat
\rho}_k)$. Accordingly, $\Sigma_k$ consists of representation functions for $G$
and has the $\prec ~,~\succ_B$-orthogonal
decomposition: \beq \label{eqn:Sigma_decomp} \Sigma_k =\oplus_{\theta\in \Rep(G)}R_\theta\otimes_\C
\Sigma_k(\theta)~~, \eeq while $Q_k$ has the form: \beq
\label{eqn:Q_decomp}
Q_k=\oplus_{\theta\in \Rep(G)}\id_{R_\theta}\otimes Q_k(\theta)~~
\eeq
for some isometries $Q_k(\theta):\Sigma_k(\theta)\rightarrow W_k(\theta)$. The
Berezin symbol map is also $\tau$-equivariant and thus takes the form:
\beq
\label{eqn:sigma_decomp}
\sigma_k=\oplus_{\theta\in \Rep(G)}\id_{R_\theta}\otimes \sigma_k(\theta)~~,
\eeq
where $\sigma_k(\theta)=Q_k(\theta)^{-1}:W_k(\theta)\rightarrow \Sigma_k(\theta)$.
A similar argument shows that the (restricted) Toeplitz quantization map takes the form:
\beq
\label{eqn:restricted_toeplitz}
T_k|_{\Sigma_k}=\oplus_{\theta\in \Rep(G)}\id_{R_\theta}\otimes T_k(\theta)
\eeq
for some bijections $T_k(\theta):\Sigma_k(\theta)\rightarrow
W_k(\theta)$. Unlike $Q_k$, the operators $T_k(\theta)$ need not be not unitary
since $T_k$ need not be unitary with respect to the scalar
products $\prec ~,~\succ_B$ and $\langle~,~\rangle_{HS}$.  Combining the
above, we find that the (restricted) Berezin transform decomposes as:
\beq
\label{eqn:beta_decomp}
\beta|_{\Sigma_k}=\oplus_{\theta\in \Rep(G)}\id_{R_\theta}\otimes \beta_k(\theta)~~,
\eeq
where $\beta_k(\theta)$ are linear automorphisms of the
subspaces $\Sigma_k(\theta)$. 

Since $X=G/H$ is a K{\"a}hler homogeneous space, its Laplace operator
$\Delta:{\cal C}^\infty(X)\rightarrow {\cal C}^\infty(X)$ is $G$-invariant:
\beq
\nn
\Delta\circ \tau=\tau\circ \Delta~~~~(\tau\in \Aut(X,\omega))~~.
\eeq

It follows that\footnote{This recovers the observation of \cite{Murray:2006pi}
that $\Delta$ preserves $\Sigma_k$ on all flag manifolds when using the
classical Berezin quantization with respect to their homogeneous K{\"a}hler
metric.} $\Delta(\Sigma_k)\subset \Sigma_k$ (thus
$\Delta_k=\Delta|_{\Sigma_k})$ and that we have a decomposition: 
\beq
\label{eqn:Delta_decomp}
\Delta_k=\oplus_{\theta\in \Rep(G)}\id_{R_\theta}\otimes \Delta_k(\theta)
\eeq
for some linear operators $\Delta_k(\theta)$ acting in the spaces
$\Sigma_k(\theta)$. It is now clear that the Berezin push (\ref{eqn:Bpush}) and the
Berezin-Toeplitz lift (\ref{lift}) of $\Delta$ take the forms:
\beq
\nn
\Delta_k^B=\oplus_{\theta\in \Rep(G)}\id_{R_\theta}\otimes \Delta_k^B(\theta)
\eeq
and:
\beq\nn
\frac{N_k+1}{\vol_\omega(X)}{\hat \Delta}_k=\oplus_{\theta\in \Rep(G)}\id_{R_\theta}\otimes {\hat \Delta}_k(\theta)~~,
\eeq
where $\Delta_k^B(\theta)=Q_k(\theta)\circ \Delta_k(\theta)\circ \sigma_k(\theta)$
and ${\hat \Delta}_k(\theta)=T_k(\theta)\circ \Delta_k(\theta)\circ
\sigma_k(\theta)$. Furthermore, we have the decomposition:
\beq
\nn
\frac{N_k+1}{\vol_\omega(X)}\Delta_{\diamond_k}= \oplus_{\theta\in \Rep(G)}\id_{R_\theta}\otimes \Delta_{\diamond_k}(\theta)~~,
\eeq
where $ {\hat \Delta}_{\diamond_k}(\theta)=\beta_k(\theta)\circ
\Delta_k(\theta)$. The relation $T_k=Q_k\circ \beta_k$ gives
$T_k(\theta)=Q_k(\theta)\circ \beta_k(\theta)$. Of course, the coherent states
in this case can be determined explicitly through Perelomov's method, and the 
operators $Q_k(\theta), T_k(\theta), \beta_k(\theta)$ etc.\ can be expressed 
in terms of the representation theory of $G$. 

\paragraph{Remark.} A particularly simple case arises when all non-zero multiplicities
$m_k(\theta)$ in the decomposition (\ref{End_decomp}) equal one. Then all non-vanishing spaces $W_k(\theta)$ are one-dimensional and can be
identified with the space $\C$ of complex numbers upon fixing a normalized
vector in each of them. The non-vanishing components
$Q_k(\theta), T_k(\theta),\sigma_k(\theta),$ $\beta_k(\theta)$ are simply
complex numbers, and we find: 
\beq \nn
\Delta_{\diamond_k}=\frac{\vol_\omega(X)}{N_k+1}\oplus_{\theta\in \Rep(G):m_k(\theta)\neq 0}{\beta_k(\theta)\Delta_k(\theta)\id_{R_\theta}}
\eeq 
and:
\beq \nn
{\hat \Delta}_k=\frac{\vol_\omega(X)}{N_k+1}\oplus_{\theta\in \Rep(G):m_k(\theta)\neq 0}{\beta_k(\theta)\Delta^B_k(\theta)\id_{R_\theta}}
\eeq
since in this case we have ${\hat \Delta}_k(\theta)=\beta_k(\theta)\Delta^B_k(\theta)$. If
one furthermore has $\beta_k(\theta)\in \R_+$ for all $k$ and $\theta$, then it
follows that both $\Delta_k$ and $\Delta_{\diamond_k}$ are $\langle ~,~\rangle_B$-Hermitian
and positive and both  ${\hat \Delta}_k$ and $\Delta_k^B$ are
$\langle~,~\rangle_{HS}$-Hermitian and positive. In such a situation, one can use the
Berezin push of $\Delta$ as a fuzzy Laplacian since it is self-adjoint with
respect to the Hilbert-Schmidt scalar product. As we shall see in the next subsection,
this very particular case arises e.g.\ for $X=\P^n$, when the Berezin
push $\Delta_k^B$ coincides with the second Casimir operator of $G=U(n+1)$ in the
representation $\End(E_k)$.

\subsection{The quantum Laplacian on  $\P^n$}

As shown in Section 4, classical Berezin quantization of $(\P^n,\omega_{FS})$ with quantum line
bundle $H$ agrees with its Berezin-Bergman quantization and with the construction of fuzzy projective
spaces used in the fuzzy geometry literature \cite{Balachandran,Balachandran2}. Since $(\P^n,\omega_{FS})$  is
the K{\"a}hler homogeneous space $U(n+1)/(U(n)\times U(1))$, and since the
hyperplane bundle is equivariant, the spaces $E_k=H^0(H^k)$ carry a 
unitary representation $\rho_k$ of $U(n+1)$. Thinking of $E_k$ as the space $R_k$ of
homogeneous polynomials of degree $k$ in $n+1$ variables, it is clear that
$\rho_k$ is the totally symmetric irreducible representation, which has Dynkin labels $(k,0,...,0)$. The
space $\End(E_k)$ thus forms the (reducible) tensor product representation
$(k,0,...,0)\otimes(0,...,0,k)$, which decomposes into irreducibles as:
\begin{equation}
 \End(E_k)\cong \bigoplus_{\ell=0}^k (\ell,0,...,0,\ell)~~.\nn
\end{equation}
Notice that all subspaces in this decomposition appear with multiplicity one. 

The usual fuzzy Laplacian is given by the second Casimir ${\hat C}_2^{(k)}$
of $U(n+1)$ in the representation ${\hat \rho}_k=\rho_k^*\otimes_\C \rho_k$ on
$\End(E_k)=E_k^*\otimes_\C E_k$. The explicit form of this operator in terms of annihilation and creation
operators follows from the Schwinger construction (cf.\ \cite{Dolan:2006tx}):
\begin{equation}
\hat{C}_2^{(k)}(C)=\sum_a[\hat{\CL}^a,[\hat{\CL}^a,C]]~~,~~~\hat{\CL}^a=\sum_{i,j}\hat{a}^\dagger_i
 \frac{\tau^a_{ij}}{2}\hat{a}_j~~.\label{eqn:second_casimir}
\end{equation}
Here $\tau^a_{ij}$ are the Gell-Mann matrices of $su(n+1)$ with normalization fixed by the Fierz identity:
\begin{equation}
\sum_a\tau^a_{ij}\tau^a_{kl}=2\left(\delta_{il}\delta_{jk}-\tfrac{1}{n+1}\delta_{ij}\delta_{kl}\right)~~.\nn
\end{equation}
Let us first show that ${\hat C}_2^{(k)}$ agrees with $\Delta_k^B$.

\paragraph{Lemma.} Let $P^{(k)}_x$ be the coherent projectors of $\P^n$ at
level $k$ and consider the vector valued function
$P^{(k)}:X\rightarrow \End(E_k)$, $P^{(k)}(x):=P^{(k)}_x$. Then $\hat{C}_2^{(k)}\circ
P^{(k)}=\Delta(P^{(k)})$, i.e.
\beq
\nn
\hat{C}_2^{(k)}(P^{(k)}_x)=\Delta P^{(k)}_x~~(x\in X)~~.
\eeq

\paragraph{Proof.} Direct computation gives:
\begin{equation} \nn
\hat{C}_2^{(k)}(C)= \hat{N}(\hat{N}+n)C-\sum_{ij}\hat{a}^\dagger_j\hat{a}_iC\hat{a}^\dagger_i\hat{a}_j~~,
\end{equation}
where $\hat{N}:=\sum_i\hat{a}^\dagger_i\hat{a}_i$ is the number operator. For
simplicity, let us now restrict to the case of $\P^1$ (the proof for
$n>1$ follows along the same lines). With the homogeneous coordinates
denoted by $z_0, z_1$, the Laplace operator on the patch\footnote{Proving the identity on a single patch is evidently sufficient, as we are missing only one point on
$\P^1$ and all the functions involved are in $\CC^\infty(\P^1)$.} $z_1\neq 0$
with local coordinate $z:=z_0/z_1$ reads as:
\begin{equation} \nn
 \Delta f:=(1+z \bz)^2\der{z}\der{\bz}f~.
\end{equation}
Introducing the quantities:
\begin{equation} \label{eqn:F_and_M}
 F^k_{rs}=\frac{z^r \bz^s}{(1+z
 \bz)^k}~~~\mbox{and}~~~M^k_{rs}=\frac{1}{k!}(\hat{a}_0^\dagger)^r(\hat{a}_1^\dagger)^{k-r}|0\rangle\langle
 0|(a_0^\dagger)^s(a_1^\dagger)^{k-s}~~,
\end{equation}
we have:
\begin{equation*}
 P^{(k)}_x=\sum_{r,s=0}^k \binom{k}{r}\binom{k}{s}F^k_{rs}M^k_{rs}~.
\end{equation*}
One easily checks the identities:
\begin{equation*}
 \Delta
 F^k_{rs}=k(k+1)F^k_{rs}-rsF^k_{r-1,s-1}-(k-r)(k-s)F^k_{rs}-rsF^k_{rs}-(k-r)(k-s)F^k_{r+1,s+1}
\end{equation*}
and:
\begin{equation*}
 \hat{C}_2^{(k)} (M^k_{rs})=k(k+1)M^k_{rs}-rs
 M^k_{r-1,s-1}-(k-r)(k-s)M^k_{rs}-rsM^k_{rs}-(k-r)(k-s)M^k_{r+1,s+1}~~.
\end{equation*}
It is also easy to check that:
\begin{equation*}
\begin{aligned}
 \sum_{r,s=0}^k\binom{k}{r}\binom{k}{s}F_{rs}^k rs
M^k_{r-1,s-1}&=\sum_{r,s=0}^{k-1}\binom{k}{r+1}\binom{k}{s+1}F_{r+1,s+1}^k
(r+1)(s+1) M^k_{rs}\\ &=\sum_{r,s=0}^{k}\binom{k}{r}\binom{k}{s}F_{r+1,s+1}^k
(k-r)(k-s) M^k_{rs}~~.\\
\end{aligned}
\end{equation*}
The same identity holds when $F_{rs}^k$ and $M_{rs}^k$ are
interchanged. Putting everything together, one finds $\hat{C}_2^{(k)}\left(P^{(k)}_x\right)=\Delta P^{(k)}_x$.

\paragraph{Proposition.} For every $k\geq 1$, the Berezin push $\Delta_k^B$ of
the truncated Laplacian $\Delta_k$ of $\P^n$ coincides
with the second Casimir of $U(n+1)$ in the representation $\End(E_k)$
\beq \nn
\Delta_k^B=\hat{C}_2^{(k)}~~.
\eeq

\paragraph{Proof.} Using the Lemma, we compute:
\begin{equation}
\begin{aligned}
\sigma\left( \Delta_k^B (C)\right)(x)=\Delta_k~\tr\left(P^{(k)}_x C\right)&= \tr\left(\hat{C}_2^{(k)}(P^{(k)}_x)
C\right)\\&=\tr\left(P^{(k)}_x \hat{C}_2^{(k)} (C)\right)=\sigma\left({\hat C_2}^{(k)}(C)\right)(x)~~.\nn
\end{aligned}
\end{equation}
The next to last equality holds due to the form
$\hat{C}_2^{(k)}(C)=\sum_a[\hat{\CL}^a,[\hat{\CL}^a,C]]$ of the Casimir in the
representation $\End(E_k)$. The conclusion of the proposition now follows by 
using injectivity of $\sigma$.

\

We next consider the Berezin-Toeplitz lift of $\Delta$.  Since $\P^n$ is a
K{\"a}hler homogeneous space, the difference between
the Berezin-Toeplitz lift and the Berezin push of $\Delta_k$ is an
$\ell$-dependent rescaling on the eigenfunctions\footnote{For brevity,
we will always denote the hyperspherical harmonics on $\P^n$ by $Y_{\ell M}$,
where $\ell$ is the angular momentum $\Delta Y_{\ell M}=\ell(\ell+n)Y_{\ell
M}$ and $M$ is a multi-index capturing all further labels. We work with the
normalization $\frac{1}{\vol_{\omega_{FS}}(P^n)}\int\frac{\omega_{FS}^n}{n!}
Y_{\ell M}Y_{\ell'M'}=\delta_{\ell\ell'}\delta_{MM'}$. A detailed discussion
can be found e.g.\ in \cite{Dolan:2006tx}.} $Y_{\ell M}$.  Let us show this
more explicitly. To be concise, we will rely
on results presented e.g.\ in \cite{Balachandran,Dolan:2006tx}, to which we
refer the reader for further details.

First, note that $\End(E_k)$ is spanned by operators $\hat{Y}_{\ell M}$, $\ell=0,...,k$, called
{\em polarization tensors}. These are the operator analogues of hyperspherical harmonics and satisfy:
\begin{equation} \label{eqn:pol_tensors}
 \hat{C}_2^{(k)}(\hat{Y}_{\ell M})=\ell(\ell+n)~~.
\end{equation}
The multi-index $M$ captures the same indices as for $Y_{\ell M}$. 
The polarization tensors are orthogonal with respect to the Hilbert-Schmidt scalar product, and we choose the normalization (cf.\ \cite{Dolan:2006tx}):
\begin{equation} \nn
 \frac{1}{\dim (\End(E_k))} \tr(\hat{Y}_{\ell M}\hat{Y}_{\ell' M'})=\delta_{\ell\ell'}\delta_{M M'}~~.
\end{equation}
Direct computation gives the relation \cite{Dolan:2006tx}:
\begin{equation} \label{eqn:T_factor}
 P^{(k)}_x=\sum_{k,\ell,M} T^{1/2}_{k,n}(\ell)Y_{\ell M}(x) \hat{Y}_{\ell M}~~,~~~
T_{k,n}(\ell):=\frac{k!(k+n)!}{n!(k-\ell)!(k+\ell+n)!}~~,
\end{equation}
from which we conclude that the Berezin symbols of the polarization tensors are:
\begin{equation} \nn \sigma(\hat{Y}_{\ell M})=\tr(P^{(k)}_x
\hat{Y}_{\ell M})=\dim(\End(E_k))T_{k,n}^{1/2}(\ell)Y_{\ell M}~~.
\end{equation} 
Using this,  one readily computes:
\begin{equation} \nn
\frac{1}{\dim(\End(E_k))}\tr(\hat{Y}_{\ell'M'}T(Y_{\ell M}))=\vol_{\omega_{FS}}(\P^n) T^{1/2}_{k,n}(\ell)\delta_{\ell\ell'}\delta_{MM'}~~, 
\end{equation}
from which it follows that the Toeplitz quantization of $Y_{\ell M}$
is given by:
\begin{equation} \nn
 T(Y_{\ell M})= \vol_{\omega_{FS}}(\P^n)T^{1/2}_{k,n}(\ell)\hat{Y}_{\ell M}~.
\end{equation}
Hence the numbers $\beta_k(\theta_\ell)$ (cf.\ Section 6.2), where
$\theta_\ell$ refers to the irrep of $SU(n+1)$ with Dynkin labels
$(\ell,0,...,0,\ell)$, are given by:
\begin{equation} \nn
 \beta_k(\theta_\ell)=\frac{\sigma(T(Y_{\ell M}))}{Y_{\ell M}}=\dim(\End(E_k))\vol_{\omega_{FS}}(\P^n)T_{k,n}(\ell)~~.
\end{equation}
Note that $\beta_k(\theta_\ell)$ are real and positive. As explained in the
previous subsection, this means that both the Berezin pull and the
Berezin-Toeplitz lift are reasonable candidates for the quantized Laplacian in
this case. It is now trivial to compute:
\begin{equation} \nn
\tr(\hat{Y}_{\ell M}\hat{C}_2^{(k)}\hat{Y}_{\ell' M'})=\dim(\End(E_k))\ell(\ell+n)\delta_{\ell
\ell'}\delta_{M M'}~
\end{equation} 
and:
\begin{equation} \nn
\tr(\hat{Y}_{\ell M}\hat{\Delta}_k\hat{Y}_{\ell' M'})=\dim(\End(E_k))^2\vol_{\omega_{FS}}(\P^n)T_{k,n}(\ell)\ell(\ell+n)\delta_{\ell \ell'}\delta_{MM'}~~.
\end{equation} 
It follows that on $\P^n$, $\hat{\Delta}_k$ and
$\Delta_k^B=\hat{C}_2^{(k)}$ are related to each other via the positive
$\ell$-dependent rescaling factors:
\begin{equation} \label{eqn:T_rescale}
 \tilde{T}_{k,n}(\ell):=\dim(\End(E_k))\vol_{\omega_{FS}}(\P^n)T_{k,n}(\ell)~~.
\end{equation}

\subsection{Approximating the spectrum of the Laplacian on Fermat curves}

In spite of its shortcomings, the truncated Laplacian (equivalently, its
Berezin push) can be used to approximate the spectrum of the classical
Laplacian. When the generalized Berezin quantization is
chosen such that $\overline{\cup_{k=0}^\infty \Sigma_k}={\cal C}^\infty(X)$,
the operators $\Delta_k$ can be viewed as approximations to the full
Laplacian and the spectrum of the latter can be approximated by
computing the spectra of $\Delta_k$. This happens, for example, when
$\Sigma_k$ are the symbol spaces of the Berezin-Bergman
quantization, since in that case the union of $\Sigma_k$ is dense in ${\cal
  C}^\infty(X)$ (see Section 5).  

To be more explicit, we will consider the
Berezin-Bergmann quantization of {\em Fermat curves}, i.e.\ the projective
algebraic curves $X_p\subset \P^2$ given by the
equation:
\begin{equation}
 f(z_0,z_1,z_2)\ :=\ z_0^p+z_1^p+z_2^p \ =\ 0~~,\label{eqn:fermat_curve}
\end{equation}
where $(z_0,z_1,z_2)$ are homogeneous coordinates on $\P^2$. These curves
are non-singular and of genus $(p-1)(p-2)/2$. In the following, we will
restrict our attention to the cases $p=2$ (the {\em conic}) and
$p=3$ (the Fermat elliptic curve).

We endow $X_p$ with the Bergman metric given by the
pull-back of the Fubini-Study metric via the inclusion map $i:X_p\embd
\P^2$. It will be convenient to cover\footnote{This covering together with
the integration method we use has been considered, e.g., in \cite{Braun:2007sn}.} $\P^2$ by the patches $U_{ijk}$:
\begin{equation}
U_{ijk}\ :=\ \{ z\in \P^2~ |~ |z_i|\geq |z_j| \geq |z_k| \}~~,\label{eqn:patches}
\end{equation}
where we choose the normalization $|z_i|=1$ and denote the resulting
coordinates by $z_m^{(ijk)}$. One easily checks that the patches intersect only on
their boundaries.

Note that  $x=z^{(ijk)}_k$ is a good local
coordinate on the patch $U_{ijk}$. The pull-back of the Fubini-Study metric is easily calculated by
noting that:
\begin{equation}
 \derr{z_j^{(ijk)}}{x}=-\left(\derr{f}{x}\right)\left(\derr{f}{z_j^{(ijk)}}\right)^{-1}~~,\nn
\end{equation}
which is a consequence of $f=0$. The patches are chosen such that the
pull-back $i^*\omega(x)=:w(x) dx\wedge d\bar{x}$ is always well defined. The
Laplacian is given by:
\begin{equation}
 \Delta\ :=\ \frac{1}{w(x)} \der{x}\der{\bar{x}}~~.\nn
\end{equation}

We approximate the integrals in (\ref{L2-projection}) by summing over the
integrand evaluated at a random sample of $N$ points on each patch and summing
over patches. To generate the sample points, we proceed as follows, cf.\
\cite{Braun:2007sn}. On the patch $U_{ijk}$ we pick a point $z_k^{(ijk)}$ in
the unit disk: $|z_k^{(ijk)}|<1$. The coordinate $z_j^{(ijk)}$ is evaluated
as:
\begin{equation}
 z_j^{(ijk)}\ =\ \phi_{\mathrm{rnd}}\left[-1-(z_k^{(ijk)})^3\right]^{\frac{1}{p}}~~,\nn
\end{equation}
where $\phi_{\mathrm{rnd}}$ is a uniformly chosen random $p$-th root of
unity. If $1\geq|z_j^{(ijk)}|\geq|z_k^{(ijk)}|$, we include the point
$(1,z_j^{(ijk)},z_k^{(ijk)})$ in the set of sample points; otherwise we pick a
new one. We then use the formula:
\begin{equation}
\nn
\int_{U_{ijk}} \omega f(x)\approx \frac{1}{6N}\sum_{n=1}^N f(x_n) w(x_n)~,
\end{equation}
The total integral is obtained by summing over all $6$ patches.

Our set $\Sigma_k$ is the set of symbols of Berezin-Bergman quantization at
level $k$. Thus we consider the polynomial ring
$B=\oplus_{k=0}^\infty B_k$ on $\P^2$ and the vanishing ideal $I=(f)=\oplus_{k=0}^\infty
I_k$ with $I_k\subset B_k$. The space of endomorphisms of $E_k=B_k/I_k$ is identified with $\Sigma_k$. A basis
$(e_i)$ for $\Sigma_k$ is constructed from a set of monomials $(\chi_\alpha)$
of degree $k$ forming a basis of $E_k$ by considering all pairs $\nu_k\,
\chi_\alpha\bar{\chi}_\beta$, where $\nu_k$ is the normalization factor:
\begin{equation}
 \nu_k\ :=\ \frac{1}{||z||^{2k}} \ = \ \frac{1}{\left(1+z_j^{(ijk)}\bz_j^{(ijk)}+z_k^{(ijk)}\bz_k^{(ijk)}\right)^k}~~. \label{eqn:normal_factor}
\end{equation}
The projector $\pi_k:\CC^\infty(X_p)\rightarrow \Sigma_k$ is defined by the
integral expression \eqref{L2-projection}.

Given an arbitrary, not necessarily orthonormal, basis $(e_i)$ of $\Sigma_k$,
we expand the eigenvalue equation $\Delta f_m = \lambda_m f_m$ in the
following manner:
\begin{equation}
\label{evprob}
\sum_i\langle e_j| \Delta |e_i\rangle\langle e_i | f'_m\rangle = \lambda_m \sum_i\langle e_j|e_i\rangle\langle e_i |f'_m\rangle~~,
\end{equation}
where $\langle~,\rangle$ denotes again the scalar product with respect to the
$L^2$-norm and $f'_m$ is the (unique) function such that $f_i=\sum_i
e_i\langle e_i |f'_m\rangle$.  After defining the vector $\vec{f}'_m=(\langle
e_i |f'_m\rangle_i$), the eigenvalue problem \eqref{evprob} turns into the
form:
\begin{equation}
 A \vec{f}'_m=\lambda_m B \vec{f}'_m~~,\nn
\end{equation}
where $A$ and $B$ are matrices, $B$ being invertible. The eigenvalues
$\lambda_m$ of the Laplace operator are therefore the eigenvalues of the
matrix $B^{-1}A$. Note that this procedure is in fact equivalent to the one used in \cite{Brauntbp}.

The numerical results\footnote{Our computations are merely
a demonstration of principle, as the algorithm is run on a laptop using
Mathematica. Switching to C and using more powerful computers, one can easily
increase the precision.} for $k\leq 2$ are presented in tables \ref{res1} and
\ref{res2}. For comparison, we ran our algorithm also for the space
$\P^2$ itself, choosing the same patches (in that case, $E_k$ is the
space of linear endomorphisms of $B_k$); the results of this are
shown in table \ref{resCP2}. In all cases, the integration was performed using
$10,000$ points per patch. The spectra of the Laplace operators
for $k\leq 4$ are displayed in figure 1.

\begin{figure}[ht]
\center
\begin{picture}(440,150)
{\epsfig{file=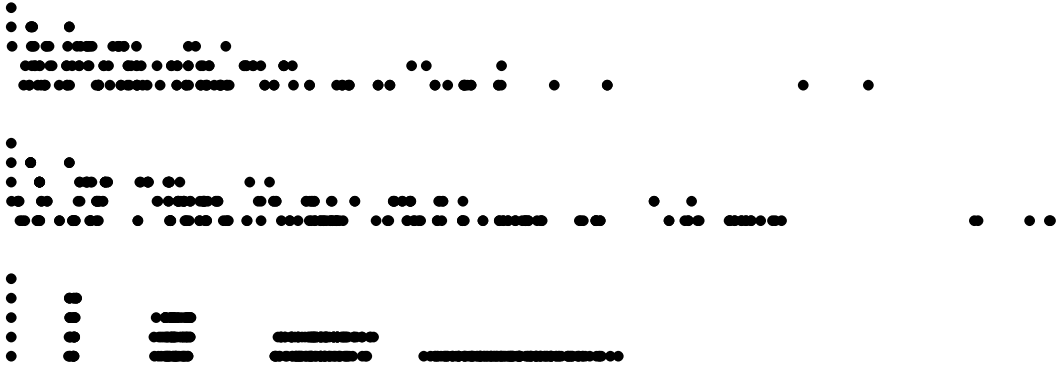, scale=1.7}}
\end{picture}
\caption{Spectra of the Laplace operator on the Fermat curves $X_2$, $X_3$ and $\P^2$ (from top to bottom) for
  $k=0,1,2,3,4$. The spreading of the eigenvalues is related to numerical
  errors.}
\end{figure}

\begin{table}[ht]
\begin{center}
\begin{tabular}{|c|l|}
 \hline
 $k$ & Eigenvalues of $\Delta$\\
\hline
 0 & ~~0\\
\hline
 1 & \begin{tabular}{llllll} 
3.00086 & 3.00081&&&&\\
1.00611 & 1.00588 & 1.00404 & 1.00404& 0.992317 & 0.992314\\ 0 &&&&&\end{tabular}
\\
\hline
2 & \begin{tabular}{lllllll}
11.0669 & 9.51623 &  9.13809\\ 
6.46006 & 5.8219 & 5.8219 & 5.5911 & 5.50849 &
5.50849 & 5.24445\\
4.17542 & 4.06504 & 3.9937 & 3.90068 & 3.90068 \\
3.58166 & 3.3962 & 2.91584\\ 
1.95561 & 1.83634 & 1.79089 & 1.19609 & 1.10872 & 1.0353\\
0.0465755
\end{tabular}\\
\hline
\end{tabular}
\end{center}
\caption{Results for the eigenvalues of the Laplace operator on $X_2$.}
\label{res1}
\end{table}
\begin{table}[ht]
\begin{center}
\begin{tabular}{|c|l|}
 \hline
 $k$ & Eigenvalues of $\Delta$\\
\hline
 0 & ~~0\\
\hline
 1 & \begin{tabular}{llllll} 
3.00086 & 3.00081&&&&\\
1.00611 & 1.00588 & 1.00404 & 1.00404& 0.992317 & 0.992314\\ 0 &&&&&\end{tabular}
\\
\hline
 2 & \begin{tabular}{lllllll}
13.3283 & 12.3062 & & & & &\\
8.70644 & 8.1688 & 8.16022 & 8.14907 & 8.13025 &
8.12288 & 8.06782\\
7.08575 & 7.05761 & 7.04119 & 7.04118 & 6.67228 & 6.63273 & \\ 
4.97054 & 4.9572 & 4.85558 & 4.84641 & 4.84596 & 4.84476 &
4.16108\\
4.10807 & 3.92685 & 3.90904 & 3.87988 & 3.85214 & 3.53332 & 3.52194\\
1.47074 & 1.46942 & 1.46242 & 1.4616 & 1.45804 & 1.45203 &\\
0.00002 & & & & & &
\end{tabular}\\
\hline
\end{tabular}
\end{center}
\caption{Results for the eigenvalues of the Laplace operator on $X_3$.}
\label{res2}
\end{table}
\begin{table}[ht]
\begin{center}
\begin{tabular}{|c|l|}
 \hline
 $k$ & Eigenvalues of $\Delta$\\
\hline
 0 & ~~0\\
\hline
 1 & \begin{tabular}{lllllllll} 
3.37386 & 3.35758& 3.34758 & 3.23647 &  
3.23203 & 3.03376 & 3.00594 & 2.99832\\ 
0 &&&&&
\end{tabular}
\\
\hline
2 & \begin{tabular}{lllllllll}
9.26534 & 9.26534 & 9.231 & 9.16842 & 9.08967 & 9.08967 & 9.08707 & 9.03964\\
8.95108 & 8.77926 & 8.60043 & 8.59167 & 8.43174 & 8.40971 & 8.37658 &
8.3692\\ 8.30439 & 8.28543 & 8.25084 & 8.25084 & 8.1738 & 7.97861 & 7.95145 & 7.95145\\
7.94024 & 7.94024 & 7.48435\\
3.30249 & 3.26888 & 3.25238 & 3.23301 & 3.22454 & 3.06392 & 3.02943 & 3.01363\\
0
\end{tabular}\\
\hline
\end{tabular}
\end{center}
\caption{Eigenvalues of the Laplace operator on $\P^2$ as computed by our
algorithm. The exact eigenvalues joining the spectrum at level $k$ are given
by $k(k+2)$ with a degeneracy of $(1+k)^3$.}
\label{resCP2}
\end{table}

\subsection{Fuzzy real scalar field theory on compact Hodge manifolds}

Ordinary real scalar field theory on $(X,\omega)$ is defined by the Euclidean action functional:
\begin{equation}
\label{S}
 S[\phi]\ =\ \frac{1}{\vol_{\omega}(X)}\int_X \frac{\omega^n}{n!}\left(\phi\Delta
   \phi+V(\phi)\right)~~(\phi\in \CC^\infty(X,\R))~~,
\end{equation}
where $\Delta$ is the Laplace operator of $(X,\omega)$, and 
$V(\phi)=\sum_{s=0}^{d} a_s\phi^s$ is a polynomial in $\phi$ of degree $d$ with real coefficients $a_k\in \R$. Notice that we
include a possible mass term for $\phi$ as a quadratic contribution to
$V$. Since $X$ is a compact space, potentials $V$ of odd degree are in
principle allowed, though the consistency of the corresponding quantum
theory depends on a detailed analysis of quantum effects. 

The discussion in the previous subsections allows us to define a ``fuzzy'' version of the action (\ref{S}) as follows: 
\beq
S_k(\Phi):=\tr\left[\Phi{\hat \Delta}_k(\Phi)+V(\Phi)\right]~~,\label{Sk}
\eeq
where $\Phi\in \End(E_k)$. The reality condition ${\bar \phi}=\phi$ is
replaced by the $\langle~,~\rangle_{HS}$-hermiticity requirement $\Phi^\dagger=\Phi$.
Using the relation $\sigma_k({\hat \Delta}_k
\Phi)=\Delta_{\diamond_k}\sigma_k(\Phi)$ (see (\ref{diamond})), we
find: 
\beq
S_k(\Phi)=S_k^\diamond(\sigma_k(\Phi))~~,\label{Skdiamond}
\eeq
where:
\beq
S_k^\diamond[\phi]:=\frac{1}{\vol_\omega(X)}\vint_k \left[\phi
  \diamond_k \Delta_{\diamond_k} \phi+V_{\diamond_k} (\phi)\right]=
\frac{1}{\vol_\omega(X)}\int_{X}\frac{\omega^n}{n!}\epsilon_k\left[\phi
  \diamond_k \Delta_{\diamond_k} \phi+V_{\diamond_k} (\phi)\right]~~~~~~\nn
\eeq
for $\phi\in \Sigma_k$ such that ${\bar \phi}=\phi$. Here
$V_{\diamond_k}(\phi):=\sum_{s=0}^{2d} a_s\phi^{\diamond_k s}$, where $\phi^{\diamond_k s}:= \phi\diamond_k \ldots \diamond_k \phi$ ($s$ times) and we used relation (\ref{vint}). 

Working with the finite dimensional space
$\Sigma_k\cong \End(E_k)$ reduces the functional integral 
$\int {\cal D}[\phi]$ in the definition of the partition function:
\begin{equation}
Z = \int {\cal D}[\phi]~~ e^{-S[\phi]}
\end{equation}
to a well-defined finite dimensional integral $Z_k$. (On $\P^n$, for example,
the functional measure ${\cal D}[\Phi]$ becomes the Dyson measure on the space
of Hermitian operators on $E_k$). Hence $Z_k$ provide {\em
regularizations} of the quantum field theory defined by (\ref{S}). These
regularized field theories are known in the literature as fuzzy scalar field
theories\footnote{See \cite{Balachandran2} for more details on this point.}.

\paragraph{Remark.} Let $\rho_k$ be the Toeplitz quantization of the function
$\frac{1}{\vol_\omega(X)\epsilon_k}$ at level $k$:
\beq
\rho_k:=T_k \left(\frac{1}{\vol_\omega(X)\epsilon_k}\right) =\frac{1}{\vol_\omega(X)}\int_{X}\frac{\omega^n}{n!}P^{(k)}_x\in \End(E_k)~~. \label{eqn:rho_toeplitz}
\eeq
Clearly $\rho_k$ is Hermitian and strictly positive on
$(E_k,\langle~,~\rangle_k)$. Furthermore $\tr(\rho_k)=1$,
so $\rho_k$ is a density operator on $E_k$. For any operator $C\in \End(E_k)$, we have: 
\beq
\tr\left(\rho_k C\right)=\frac{1}{\vol_\omega(X)}\int_{X}\frac{\omega^n}{n!}\sigma_k(C)~~. \nn
\eeq
Hence the operator $\rho_k$ allows us to remove the epsilon function from the integral.

\section{Directions for further research}

Generalized Berezin quantization raises a series of natural questions about
the asymptotic behavior of the quantization maps $Q_k$ for large $k$ as a
function of the defining sequence of scalar products $(~,~)_k$. In particular,
one would like to know what conditions should be imposed on the large $k$
behavior of these scalar products in order to ensure that the generalized
quantization prescription induces a formal star product on ${\cal
C}^\infty(X)$ and thus defines a formal deformation quantization. Other
natural questions involve the relation with Chow-Mumford stability and
K-stability and with approximation theorems for K{\"a}hler metrics of constant
scalar curvature.  An important set of applications concerns the quantization
of toric varieties, and the extension to the singular case.

The definition of the fuzzy Laplace operator as the Berezin-Toeplitz lift of the
classical Laplacian remains somewhat ad hoc. A better understanding of the
quantization of the classical Laplacian $\Delta$ seems to require the
quantization of differential forms and the construction of a quantum analogue
of a volume form. 

It would also be interesting to examine the relevance of our general quantized spaces
within string theory. In particular, one could study the extension of the Myers effect
\cite{Myers:1999ps} to  more general Hodge manifolds.

\acknowledgments
DM is supported by an IRCSET (Irish Research Council for Science, Engineering
and Technology) postgraduate research scholarship. CS is supported by an IRCSET postdoctoral fellowship.

\section{Glossary of notation}

Notation is in page order of definition (rather than first appearance) excluding the introduction. Note that $\langle~,~\rangle_k$ indicates an \emph{induced} Hermitian scalar product on $E_k$ depending on the situation (usually the $L^2$-scalar products $\langle~,~\rangle^h_k$) while $(~,~)_k$ is an arbitrary sequence of Hermitian scalar products on $E_k$ used in the generalized quantization procedures.

\scriptsize
\renewcommand{\arraystretch}{1.25}
\begin{center}
\noindent \begin{longtable}{cp{3.5in}@{\ \ }@{\extracolsep{\fill}}c} 
\toprule[1pt]
\textbf{Notation} & \textbf{Explanation} & \textbf{Page (Eqn)} \\
\midrule
$X$ & compact complex manifold, usually K\"ahler and/or Hodge & 5 \\
$(X,L)$ 	& polarized complex manifold 	& 7 \\
$R(X,L)$ & homogeneous coordinate ring of $(X,L)$ embedded in $\P V$ & 10 \\
$\omega$	& K\"ahler form, usually $L$-polarized & 5\\
$\omega_{FS}$ & K\"ahler form of Fubini-Study metric & 11, 41 \\
$(X, L, \omega)$ & polarized Hodge manifold & 5 \\
$L^k$		& $L^k := L^{\otimes k}$ & 5 \\
$\Gamma (L^k)$	& space of smooth sections of $L^k$ (contains $E_k$) & 6 \\
$(L, h)$	& Hermitian holomorphic line bundle on $(X, \omega)$, ``prequantum bundle'' & 6 \\
$(X, \omega, L, h)$ & prequantized Hodge manifold & 6 \\
$\Aut(X,\omega,L,h)$ & automorphism group of a prequantized Hodge manifold & 6 \\
$\Aut_{L,h}(X,\omega,L,h)$ & $\Aut(X,\omega,L,h)/U(1)$, subgroup of isometries admitting a lift & 7 \\
$\nabla$ 	& Chern connection associated to $(L,h)$ & 5 \\
$\nabla_k$ 	& Chern connection associated to $(L^k,h_k)$ & 6 \\
$F$ 		& curvature of $\nabla$ & 5 \\
$F_k$ 		& curvature of $\nabla_k$ & 6 \\
$h_k$		& $h_k :=h^{\otimes k}$, Hermitian scalar product on $L^k$ & 6 \\
$h_B$ & Bergman metric & 8 (\ref{hatbergman}) \\
$h_{FS}$ & Hermitian metric on $H$, $h_{FS}^k:= h_{FS}^{\otimes k}$ & 41 (\ref{eqn:hypermetric}) \\
$h$ & induced Bergman Hermitian scalar product on $L$ & 46 \\
$\mu$ 		& positive (Radon) measure on $X$ & 6 \\
$\mu_\epsilon$ & $\mu_\epsilon:= \mu \epsilon$ & 22 \\
$\mu_h$ & $\mu_h := \mu_{\omega_h}$, Liouville measure defined by $\omega_h$ & 23 \\
$L^2 (X,h,\mu)$ & $L^2$-completion of $\Gamma (L^k)$ with respect to $\langle~,~\rangle_k^{\mu,h}$ & 6 \\
$\rho_k$ 	& group action of $\Aut(X,\omega,L,h)$ on $\End(H^0 (L^k))$ & 6 (\ref{rho}) \\
$E_k$ & space of holomorphic sections of $L^{\otimes k}$ & 12 \\
$\tau$ & action of $\Aut_{L,h}(X,\omega)$ on $C^\infty (X)$ & 17 (\ref{eqn:tau}) \\
${\cal E}_X$ & Hilbert direct sum of $E_k$, ${\cal E}_X:= \overline{\oplus}^\infty_{k=0} (E_k, \langle~,~\rangle_k)$ & 31 (\ref{eqn:hilbert_direct_sum}) \\
$B$ & symmetric algebra associated to $E$, $B:=\oplus_{k=0}^\infty E^{\odot k}$ & 35 (\ref{eqn:symmetric_algebra}) \\
${\cal B}(V)$ & weighted Bargmann space ${\cal B}(V):=L^2_{\rm hol}(V, d\nu)$ of $\nu$-square integrable entire functions on $V$ & 37 \\
${\cal L}({\cal B})$ & algebra of bounded operators on ${\cal B}$ & 39 \\
$\tau$ & tautological bundle, $\tau={\cal  O}_{\P(V)}(-1)$ & 40 \\
$H$ & hyperplane bundle $H={\cal  O}_{\P(V)}(1)$ dual to $\tau$ & 41 \\
$I$ & graded ideal in $B$ defined by $\phi:R\stackrel{\sim}{\rightarrow}B/I$ & 45 (\ref{phi}) \\
$\hat{C}_2^{(k)}$ & fuzzy Laplacian, second Casimir of $U(n+1)$ in $\hat{\rho}_k$ representation & 55 (\ref{eqn:second_casimir}) \\
$\tau^a_{ij}$ & Gell-Mann matrices of $su(n+1)$ & 55 \\
$Y_{\ell M}$ & hyperspherical harmonics on $\P^n$ & 57 \\
$\hat{Y}_{\ell M}$ & polarization tensors & 57 (\ref{eqn:pol_tensors}) \\
$T_{k,n} (\ell)$ & $T_{k,n}(\ell):=\frac{k!(k+n)!}{n!(k-\ell)!(k+\ell+n)!}$ & 57 (\ref{eqn:T_factor}) \\
 & & \\
\midrule
 & & \\
$\sigma$ & lower Berezin symbol map & 14 (\ref{symbol}) \\
$\sigma_k$ & Berezin symbol maps & 12 \\
$Q$ & generalized Berezin quantization map & 14 (\ref{eqn:gen_berezin_q}) \\
$Q_k$  & Berezin quantization maps & 12 \\
$T$ & generalized Toeplitz quantization map $T:C^\infty (X) \rightarrow \End(E)$ & 24 (\ref{eqn:gen_toep}) \\
$T_k$ & Toeplitz quantization map $T$ at level $k$ & 30 \\
$\beta$ & Berezin transform with respect to $(~,~)$ & 25 \\
$\beta_k$ & Berezin transform $\beta$ at level $k$ & 30 \\
$\beta_{mod}$ & modified Berezin transform & 28 (\ref{mod_Berezin_tf}) \\
$\bbeta$ & formal Berezin transform & 33 (\ref{eqn:formalberezin}) \\
${\cal O}^B$ & Berezin push of ${\cal O}$ & 16 (\ref{eqn:Bpush}) \\
${\cal V}_B$ & Berezin pull of ${\cal V}$ & 16 (\ref{eqn:Bpull}) \\
$\hat{{\cal D}}$ & Berezin-Toeplitz lift of ${\cal D}$ & 27 (\ref{lift}) \\
${\cal D}_\diamond$ & Berezin-Toeplitz transform of ${\cal D}$ & 28 (\ref{diamond}) \\
 & & \\
\midrule
 & & \\
$e_q$ & Rawnsley coherent vector corresponding to $q\in \L_0$ & 13 \\
$e_x$ & Rawnsley coherent state at $x=\pi(q)$ & 14 \\
$e^{(k)}_{v}$ & Rawnsley's coherent vectors in projective case & 43 (\ref{eqn:rawnsleyproj}) \\
$P_x$ & coherent projector & 14 (\ref{eqn:Rawnsley_proj}) \\
$P^{(k)}_{[v]}$ & Perelomov's coherent projectors in projective case & 43 (\ref{eqn:perelomov}) \\
$P^{(k)}_{[v]}$ & Rawnsley coherent projector in Berezin-Bergman quantization & 48 (\ref{eqn:rawnsleyproj_bb}) \\
$P_x^{(k)}$ & Coherent projectors of $\P^n$ at level $k$ & 55 \\
$\Psi$ & squared two-point function & 16 (\ref{two_point}) \\
 & & \\
\midrule
 & & \\
$\langle~,~\rangle_k^{\mu,h}$ & Hermitian scalar product on $\Gamma (L^k)$ with respect to measure $\mu$ & 6 (\ref{L2mu}) \\
$\langle~,~\rangle_k^{h}$ & Hermitian scalar product on $\Gamma (L^k)$ with respect to Liouville measure $\mu_\omega:=\frac{\omega^n}{n!}$ & 6 (\ref{L2}) \\
$\langle~,~\rangle_{k,\sigma}$ & scalar product on space of smooth functions on $U_\sigma$ & 7 (\ref{eqn:Usigma_product}) \\
$\langle~,~\rangle$ & $L^2$-scalar product on $E=H^0(L)$ defined by $h$, i.e. $\langle~,~\rangle := \langle~,~\rangle_1^h$ & 9 \\
$\langle~,~\rangle_{HS}$ & Hilbert-Schmidt operator product & 15 \\
$\langle~,~\rangle_X$ & $\langle~,~\rangle_X:= \sum_{k=0}^\infty \langle~,~\rangle_k$ & 31 (\ref{eqn:direct_sum_sp}) \\
$\langle~,~\rangle_k$ & scalar product on $B_k$ associated to $h_{FS}^k$ & 41 (\ref{eqn:hyperproduct}) \\
$\langle~,~\rangle_B$ & Bargmann product & 42 (\ref{products}) \\
$\epsilon$ & $\epsilon:=\frac{\hat{h}}{\hat{h}_B}$, epsilon function of $h$ relative to $(~,~)$ & 8
(\ref{eqn:relepsilon}) \\
$\epsilon$ & may also refer to \emph{absolute} epsilon function of $h$ & 9 \\
$(~,~)'$ & arbitrary Hermitian scalar product on $E$ distinct from $(~,~)$ & 18 \\
$\prec~,~\succ_B$ & Berezin scalar product & 15 (\ref{diamond_product}) \\
$\prec~,~\succ$ & scalar product on $C^\infty (X)$ induced by $\mu$ & 22 (\ref{mu_product}) \\
$\prec~,~\succ_\epsilon$ & scalar product on $C^\infty (X)$ induced by $\mu_\epsilon$ & 22 (\ref{epsilon_product}) \\
$\epsilon_k^{\P V}$ & epsilon function in projective case & 43 (\ref{proj_epsilon}) \\
 & & \\
\midrule
 & & \\
$\star$ & formal star product & 32 \\
$\star_T$ & Toeplitz star product & 32 \\
$\star_B$ & Berezin star product & 33 \\
$\diamond$ & Berezin product (aka coherent state star product) & 15 (\ref{Ber_product}) \\
$\diamond_k$ & Berezin product (aka coherent state star product) at level $k$ & 34 (\ref{eqn:Bproduct_k}) \\
\bottomrule[1pt]
\end{longtable}
\end{center}
\renewcommand{\arraystretch}{1.00}

\end{document}